\documentclass[%
aip,jap,
 amsmath,amssymb,floatfix,
 reprint,%
]{revtex4-1}

\usepackage[utf8]{inputenc}
\usepackage{graphicx}
\usepackage{dcolumn}
\usepackage{bm}
\usepackage[T1]{fontenc}
\usepackage{mathptmx}
\usepackage{hyperref}
\usepackage{epstopdf}
\usepackage{multirow}
\usepackage{mathtools}
\usepackage{xcolor}
\usepackage{textcomp}
\usepackage{multirow}

\begin{document}

\preprint{AIP/123-QED}

\title{Introduction to Spin Wave Computing}

\author{Abdulqader Mahmoud}
\affiliation{Delft University of Technology, Department of Quantum and Computer Engineering, 2628 CD Delft, The Netherlands}

\author{Florin Ciubotaru}
\affiliation{Imec, 3001 Leuven, Belgium}

\author{Frederic Vanderveken}
\affiliation{KU Leuven, Department of Materials, SIEM, 3001 Leuven, Belgium}
\affiliation{Imec, 3001 Leuven, Belgium}

\author{Andrii V. Chumak}
\affiliation{Faculty of Physics, University of Vienna, 1090 Wien, Austria}

\author{Said Hamdioui}
\affiliation{Delft University of Technology, Department of Quantum and Computer Engineering, 2628 CD Delft, The Netherlands}

\author{Christoph Adelmann}
\email{Christoph.Adelmann@imec.be}
\affiliation{Imec, 3001 Leuven, Belgium}

\author{Sorin Cotofana}
\email{S.D.Cotofana@tudelft.nl}
\affiliation{Delft University of Technology, Department of Quantum and Computer Engineering, 2628 CD Delft, The Netherlands}

\begin{abstract}
This paper provides a tutorial overview over recent vigorous efforts to develop computing systems based on spin waves instead of charges and voltages. Spin-wave computing can be considered as a subfield of spintronics, which uses magnetic excitations for computation and memory applications. The tutorial combines backgrounds in spin-wave and device physics as well as circuit engineering to create synergies between the physics and electrical engineering communities to advance the field towards practical spin-wave circuits. After an introduction to magnetic interactions and spin-wave physics, the basic aspects of spin-wave computing and individual spin-wave devices are reviewed. The focus is on spin-wave majority gates as they are the most prominently pursued device concept. Subsequently, we discuss the current status and the challenges to combine spin-wave gates and obtain circuits and ultimately computing systems, considering essential aspects such as gate interconnection, logic level restoration, input-output consistency, and fan-out achievement. We argue that spin-wave circuits need to be embedded in conventional CMOS circuits to obtain complete functional hybrid computing systems. The state of the art of benchmarking such hybrid spin-wave--CMOS systems is reviewed and the current challenges to realize such systems are discussed. The benchmark indicates that hybrid spin-wave--CMOS systems promise ultralow-power operation and may ultimately outperform conventional CMOS circuits in terms of the power-delay-area product. Current challenges to achieve this goal include low-power signal restoration in spin-wave circuits as well as efficient spin-wave transducers. 
\end{abstract}

\maketitle

\section{Introduction}
\label{sec:introduction}
	
Current computing systems rely on paradigms, in which information is represented by electric charge or voltage, and computation is performed by charge movements. The fundamental circuit element in this framework is the transistor, which can serve both as a switch and an amplifier. Today's large-scale integrated circuits are based on complementary metal-oxide-semiconductor (CMOS) field-effect transistors because of their high density, low power consumption, and low fabrication cost.\cite{device1,device2,lojek_history_2007} Using CMOS transistors, logic gates can be built that represent a full set of Boolean algebraic operations. Such basic Boolean operations are fundamental for the design of mainstream logic circuits and, together with charge-based memory devices, of computing systems.\cite{bool1,bool2}
	
In the first decades after its introduction into the mainstream in 1974, the device density and the performance of the CMOS technology have been steadily improved by geometric Dennard scaling,\cite{Dennard} following the famed Moore's law.\cite{moore_cramming_1998} This progress has been orchestrated first in the USA by the national technology roadmap for semiconductors, and, after 1998, worldwide by the international technology roadmap for semiconductors (ITRS).\cite{ITRS_web} This has allowed CMOS technology to simultaneously drive and respond to an exploding information technology market. Today, CMOS has clearly consolidated its leading position in the digital domain. In the last two decades, CMOS scaling has increasingly required the introduction of disruptive changes in the CMOS transistor and circuit architecture beyond Dennard scaling to sustain Moore's law,\cite{bohr_moores_2011,kuhn_considerations_2012} including \emph{e.g.}~Cu interconnects,\cite{murarka_copper_1995} high-$\kappa$ dielectrics,\cite{bohr_high-k_2007} or the FINFET architecture.\cite{auth_22nm_2012} In the future, CMOS scaling is expected to decelerate\cite{waldrop_chips_2016} mainly due to unsustainable power densities, high source--drain and gate leakage currents,\cite{cmosscaling2,cmosscaling3} reduced reliability,\cite{cmosscaling1} and economical inefficiency.\cite{cmosscaling1,cmosscaling2} Yet, despite the slowdown, Moore's law and CMOS scaling is not expected to end in the next decade and even beyond. The roadmap for future developments is summarized in the International Roadmap for Devices and Systems (IRDS).\cite{IRDS_web}
	
For many years, Moore's law (especially the threat of its end) has been accompanied by research on alternative computing paradigms beyond the CMOS horizon to further improve computation platforms.\cite{feitelson_optical_1988,streibl_digital_1989,hutchby_extending_2002,bourianoff_future_2003,bourianoff_research_2007,miller_are_2010,survey1,survey2,nikonov_benchmarking_2014,survey4,IRDS_web} Recently, this has accelerated due to a surge of interest in non-Boolean computing approaches for machine learning applications.\cite{mitchell_machine_1997,murphy_machine_2012,alpaydin_introduction_2020} Such computing paradigms can be based on devices with transistor functionality (\emph{e.g.}~tunnel FETs)\cite{boucart_double-gate_2007} or alternatives (\emph{e.g.}~memristors).\cite{jeong_memristors_2016,li_review_2018} Amongst all beyond-CMOS approaches, spintronics, which uses magnetic degrees of freedom instead of electron charge for information coding,\cite{wolf_spintronics:_2001,felser_spintronics_2013, bandyopadhyay_introduction_2015,xu_handbook_2016,sander_2017_2017,dieny_opportunities_2019} has been identified as particularly promising due to the low intrinsic energies of magnetic excitations as well as their collective nature.\cite{survey1,survey2,nikonov_benchmarking_2014,atulasimha_nanomagnetic_2016,manipatruni_beyond_2018} Numerous implementations of spintronic Boolean logic devices have been investigated based on magnetic semiconductors,\cite{zutic_spintronics:_2004} individual atomic spins,\cite{khajetoorians_realizing_2011} spin currents,\cite{datta_electronic_1990,hall_performance_2006,dery_spin-based_2007} nanomagnets,\cite{cowburn_room_2000,ney_programmable_2003, imre_majority_2006,behin-aein_proposal_2010,manipatruni_scalable_2019} domain walls,\cite{allwood_magnetic_2005, xu_all-metallic_2008,nikonov_proposal_2011} skyrmions,\cite{zhang_magnetic_2015,koumpouras_majority_2018} or spin waves.\cite{logic21,logic1,Magnon_spintronics} While some approaches try to provide transistor-like functionality,\cite{zutic_spintronics:_2004,datta_electronic_1990,hall_performance_2006,dery_spin-based_2007,chumak_magnon_2014} others aim at replacing logic gates rather than individual transistors.\cite{logic21,logic11,logic1,Magnon_spintronics,zografos_exchange-driven_2017} Among the latter group of spintronic logic gates, majority gates have received particular attention due to the expected simplification of logic circuits.\cite{logic1,nikonov_benchmarking_2014,radu_spintronic_2015,zografos_spin-based_2019} While majority gates have been researched for decades,\cite{hampel_threshold_1971} their CMOS implementation is inefficient and therefore has not been widely used in circuit design. However, the advent of compact (spintronic) majority gates has recently led to a revival of majority-based circuit synthesis.\cite{MAJ1,MAJ2,radu_spintronic_2015}
	
A group of disruptive spintronic logic device concepts have been based on spin waves as information carriers.\cite{Magnonics,Serga10, lenk_building_2011,demokritov_magnonics_2013,krawczyk_review_2014,Magnon_spintronics,chumak_magnon_2014,karenowska_magnon_2016,csaba_perspectives_2017,gubbiotti_three-dimensional_2019} Spin waves are oscillatory collective excitations of the magnetic moments in ferromagnetic or antiferromagnetic media \cite{Spin_wave,Bloch,herring_theory_1951} and are introduced in more detail in Sec.~\ref{sec:Basics and background of SW technology}. As their quanta are termed \emph{magnons}, the field of is also often referred to as \emph{magnonics}. The frequency of spin waves in ferromagnets is typically in the GHz range, their intrinsic energy is low ($\sim$ \textmu{}eV for individual magnons), and their propagation velocity can reach values up to several km/s (\textmu{}m/ns). At low amplitude, spin waves are noninteracting, enabling multiplexing and parallelism in logic devices and interconnections.\cite{parallel_data_processing1} By contrast, spin waves can exhibit nonlinear behavior at high amplitudes (Sec.~\ref{Sec_NL_SW_phys}), which can be exploited in spintronic devices and circuits (Secs.~\ref{sec:General Spin Wave device structure} and \ref{sec:Requirements for Spin Wave Circuit Design}). As shown in Sec.~\ref{Sec_SW_Logic_gates}, spin waves are especially suitable for the implementation of compact majority gates due to their wave-like nature. Their short wavelengths down to the nm range at microwave (GHz) frequencies allow for the miniaturization of the devices while keeping operating frequencies accessible.

\begin{figure}[t]
\includegraphics[width=8.5cm]{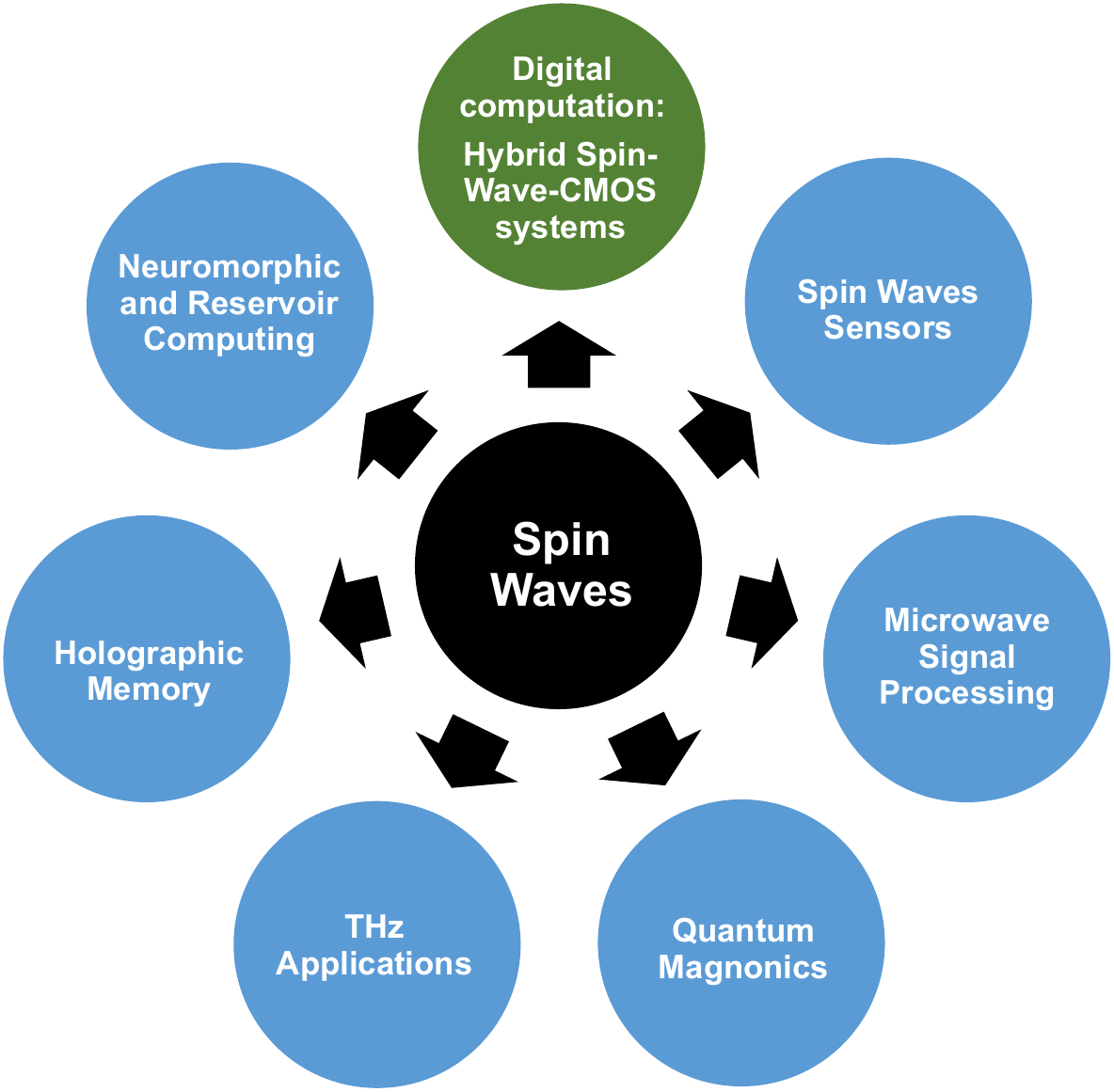}
\caption{Overview of different envisaged applications of spin waves. This tutorial focuses on applications in digital logic based on hybrid spin-wave--CMOS computing systems. Other applications fields are reviewed briefly in Sec.~\ref{Sec_beyond}.}
\label{fig:overview}
\end{figure}

In the last two decades, magnetic devices have been successfully commercialized for nonvolatile memory applications (magnetic random-access memory, MRAM)\cite{gallagher_development_2006,chappert_emergence_2007,kent_new_2015,dieny_introduction_2016,bhatti_spintronics_2017} and as magnetic sensors.\cite{lenz_review_1990,parkin_magnetically_2003,ripka_advances_2010} Yet, despite tremendous progress in the theory and numerous proof-of-concept realizations of spintronic and magnonic logic devices, no competitive spintronic or magnonic logic circuits have been demonstrated to date. It is clear that the step from individual basic spintronic device concepts to operational circuits and systems is large and an additional complementary effort is still required to successfully compete with CMOS in practice. Such an effort is inherently multidisciplinary and needs to involve both spin-wave and device physics as well as circuit and systems engineering. This paper provides a tutorial introduction to spin-wave computing technology and its potentials from a circuit and computation viewpoint. The focus is on the achievements but also on the gaps in the current understanding that still inhibit the realization of practical competitive spin-wave circuits. The main goal of the tutorial is to provide simultaneous insight in the underlying physics and the engineering challenges to facilitate mutual synergistic interactions between the fields. The paper starts with an introduction to the physics of spin waves (Sec.~\ref{sec:Basics and background of SW technology}). Subsequently, the computation paradigm based on spin waves is introduced and the fundamental requirements for the realization of spin-wave circuits are discussed (Sec.~\ref{sec:SW Computing Background}). Next, we provide an overview of different spin-wave transducers (Sec.~\ref{Sec_transducers}) and devices (Sec.~\ref{sec:General Spin Wave device structure}). This is followed by a discussion of the current understanding of spin-wave circuits (Sec.~\ref{sec:Requirements for Spin Wave Circuit Design}) and computing platforms (Sec.~\ref{sec:Hybrid_Systems}). Beyond digital computation, spin waves have also the potential to be used in a number of additional applications fields in electronics (Fig.~\ref{fig:overview}). This is briefly reviewed in Sec.~\ref{Sec_beyond}. Finally, Sec.~\ref{sec:Conclusions} concludes the paper with an overview of the the state of the art of spin-wave technology and identifies the challenges ahead towards the design and realization of competitive spin-wave-based computing systems.

\section{Physics of spin waves}
\label{sec:Basics and background of SW technology}
	
This section provides an introduction to spin waves and their characteristics. We first start by explaining the relevant basic magnetic interactions, followed by a discussion of the resulting magnetization dynamics. 
	
\subsection{Magnetization and magnetic interactions}

\begin{table*}[t]
\caption{Material properties of representative ferromagnetic materials, as well as propagation properties (group velocity, lifetime, and propagation distance) of surface spin waves with a wavelength of $\lambda = 1$ \textmu{}m in a 500 nm wide and 20 nm thick waveguide (external magnetic bias field $\mu_0H = 100$ mT).}. 
\label{table:1}
\begin{tabular}{lccccccc}
\hline\hline
\multirow{2}{*}{Material} & $M_\mathrm{s}$ & Gilbert damping & Exchange length & Group velocity & Lifetime & Propagation & \multirow{2}{*}{References} \\
 & (MA/m) & $\alpha$ ($\times 10^{-3}$)  & $l_\mathrm{ex}$ (nm) & (\textmu{}m/ns) & (ns) & distance (\textmu{}m) &  \\ \hline
Fe & 1.7 & 60 & 3.4 & 5.8 & 0.08 & 0.5 & \onlinecite{Danan68,Kim93,Kin15,Vavassori15,Bishop77} \\
Co & 1.4 & 5 & 4.8 & 4.6 & 1.2 & 5.5 & \onlinecite{Aus98,Vernon84,Michels00a,Schoen16,Heinrich91} \\
Ni & 0.5 & 45 & 7.4 & 1.1 & 0.3 & 0.3 & \onlinecite{Sun05,Talagala02, Michels00, Walowski08,Danan68} \\
YIG (Y$_3$Fe$_5$O$_{12}$, \textmu{}m films) & 0.14 & 0.05 & 17 & 42 & 600 & 25000 & \onlinecite{Serga10,Cherepanov93,Glass88,Geller57,Klingler15,Bertaut56}\\
YIG (Y$_3$Fe$_5$O$_{12}$, nm films) & 0.14 & 0.2 & 17 & 0.3 & 150 & 44 & \onlinecite{Pirro14,Hahn14,Dubs17,Sun12,Yu14,Onbasli14,Liu14}\\
Permalloy (Ni$_{80}$Fe$_{20}$) & 0.8 & 7 & 6.3 & 2.2 & 1.4 & 3.2 & \onlinecite{Demidov15,Kalarickal06,Sebastian15,Patton68} \\
CoFeB & 1.3 & 4 & 3.9 & 3.9 & 1.7 & 6.6 & \onlinecite{Brunsch79,Liu11,Conca13}\\
Co$_2$(Mn$_\mathrm{x}$Fe$_{1-\mathrm{x}}$)Si & 1.0 & 3 & 4.5 & 2.8 & 2.7 & 7.9 & \onlinecite{Liu09,Trudel10,Oogane10,Sebastian12}\\
\hline\hline    
\end{tabular}
\end{table*}

Magnetic materials contain atoms with a net magnetic dipole moment $\boldsymbol{\mu}$. Therefore, they can be considered as a lattice of magnetic dipoles with specific amplitude and orientation at every lattice site. At dimensions much larger than the interatomic distances, it is more convenient to work with a continuous vector field than with discrete localized magnetic dipoles, \emph{i.e.}~with the so-called semiclassical approximation. The continuous vector field is called the magnetization and is defined as the magnetic dipole moment per unit volume \cite{Magnetism1}
\begin{equation}
\mathbf{M} = \frac{\sum_i \boldsymbol{\mu}_i}{\delta V}\,.
\end{equation}
\noindent At temperatures far below the Curie temperature, the magnetization \emph{norm} is constant throughout the material and is called the saturation magnetization $M_\mathrm{s}$. On the other hand, the magnetization \emph{orientation} can be position dependent and is determined by various magnetic interactions. In the following, the most important magnetic interactions are briefly explained. 

The Zeeman interaction describes the influence of an external magnetic field $\mathbf{H}_\mathrm{ext}$ on the magnetization. The Zeeman energy density (energy per unit volume) is given by
\begin{equation}
\mathcal{E}_\mathrm{Z} = -\mu_0 \mathbf{M}\cdot \mathbf{H}_\mathrm{ext}\,,
\end{equation}
\noindent with $\mu_0$ the vacuum permeability. Hence, the energy is minimal when the magnetization is parallel to the external magnetic field.
	
Apart from external magnetic fields, the magnetization itself also generates a magnetic field, termed the dipolar magnetic field. For a given magnetization state, it is found by solving Maxwell's equations.\cite{Spin_wave} The dipolar magnetic field inside the magnetic material is called the demagnetization field, whereas the field outside is called the stray field. The energy density of the self-interaction of the magnetization with its own demagnetization field is given by 
\begin{equation}
\mathcal{E}_\mathrm{d} = -\frac{\mu_0}{2} \mathbf{M}\cdot \mathbf{H}_\mathrm{d}\,,
\end{equation}
\noindent with $\mathbf{H}_\mathrm{d}$ the demagnetization field. The demagnetization field itself strongly depends on the \emph{shape} of the magnetic element.\cite{Magnetostatics_ref1,Magnetism1} The demagnetization energy is minimal when the magnetization is oriented along the longest dimension of the magnetic object. This magnetization anisotropy is therefore often called shape anisotropy.
	
The crystal structure of the magnetic material can also introduce an anisotropic behavior of the magnetization. This is called magnetocrystalline anisotropy and originates from the spin--orbit interaction, which couple the magnetic dipoles to the crystal orientation.\cite{Ferromagnetism} As a result, the magnetization may have preferred orientations with respect to the crystal structure. Magnetization directions that correspond to minimum energy are called easy axes, whereas magnetization orientations with maximum energy are called hard axes. Different types of magnetocrystalline anisotropy exist, depending on the crystal structure.\cite{Ferromagnetism} As an example, the energy density for uniaxial magnetocrystalline anisotropy can be expressed by
\begin{equation}
\mathcal{E}_\mathrm{ani} = -K_1 (\mathbf{u} \cdot \boldsymbol{\zeta})^2 - K_2 (\mathbf{u}\cdot \boldsymbol{\zeta})^4\,,
\end{equation}
\noindent with $\mathbf{u}$ the easy axis, $\boldsymbol{\zeta} = \mathbf{M}/M_\mathrm{s}$ the magnetization direction, and $K_1$ and $K_2$ the first and second order anisotropy constants, respectively. 
	
It is often convenient to describe magnetic interactions by corresponding effective magnetic fields. The general relation between a magnetic energy density and its corresponding effective field is given by 
\begin{equation}
\label{eq:H_eff}
\mathbf{H}_\mathrm{eff} = -\frac{1}{\mu_0} \frac{d \mathcal{E}(\mathbf{M})}{d\mathbf{M}} \,.
\end{equation}
\noindent For the magnetocrystalline interaction, this becomes
\begin{equation}
\mathbf{H}_\mathrm{ani} = \frac{2K_1}{\mu_0 M_\mathrm{s}}(\mathbf{u} \cdot \boldsymbol{\zeta})\mathbf{u} + \frac{4K_4}{\mu_0 M_\mathrm{s}}(\mathbf{u} \cdot \boldsymbol{\zeta})^3 \mathbf{u} \,.
\end{equation}
\noindent In the case of polycrystalline materials, every grain may possess a different easy axis orientation. Therefore, the average magnetocrystalline anisotropy in macroscopic polycrystalline materials is zero and can be neglected, as it can be for amorphous materials.
	
Another important magnetic interaction is the exchange interaction. It describes the coupling between neighboring magnetic dipoles and has a quantum-mechanical origin. In continuum theory, the exchange energy density is given by 
\begin{equation}
\mathcal{E}_\mathrm{ex} = \frac{A_\mathrm{ex}}{M_\mathrm{s}^2} \left[(\nabla M_\mathrm{x})^2+(\nabla M_\mathrm{y})^2+(\nabla M_\mathrm{z})^2\right]\,,
\end{equation}
\noindent with $A_\mathrm{ex}$ the exchange stiffness constant. In ferromagnetic materials, the exchange stiffness constant is positive, which means that the exchange energy is minimum when the magnetization is uniform. In antiferromagnetic materials, the exchange stiffness constant is negative, and the exchange energy is minimum when neighboring atomic dipoles are antiparallel. The corresponding exchange field is given by
\begin{equation}
\mathbf{H}_\mathrm{ex} = \frac{2A_\mathrm{ex}}{\mu_0 M_\mathrm{s}^2} \Delta \mathbf{M} = l^2_\mathrm{ex} \Delta \mathbf{M} \equiv \lambda_\mathrm{ex} \Delta \mathbf{M}\,,
\end{equation}
\noindent with $\Delta$ the Laplace operator, $\lambda_\mathrm{ex}$ is the exchange constant, and $l_\mathrm{ex}$ the exchange length. This length is typically a few nm (Tab.~\ref{table:1}) and characterizes the competition between the exchange and dipolar interaction. At length scales below $l_\mathrm{ex}$, the exchange interaction is dominant, and the magnetization is uniform. At larger length scales, the dipolar interaction dominates and domains with different magnetization orientations can be formed. 
	
In addition to the previously described interactions, various other interactions exist, such as the Dzyaloshinskii–Moriya interaction or the magnetoelastic interaction. Detailed discussions of the physics of these different interactions can be found in Refs.~\onlinecite{Ferromagnetism,Magnetism1,Magnetostatics_ref1}. Basic notions of the magnetoelastic interaction are also discussed in Sec.~\ref{Sec:ME_transducers}.
	
\subsection{Magnetization dynamics and spin waves}
\label{Magnetization dynamics and spin waves}
	
The dynamics of the magnetization in presence of one or several of effective magnetic fields are described by the Landau–-Lifshitz–-Gilbert (LLG) equation \cite{LL_eq,G_eq}
\begin{equation}
\label{eq:llg}
\frac{d\mathbf{M}}{dt} = -\gamma \mu_0(\mathbf{M} \times \mathbf{H}_{\mathrm{eff}}) + \frac{\alpha}{M_\mathrm{s}} \left( \mathbf{M} \times \frac{d\mathbf{M}}{dt} \right)\,,
\end{equation}
\noindent where $\gamma$ the absolute value of the gyromagnetic ratio, $\mu_0$ is the vacuum permeability, $\alpha$ the Gilbert damping constant, and $\mathbf{H}_{\mathrm{eff}}$ the effective magnetic field. This effective field is the sum of all effective fields due to magnetic interactions and the external magnetic field. Hence, every magnetic interaction contributes to the magnetization dynamics via the cross product of the magnetization with its corresponding effective field. 
	
\begin{figure}[t]
\includegraphics[width=7.5cm]{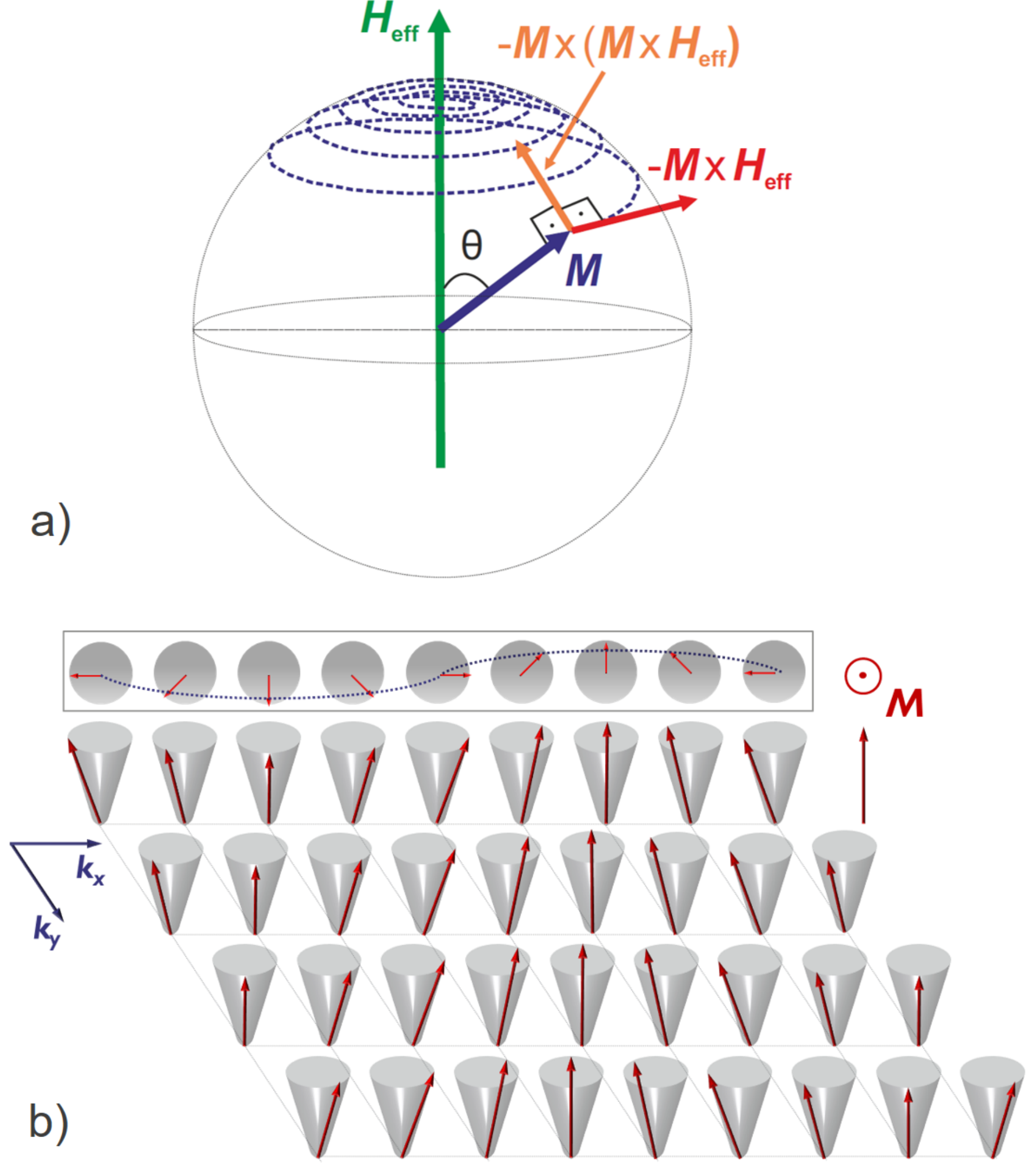}
\caption{Schematic of the magnetization dynamics described by the LLG equation. (a) The trajectory of the magnetization is determined by the combination of two torques [Eq.~(\ref{eq:llg})]: (i) the precessional motion stems from $ \mathbf{M} \times \mathbf{H}_\mathrm{eff}$, whereas (ii) the damping term $\mathbf{M} \times \frac{d\mathbf{M}}{dt} = \mathbf{M} \times (\mathbf{M} \times \mathbf{H}_\mathrm{eff})$ drives the magnetization towards the direction of $\mathbf{H}_\mathrm{eff}$. (b) Schematic representation of a spin wave in a two-dimensional lattice of magnetic moments: top view of the first lattice row (top) and side view of the two-dimensional lattice (bottom).}
\label{fig:spin_precession}
\end{figure}

In equilibrium, the magnetization is parallel to the effective field. However, when the magnetization is not parallel to the effective field, it precesses around this field, as described by the first term in the LLG equation. The second term describes the attenuation of the precession and represents the energy loss of the magnetic excitations into the lattice (phonons) and the electronic system (electrons, eddy currents). All these effects are subsumed in the phenomenological Gilbert damping constant $\alpha$. The combined effect of both terms in the LLG equation results in a spiral motion of the magnetization around the effective magnetic field towards the equilibrium state, as graphically depicted in Fig.~\ref{fig:spin_precession}(a).
	
The LLG equation indicates that small oscillations of the effective magnetic field in time result in a precession of the magnetization. The precession can be either uniform or nonuniform over the magnetic volume. The case of uniform precession with a spatially constant phase is called ferromagnetic resonance. For nonuniform precession, the phase of the precession is position dependent and wave-like excitations of the magnetization exist, called spin waves [see Fig.~\ref{fig:spin_precession}(b)]. Spin waves can thus be considered as stable wave-like solutions of the LLG equation. The ansatz for the magnetization dynamics of a spin wave in a bulk ferromagnet can be written as
\begin{equation}
\label{Eq:magwave}
\mathbf{M}(\mathbf{r},t) = \mathbf{M}_0 + \mathbf{m} = \mathbf{M}_0 + \tilde{\mathbf{m}}e^{i(\omega t + \mathbf{k} \cdot \mathbf{r})} \,,
\end{equation}
\noindent with $M_0$ the static magnetization component, $\omega$ the angular frequency, and $\mathbf{k}$ the wavenumber. The effective magnetic field is then given by
\begin{equation}
\mathbf{H}_\mathrm{eff}(\mathbf{r},t) = \mathbf{H}_0 + \mathbf{h} = \mathbf{H}_0 + \tilde{\mathbf{h}}e^{i(\omega t + \mathbf{k} \cdot \mathbf{r})} \,,
\end{equation}
\noindent with $\mathbf{H}_0$ and $\mathbf{h}$ the static and dynamic components of the effective magnetic field, respectively. As discussed above, this effective magnetic field is the sum of the different effective fields due to the relevant magnetic interactions. 
	
For weak excitations, \emph{i.e.} $||\mathbf{m}|| \ll ||\mathbf{M}_0|| \approx M_\mathrm{s} $, the LLG equation can be linearized by neglecting terms quadratic in $\mathbf{m}$. After a temporal Fourier transform, we obtain
\begin{equation}
\label{eq:lin_LLG}
i\omega \mathbf{m} = -\gamma \mu_0 (\mathbf{M}_0 \times \mathbf{h} + \mathbf{m}\times \mathbf{H}_0) + \frac{i \omega \alpha}{M_\mathrm{s}} (\mathbf{M}_0 \times \mathbf{m}) \,.
\end{equation}
\noindent For specific values of $\mathbf{k}$ and $\omega$, this linearized LLG equation has nontrivial solutions, which represent stable collective magnetization excitations of the form $\tilde{\mathbf{m}}e^{i(\omega(k) t + \mathbf{k} \cdot \mathbf{r})}$, \emph{i.e.}~spin waves. The function $\omega = f(\mathbf{k})$ that relates the spin-wave oscillation frequency to the wavevector is called the dispersion relation. The group velocity of a (spin) wave is defined by the gradient of the dispersion relation, $\mathbf{v}_\mathrm{g} = \nabla_\mathbf{k}\omega$ and represents the direction and the speed of the wave energy flow. By contrast, the phase speed, $\mathbf{v}_\mathrm{p}=\mathbf{k}\omega/||\mathbf{k}||^2$, describes the direction and speed of the wave phase front propagation.
	
As discussed in detail in Sec.~\ref{sec:SW Computing Background}, waveguide structures are of crucial importance for spin-wave devices and circuits. Therefore, in the following, we briefly discuss the behavior of spin waves in waveguides with dimensions comparable or smaller to the wavelength. In such waveguides, the behavior and specifically the dispersion relation of spin waves are strongly affected by waveguide boundaries and lateral confinement effects. Considering a waveguide with a thickness $d$ that is much smaller than its width $w$ and with a rectangular cross section, the spin-wave dispersion relation is given by \cite{dipole_exchange_spin_wave}
\begin{equation}
\label{eq:Disp_Rel}
\omega_\mathrm{n} = \sqrt{(\omega_0 + \omega_\mathrm{M}\lambda_\mathrm{ex}k_\mathrm{tot}^2) (\omega_0+\omega_\mathrm{M}\lambda_\mathrm{ex} k_\mathrm{tot}^2 + \omega_M F)}\,,
\end{equation}
\noindent with $\omega_0=\gamma \mu_0 H_0$, $\omega_\mathrm{M}=\gamma \mu_0 M_0$, and the abbreviations
\begin{equation}
\begin{split}
\label{eq:F}
F = P + \sin^2\phi\times \bigg( 1-P(1+\cos^2(\theta_\mathrm{k} - \theta_\mathrm{M})) + \\
\left. \frac{\omega_\mathrm{M} P(1-P)\sin^2(\theta_\mathrm{k} - \theta_\mathrm{M})}{\omega_0 + \omega_\mathrm{M}\lambda_\mathrm{ex} k_\mathrm{tot}^2} \right)  \,,
\end{split}
\end{equation}
\noindent and
\begin{equation}
\label{eq:g}
P = 1- \frac{1-e^{-dk_\mathrm{tot}}}{dk_\mathrm{tot}}\,.
\end{equation} 
\noindent Here, $k^2_\mathrm{tot} = k^2 + k_\mathrm{n}^2$ with $k_\mathrm{n} =  n\pi/w$ the quantized wavenumber, $n$ is the mode number, $k$ is the wavenumber in the propagation direction, $\theta_\mathrm{k} = \arctan(k_\mathrm{n}/k)$, $\phi$ is the angle between the magnetization and the normal to the waveguide, and $\theta_\mathrm{M}$ is the angle between the magnetization and the longitudinal waveguide axis. Note that this equation is only valid if the waveguide is sufficiently thin, \emph{i.e.}~$kd \ll 1$, and the dynamic magnetization is uniform over the waveguide thickness. We also remark that, depending on the magnetization distribution and the demagnetization field at the waveguide edges, it may be necessary to use an effective width instead of the physical width to accurately describe the dispersion relations.\cite{Guslienko05,Wang19}

\begin{figure}[t]
\includegraphics[width = 7cm]{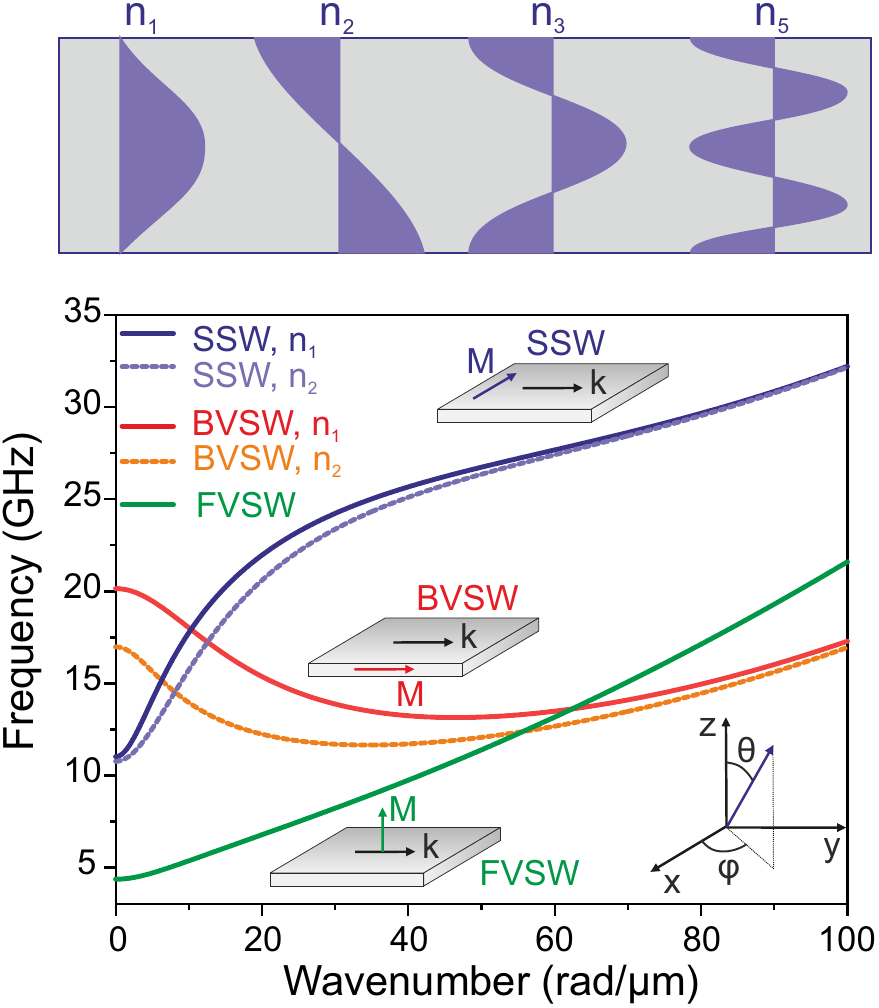}
\caption{Dispersion relation of backward volume spin waves (BVSW), surface spin waves (SSW), and forward volume spin waves (FVSW) in a 500 nm wide and 30 nm thick CoFeB waveguide. For BVSW and SSW, the dispersion relations of the first two laterally confined width modes ($n_1$ and $n_2$) are shown. The material parameters are listed in Tab.~\ref{table:1} and the external magnetic field was $\mu_0H = 100$ mT. The top panel depicts the mode profiles (top view) for confined width modes with mode numbers as indicated. }
\label{fig:spin wave dispersion}
\end{figure}

For short wavelengths (for large $k$), the exchange interaction is dominant. In this limit, the dispersion relation shows a quadratic behavior $\omega_\mathrm{n,ex} = \omega_\mathrm{M}\lambda_\mathrm{ex}k_\mathrm{tot}^2$, independent of the  magnetization orientation. By contrast, for long wavelengths (for small $k$), the dipolar interaction becomes dominant. Then, the dispersion relation is given by $\omega_\mathrm{n,dip} = \sqrt{\omega_0  (\omega_0+ \omega_M F)}$. The factor $F$ strongly depends on the magnetization orientation, indicating that the dipolar interaction leads to anisotropic spin-wave properties. In the limit of infinite wavelengths, the frequency approaches the ferromagnetic resonance frequency, which can be considered as a spin wave with $k = 0$. 
	
Figure~\ref{fig:spin wave dispersion} represents the spin-wave dispersion relations for different geometries in a 500 nm wide CoFeB waveguide (see Tab.~\ref{table:1} for material parameters) for an external magnetic field of $\mu_0H = 100$ mT. In general, the dispersion relation of long-wavelength dipolar spin waves depends on the direction of the wavevector (the propagation direction) and the static magnetization, as described by Eq.~(\ref{eq:Disp_Rel}). It is however instructive to discuss three limiting cases of dipolar spin waves that are often called surface spin waves, forward volume waves, and backward volume waves. 

The first case corresponds to the geometry, in which both the static magnetization and the propagation direction (the wavevector) lie in the plane of the waveguide and are perpendicular to each other, \emph{i.e.}~$\phi=\frac{\pi}{2}$ and $\theta_\mathrm{M}=\frac{\pi}{2}$. Such spin waves are called surface spin waves (SSW) since they decay exponentially away from the surface.\cite{Magnetostatics_ref2} Despite their name, the magnetization can still be considered uniform across the film for sufficiently thin films with $kd \ll 1$. The dispersion relations of the first two SSW width modes ($n_1$ and $n_2$) in a 500 nm wide CoFeB waveguide are depicted in Fig.~\ref{fig:spin wave dispersion} for an external field of $\mu_0H = 100$ mT. The curves indicate that the group and phase velocities are parallel and point in the same direction. 
	
In the second geometry, the static magnetization is both perpendicular to the propagation direction and the waveguide plane, \emph{i.e.}~$\theta_\mathrm{M}=\frac{\pi}{2}$ and $\phi=0$. The spin waves in this geometry have dynamic magnetization components in the plane of the waveguide and a group velocity parallel to the phase velocity. Such spin waves are called forward volume spin waves (FVSW) and their dispersion relation is also represented in Fig.~\ref{fig:spin wave dispersion}. 
	
In the third geometry, the static magnetization is parallel to the propagation direction, both lying in the plane along the waveguide, \emph{i.e.}~$\phi=\frac{\pi}{2}$ and $\theta_\mathrm{M}=0$. In this case, dipolar spin waves have a negative group velocity, which is antiparallel to the positive phase velocity, \textit{i.e.} group and phase velocities point in opposite directions. Therefore, such waves are referred to as backward volume spin waves (BVSW). Their dispersion relation is also depicted in Fig.~\ref{fig:spin wave dispersion} for the first two width modes ($n_1$ and $n_2$). 
	
\begin{figure}[t]
\includegraphics[width=7.2cm]{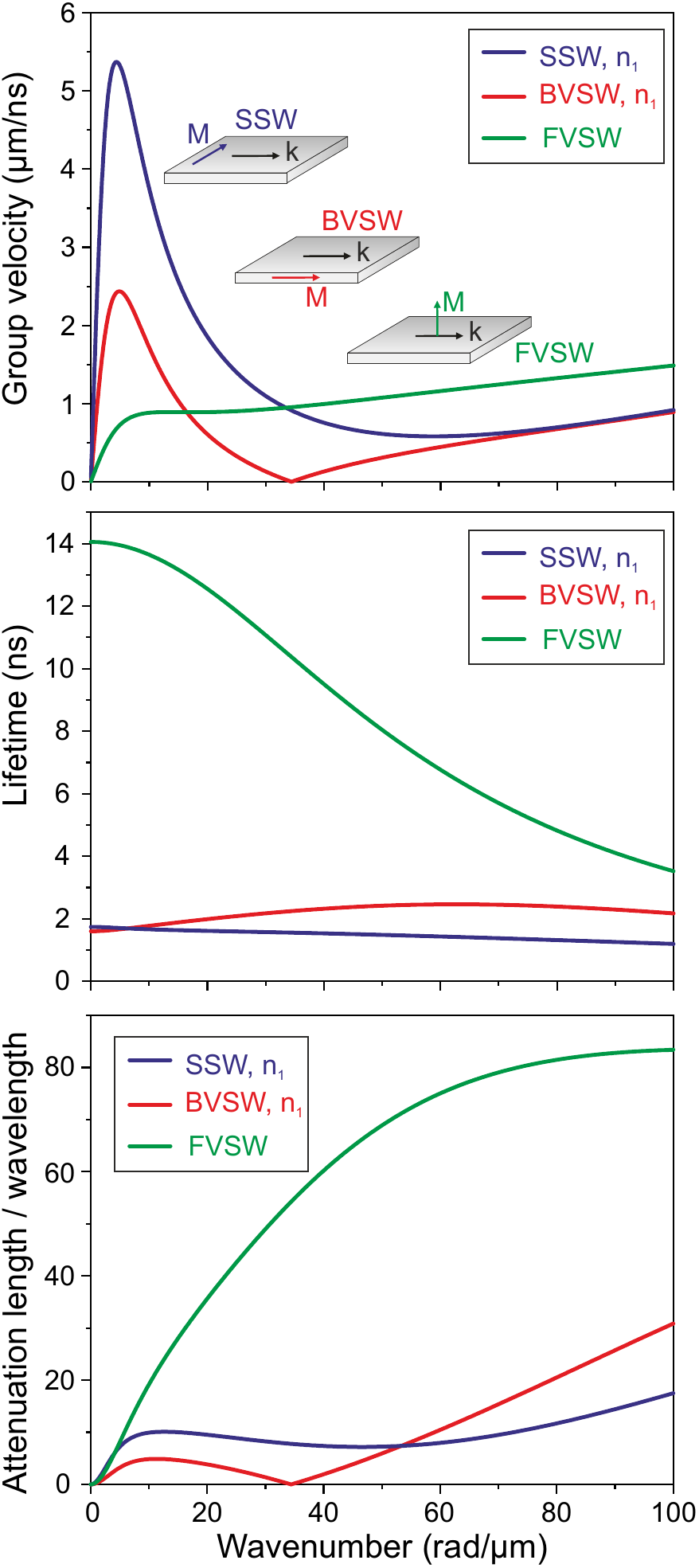}
\caption{Propagation characteristics of backward volume spin waves (BVSW), surface spin waves (SSW), and forward volume spin waves (FVSW) in a 500 nm wide and 30 nm thick CoFeB waveguide, derived from the dispersion relations in Fig.~\ref{fig:spin wave dispersion}. (a) Group velocity, (b) lifetime, and (c) attenuation length of the spin waves normalized by the wavelength as a function of their wavevector. For BVSW and SSW, data are shown for the first laterally confined width mode ($n_1$). In all cases, the material parameters were those of CoFeB (see Tab.~\ref{table:1}) and the external magnetic field was $\mu_0H = 100$ mT.}
\label{fig:SW_characterstics}
\end{figure}

When the external driving magnetic fields are removed, the spin-wave amplitude decreases exponentially with a characteristic lifetime given by \cite{Spin_wave}
\begin{equation}
\label{Eq:lifetime}
\tau = \left( \alpha \omega_\mathrm{n} \frac{\partial\omega_\mathrm{n}}{\partial\omega_0} \right)^{-1} \,.
\end{equation}
\noindent The spin-wave attenuation length represents the distance that a spin wave can travel until its amplitude has been reduced by $1/e$. It is given by the product of the lifetime and the group velocity $\delta=\tau\times v_\mathrm{g}$. As shown in Tab.~\ref{table:1}, spin-wave lifetimes are on the order of ns in metallic ferromagnets, such as CoFeB or Ni, whereas they can reach values close to the \textmu{}s range in low-damping insulators, such as Y$_3$Fe$_5$O$_{12}$ (yttrium iron garnet, YIG). Since spin-wave group velocities are typically a few \textmu{}m/ns (km/s), attenuation lengths are on the order of \textmu{}m for metallic ferromagnets to mm for YIG. 

The spin-wave group velocity, lifetime, and attenuation length (normalized to the wavelength) for the three cases of SSW, FVSW, and BVSW are plotted in Fig.~\ref{fig:SW_characterstics} as a function of the wavenumber for a CoFeB waveguide and an external magnetic field of $\mu_0H = 100$ mT. Note that, when the static magnetization orientation is intermediate between the three limiting cases, the spin-wave properties also show intermediate characteristics. As a final remark, the BVSW and FWSV geometries both lead to volume waves, which means that increasing the waveguide thickness may lead to the formation of quantized spin-wave modes along the thickness of the film at higher frequencies.

For SSW and BVSW, the group velocity reaches a maximum at small wavenumbers, which stems from the dipolar interaction. For BVSW, the group velocity becomes zero at a finite wavenumber (frequency) beyond the maximum, due to the competition between the dynamic dipolar and exchange fields. In the exchange regime, the group velocities of SSW and BVSW become equal and further increase with the wavenumber. For logic applications, it is desirable to use spin waves with large group velocities that ensure fast signal propagation and thus reduced logic gate delays. Moreover, large attenuation lengths reduce losses during spin wave propagation and are therefore also favorable for spin-wave devices. This will be further discussed below.

Group velocities depend in general on the properties of the ferromagnetic medium, as shown in Tab.~\ref{table:1}. The group velocity decreases typically strongly with decreasing film (or waveguide) thickness. This can be compensated by using magnetic materials with larger saturation magnetization $M_s$. The spin-wave lifetime in Eq.~(\ref{Eq:lifetime}) depends on the Gilbert damping $\alpha$. As the attenuation length is given by the product of the group velocity and the lifetime, the largest values are obtained for low-damping magnetic materials with large $M_s$. In practice, the two parameters $\alpha$ and $M_s$ may need to be traded off against each other, as indicated by Tab.~\ref{table:1}. Additional material properties for ideal magnetic materials for logic computing applications are the possibility for co-integration along CMOS as well as a high Curie temperature to ensure temperature insensitivity. This renders the complexity of the materials selection process and currently no clearly preferred materials has emerged yet. Future material research in this field is thus of great interest to optimize conventional materials or to establish novel magnetic materials for spin-wave applications.

\subsection{Nonlinear spin-wave physics}
\label{Sec_NL_SW_phys}

The previous section has discussed spin-wave physics using the linearized LLG equation (\ref{eq:lin_LLG}). Such an approach is valid for small amplitudes and describes noninteracting spin waves. However, the full LLG equation (\ref{eq:llg}) is nonlinear and thus nonlinear effects can arise for large spin-wave amplitudes. Since nonlinear effects are central for several spin-wave device concepts, this section provides a brief overview over the topic. More details can be found in Refs.~\onlinecite{cottam_linear_1994,mag1,Magnetostatics_ref1,Spin_wave,demokritov_magnonics_2013}.

The theoretical model for nonlinear spin-wave interactions was originally developed by Suhl, and thus nonlinear spin-wave processes are often referred to as Suhl instabilities of the first and second order.\cite{suhl_nonlinear_1956,suhl_theory_1957,Magnetostatics_ref1} Later, a generalized quantum-mechanical description of nonlinear magnons (quantized spin waves), termed S-theory, was developed by Zakharov, L'vov, and Starobinets.\cite{zakharov_spin-wave_1975,lvov_wave_1994} Today, these models are primarily used to describe a variety of different nonlinear and parametric spin-wave phenomena.\cite{krivosik_hamiltonian_2010,serga_parametrically_2007,wang_realization_2019,pirro_non-gilbert-damping_2014,Serga10,SWA5}

In general, the diverse nonlinear effects can be categorized into two groups: (i) multimagnon scattering\cite{Magnetostatics_ref1,lvov_wave_1994} and (ii) the reduction of saturation magnetization at large precession angles.\cite{krivosik_hamiltonian_2010,wang_realization_2019} However, (ii) can also be described by four-magnon scattering, so the separation into groups is not strict. Multimagnon scattering effects (i) primarily include three-magnon splitting (\emph{i.e.}~the decay of a single magnon into two), which can be used for the amplification of spin waves as a parametric process of the first order,\cite{Magnetostatics_ref1,ordonez-romero_three-magnon_2009} three-magnon confluence (\emph{i.e.} the combination of two magnons forming a single one), and four-magnon scattering (\emph{i.e.} the inelastic scattering of two magnons) that is fundamental for some spin-wave transistor concepts in Sec.~\ref{Sec:SW_transistors}.\cite{chumak_magnon_2014} 

In all nonlinear scattering processes, the total energy and momentum are conserved. The magnon spectra in macroscopic structures always consist of a practically infinite number of modes with different wavevector directions. Hence, an initial pair of magnons, which participates \emph{e.g.}~in a four-magnon scattering process, can always find a pair of secondary magnons.\cite{Magnetostatics_ref1,lvov_wave_1994} However, in magnetic nanostructures,\cite{Wang19,heinz_propagation_2020} the magnon density of states (scaling with the the structure size) also decreases, which makes the ``search'' for secondary magnon pairs more complex.\cite{jungfleisch_thickness_2015,mohseni_controlling_2020} Thus, the downscaling of magnonic nanostructures leads to a strong modification of nonlinear spin-wave physics, which offers the possibility to control (in the simplest case, switch on or off) nonlinear processes by the selection of the operating frequency and the external magnetic field. 

By contrast, processes (ii), which describe nonlinear frequency shifts of the spin-wave dispersion with increasing spin-wave amplitude, are typically more pronounced at the nanoscale.\cite{Wang19} These phenomena do not require any specific adjustment of the operating point and can thus be useful for spin-wave devices. In particular, the nonlinear shift of the spin-wave dispersion relation allows for the realization of nonlinear directional couplers, as discussed in Sec.~\ref{Sec_Dir_Couplers}.\cite{wang_realization_2019}

\section{Fundamentals of spin-wave computing}
\label{sec:SW Computing Background}
	
In this section, we discuss the fundamental principles of different disruptive computation paradigms based on spin waves to establish a framework for the architecture of a spin-wave-based computer. We start by introducing the basic components of a computing system, their implementations using spin waves, and the limitations of an all-spin-wave system. 
	
\subsection{Basic computer architectures}
	
\begin{figure}
\includegraphics[width=8cm]{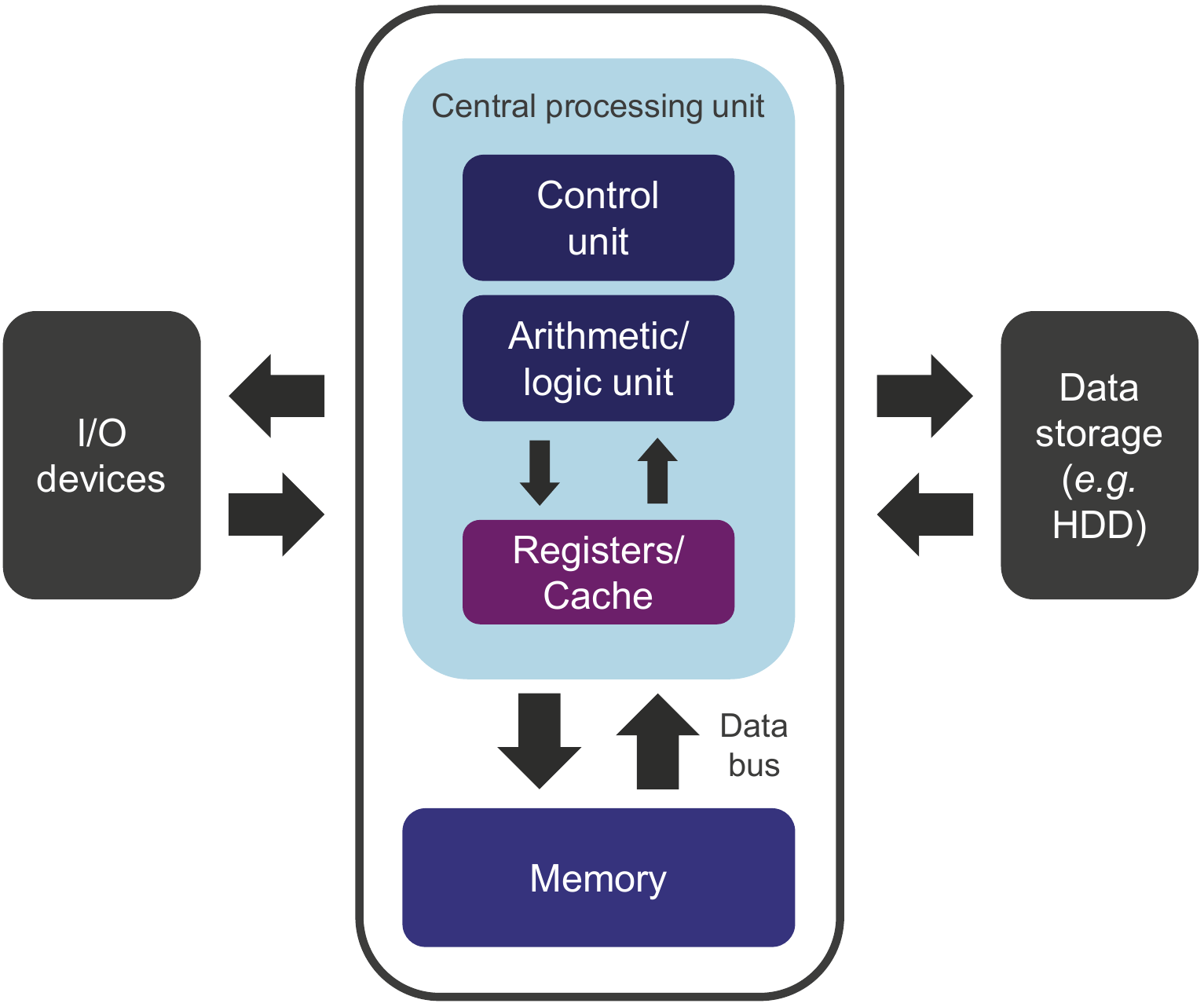}
\caption{Schematic of a Von Neumann computer consisting of a central processing unit and a memory, interconnected by a data bus.}
\label{fig:von Neumann}
\end{figure}

Despite many advances in computer architecture, the majority of today's computing systems can still be considered to be conceptually related to the Von Neumann architecture that was developed originally in the 1940s.\cite{patterson_computer_2011} Such a system consists of three essential parts: (i) a central processing unit that processes the instructions of the computer program and controls the data flow, (ii) a memory to store data and instructions, and (iii) a data bus as interconnection that links the the various parts within the processor and the memory and provides communication with the outside world. A schematic of such a system is shown in Fig.~\ref{fig:von Neumann}. Hence, to design a computer system that operates entirely with spin waves, spin-wave processors, spin-wave memory, as well as spin-wave interconnects need to be developed. Moreover, interfaces between the spin-wave processor and the outside periphery---presumably charge-based---are required, including a power supply.
	
The performance of a computing system is generally limited by the weakest component. Its computing throughput is restricted by the slowest part and the power consumption is determined by the most power-hungry subsystem. As detailed below, there is currently no comprehensive concept for a full spin-wave computer. In the following, we discuss requirements, basic approaches, and potential spin-wave-based implementations of the main components of a computer and finally suggest how a spin-wave-based computing system may resemble.
	
Recently, there has been growing interest in alternative computing paradigms beyond Von Neumann architectures, especially in the field of machine learning.\cite{mitchell_machine_1997,murphy_machine_2012,alpaydin_introduction_2020} Whereas the implementation of such architectures by spin waves is an intriguing prospect, research on this topic is still in its infancy.\cite{khitun_magnetic_2010,macia_spin-wave_2011,memory4,arai_neural-network_2018,reservoir_computing,tanaka_recent_2019,watt_reservoir_2020} A detailed discussion of such systems is beyond the scope of the tutorial. Nonetheless, it is clear that many of the arguments below remain relevant. Further information can be found in Sec.~\ref{Sec_analog_comp}.
	
\subsection{Information Encoding}
\label{Information_encoding}
	
\begin{figure}[t]
\includegraphics[width=7.4cm]{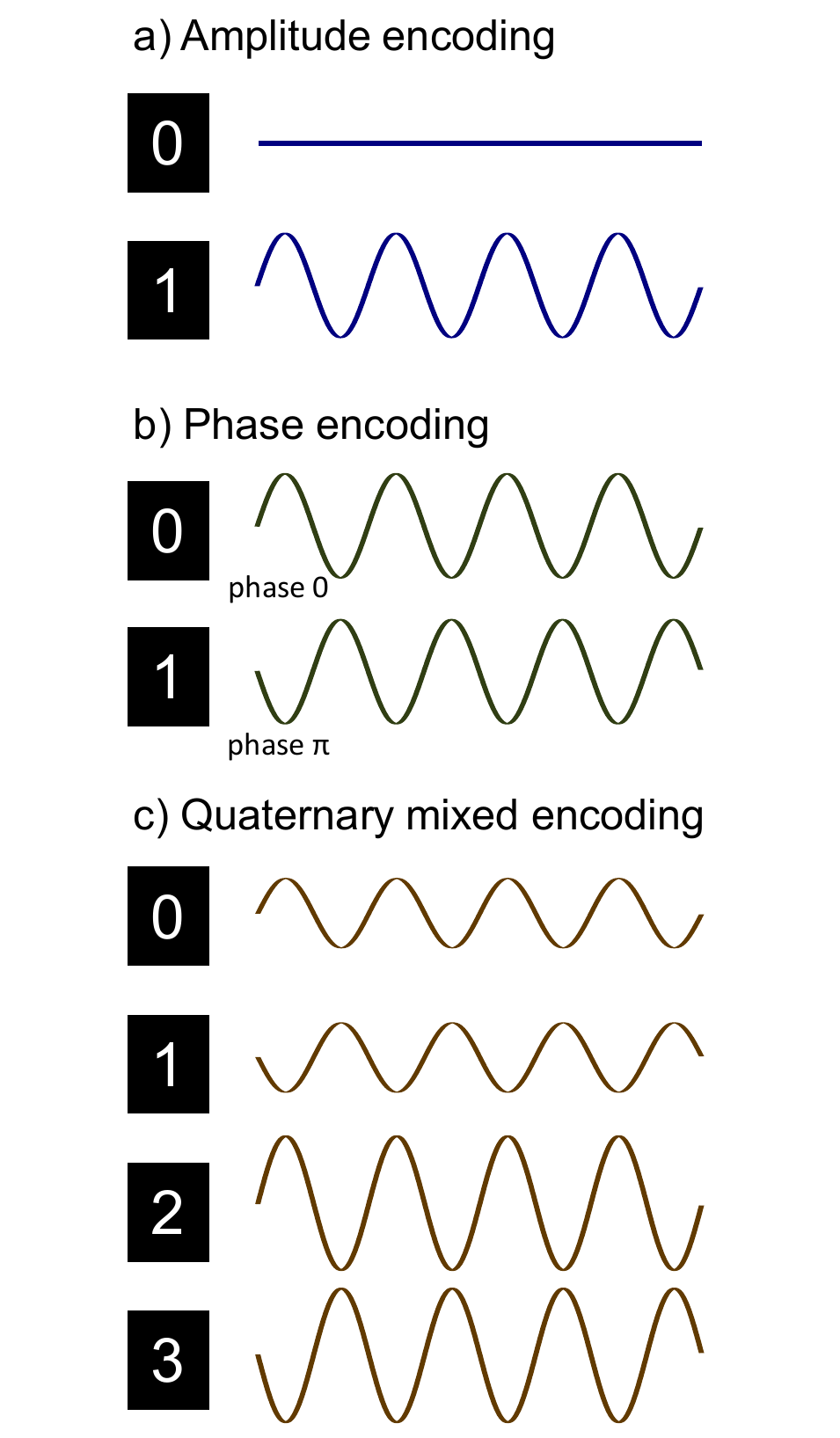}
\caption{Different schemes to encode information in (spin) waves: (a) binary amplitude encoding, (b) binary phase encoding, and (c) quaternary (nonbinary) mixed amplitude and phase encoding.}
\label{fig:encoding}
\end{figure}

Before discussing spin-wave computing concepts, we need to define how information can be encoded in a spin wave. Waves are characterized by amplitude (intensity), phase, wavelength, and frequency, which can all be used for information encoding. It is clear that the encoding scheme determines the interactions that can be employed for information processing and computation. Presently, device proposals typically rely on information encoded in spin-wave amplitude and/or phase (see Fig.~\ref{fig:encoding}). Moreover, the usage of different frequency channels has been proposed to enable parallel data processing based on frequency-division multiplexing.\cite{parallel_data_processing1,counter}
	
In amplitude-based information encoding, two main schemes can be pursued: (i) amplitude level encoding, and (ii) amplitude threshold encoding. In amplitude level encoding, the presence of a spin wave in a waveguide is referred to as a logic $1$ and no spin wave as a logic $0$ [Fig.~\ref{fig:encoding}(a)]. By contrast, in amplitude threshold encoding, a logic $1$ is represented by a spin wave with an amplitude above a certain threshold and a logic $0$ otherwise (or \emph{vice versa}). Multiple thresholds can be defined to represent nonbinary information and enable multivalued logic and computing. For example, if $\left\{X, Y\right\}$ with $X < Y$ are defined as a set of thresholds, a spin-wave amplitude greater than $Y$ can represent a $1$, an amplitude between $X$ and $Y$ a $0$, and an amplitude below $X$ a $-1$. 

Alternatively, information can be encoded in the (relative) spin-wave phase, such that \emph{e.g.}~a relative phase of $0$ (\emph{i.e.}~a spin wave in phase with a reference) refers to a logic $1$, while a relative phase of $\pi$ refers to a logic $0$ [Fig.~\ref{fig:encoding}(b)]. Furthermore, additional phases can be utilized for multivalued logic, \emph{e.g.}~$\left\{1, 0, -1\right\}$ can be represented by the set of phases $\left\{0, \frac{\pi}{2},\pi\right\}$. Such ternary computing schemes can have advantages over binary ones and the implementation of ternary logic circuits using (spin) waves may be an interesting future research topic, \emph{e.g.}~for computer arithmetics or neural networks.

Combinations of amplitude and phase encoding schemes are also possible and open further pathways towards effective nonbinary data processing [Fig.~\ref{fig:encoding}(c)]. For example, the data set $\left\{0, 1, 2, 3\right\}$ can be encoded using two amplitude levels $\left\{A, 2A\right\}$ and two phases $\{0,\pi\}$ by $0 := \{A,0\}$, $1 := \{A,\pi\}$, $2 := \{2A,0\}$, and $3 := \{2A,\pi\}$. Such schemes can be easily generalized to larger sets of nonbinary information.
	
The different encoding schemes have specific advantages and drawbacks when implemented in spin waves. Spin waves have typical propagation distances of \textmu{}m to mm, depending on the host material. For amplitude coding, the maximum size of a spin-wave circuit needs to be much smaller than the spin-wave attenuation length, since the logic level may otherwise change during propagation. By contrast, the phase of a wave is not affected by attenuation. While computing schemes may still require well-defined amplitudes, as further outlined below, the logic value encoded in the spin wave is nonetheless stable during propagation. Moreover, the phase coherence times of spin waves are long and phase noise can be kept under control even for nanofabricated waveguides with \emph{e.g.}~considerable line width roughness,\cite{collet_spin-wave_2017} rendering phase encoding rather stable. However, the largest differences between encoding schemes lie in the different interactions and processes required for computation, which is the topic of the next section.
	
We finally note that spin waves are noninteracting in the small signal approximation, \emph{i.e.}~for small amplitudes. Therefore, parallel data processing is possible using \emph{e.g.}~frequency-division or wavelength-division multiplexing. An information encoding scheme can then be defined at each frequency or wavelength and computation can occur in parallel in the same processor. Multiplexing in spin-wave systems is discussed further in Sec.~\ref{sec:Hybrid_Systems}.
	
\subsection{How to compute with (spin) waves?}
\label{Computing_schemes_interference}

\begin{figure*}[t]
\includegraphics[width=15cm]{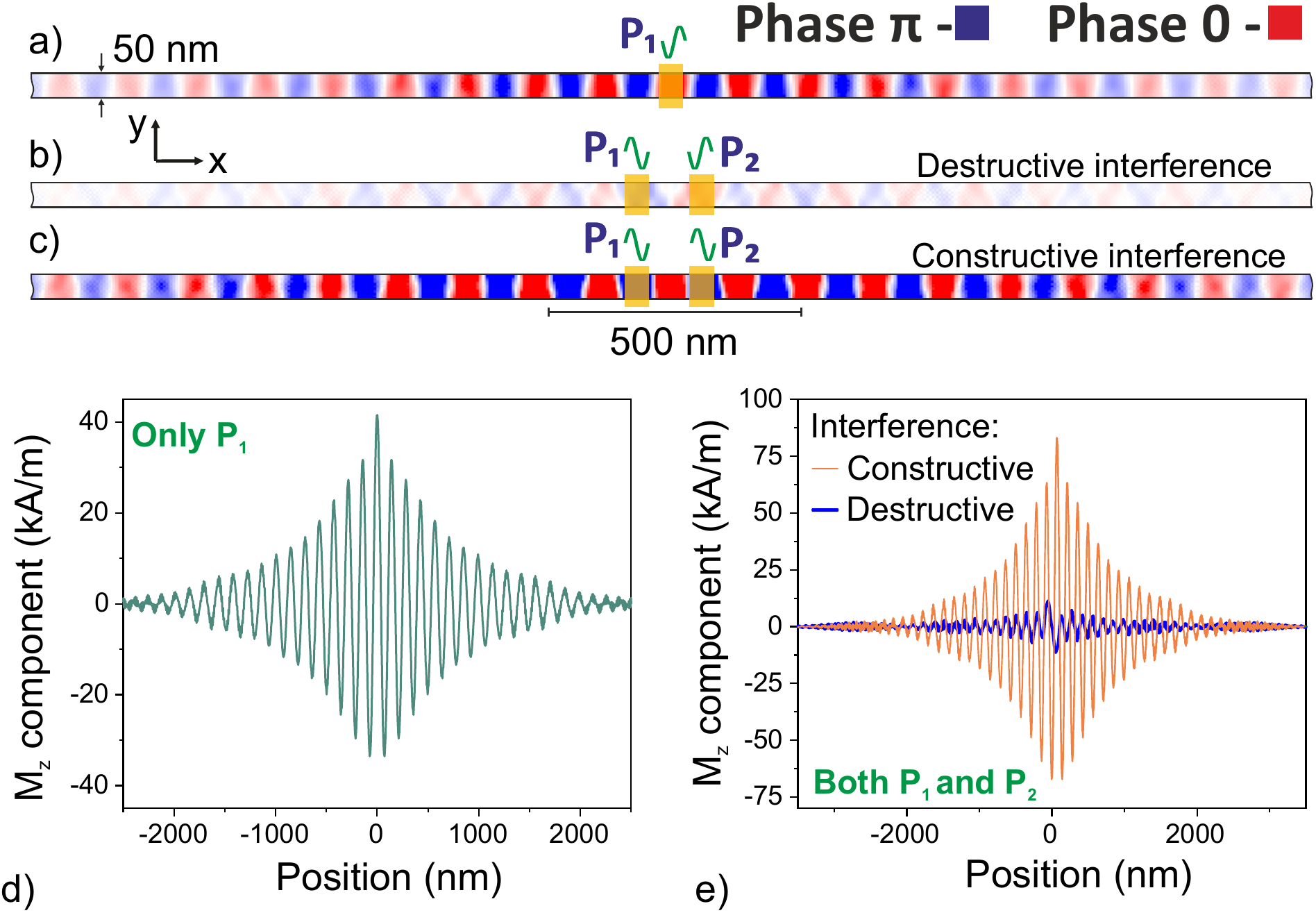}
\caption{Out-of-plane component of magnetization ($M_{z}$) in a 50 nm wide and 5 nm thick CoFeB waveguide obtained by micromagnetic simulations: Snapshots images of the spin waves emitted by a single port (a), and two in-phase (b) or anti-phase (c) ports at a frequency of 15 GHz. The corresponding amplitudes along the magnetic waveguide are shown in panels (d) and (e), respectively. The material parameters considered in simulations were taken from table~\ref{table:1}. The magnetic waveguide was initially magnetized longitudinally, whereas the simulations of spin-wave propagation were carried out in zero magnetic bias field.  Spin waves were excited by a uniform out-of-plane magnetic field at positions P$_1$ and P$_2$ in the waveguide center.}
\label{fig:interference}
\end{figure*}

When logic levels are encoded in spin-wave amplitude or phase, performing a logic operation requires the combination of different input waves and the generation of an output wave with an amplitude or phase corresponding to the desired logic output state. In principle, the superposition of waves can lead to the addition of either their intensity or their amplitude, depending whether the waves are incoherent or coherent.\cite{born_principles_1999} Since practical spin-wave signals typically have a large degree of phase coherence, further discussion can be limited to coherent superposition. In absence of nonlinear effects, the interaction of coherent waves is described by interference, \emph{i.e.}~the addition of their respective amplitudes at each point in space and time. We also limit the discussion to the superposition of waves with identical frequency and wavelength. Whether the interference of waves with different frequency or wavelength can also be (efficiently) utilized to evaluate logic functions is still an open research question with the potential for additional avenues towards novel computation paradigms. 
	
For in-phase waves with equal frequency, constructive interference leads to a peak-to-peak amplitude of the generated wave that is equal to the sum of the peak-to-peak amplitudes of the input waves. By contrast, destructive interference leads to a subtraction of the peak-to-peak amplitudes of input waves when their phase difference is $\pi$. For spin waves, the corresponding magnetization dynamics are depicted in Fig.~\ref{fig:interference}. In narrow waveguides, the spin-wave modes [see Fig.~\ref{fig:interference}(a) for the mode pattern of the first width mode] may deviate from plane waves due to lateral confinement and the effect of the demagnetizing field, as discussed in Sec.~\ref{Magnetization dynamics and spin waves}. Nonetheless, micromagnetic simulations, which rely on solving the LLG equation numerically,\cite{oommf,mumax} for a CoFeB waveguide [Figs.~\ref{fig:interference}(b) and \ref{fig:interference}(c)]  indicate that confined spin waves still show the expected interference. By placing two spin-wave sources on the same waveguide, destructive [Fig.~\ref{fig:interference}(b)] or constructive [Fig.~\ref{fig:interference}(c)] interference is obtained for a relative phase of $\pi$ or 0, respectively. The observation of incomplete destructive interference in Fig.~\ref{fig:interference}(b) can be linked to spin-wave attenuation, which leads to slightly different amplitudes of the two waves at both sides of the spin-wave sources.
	
Wave interference can be exploited to compute basic Boolean operations using the different encoding schemes. For example, using amplitude level encoding, it is easy to see that the constructive interference of two waves generates output of an OR operation, whereas their destructive interference (with a phase shift of $\pi$ between the waves) produces the output of an XOR operation. Many proposals and experimental studies have focused on phase encoding and the calculation of the majority function, MAJ.\cite{logic1,logic13,logic14,logic10,radu_spintronic_2015,SWA8,ganzhorn_magnon-based_2016,logic20,mahmoud_fan-out_2020} This stems from the fact that the phase of the output wave, ensuing from the interference of three input waves, is simply the majority of the phases of the input waves when logic 1 is encoded in phase 0 and logic 0 in phase $\pi$ (or \emph{vice versa}). Together with recent advances in MAJ-based circuit design,\cite{MAJ1,MAJ2,MAJ3,Pipeline1} this has led to a strong interest in spintronics\cite{imre_majority_2006,nikonov_proposal_2011,radu_spintronic_2015,manipatruni_beyond_2018,manipatruni_scalable_2019} and in particular spin-wave majority gates.\cite{logic1,hybrid_spin_CMOS,radu_spintronic_2015,radu_overview_2016} As an example, the carry out bit in a full adder (a fundamental building block in processor design) is directly computed by a three-input majority function [\emph{cf.}~Eq.~(\ref{Eq_MAJ})]. In addition, many error detection and correction schemes rely on $n$-input majority logic.\cite{lucas_iterative_2000,palanki_iterative_2007} 
	
For novel computation paradigms, including (spin) wave computing, a main requirement is the possibility to implement any arbitrary logic function that can be defined within its basic formalism by means of a universal gate set. For example, within Boolean algebra, any logic function can be expressed as a sum of products or as a product of sums. Using double complements and De Morgan's laws, it can be demonstrated that any logic function can be implemented by either NAND or NOR gates only. Therefore, NAND or NOR constitute each a universal gate with efficient CMOS implementations. As mentioned above, (spin) wave interference provides a natural support to implement majority gates, MAJ, which form a universal gate set in combination with inverters, INV. In phase encoding, an inverter can be realized by a passive delay line of length $(n-\frac{1}{2})\times\lambda$ (with $\lambda$ the spin-wave wavelength and $n = 1,2,3,\ldots$ an integer) that leads to a phase shift of $\pi$ during propagation. In amplitude encoding, inverters are more complex and typically require active components. In this case, an inverter can be realized by interference with a reference wave with a phase of $\pi$. As an example, XOR, XNOR, and a full adder (sum $\Sigma$ and carry out $C_\mathrm{out}$) can then be implemented with majority gates and inverters as follows:  
\begin{equation}
\label{Eq_MAJ}
\begin{split}
A \oplus B &= \mathrm{MAJ}\left(\mathrm{MAJ}\left(A,\Bar{B},0\right), \mathrm{MAJ}\left(\Bar{A},B,0\right),1\right) \\ 
\overline{A \oplus B} & =\mathrm{MAJ}\left(\mathrm{MAJ}\left(\Bar{A},\Bar{B},0\right), \mathrm{MAJ}\left(A,B,0\right),1\right) \\
\Sigma & = \mathrm{MAJ}\left(\mathrm{\overline{MAJ}}\left(A,B,C_{in}\right), \mathrm{MAJ}\left(A,B,\Bar{C}_\mathrm{in}\right), C_\mathrm{in}\right)\\
C_\mathrm{out} & = \mathrm{MAJ}\left(A,B,C_\mathrm{in}\right)
\end{split}
\end{equation}

It should be mentioned that wave-based computing is not limited to the usage of spin waves. Similar concepts have been proposed for surface plasmon polaritons,\cite{wei_cascaded_2011,fu_all-optical_2012,lal_noble_2012,dutta_proposal_2017} or acoustic waves/phonons.\cite{maldovan_sound_2013,sklan_splash_2015} A discussion of the advantages and disadvantages of different physical implementations of wave computing is beyond the scope of this tutorial but it is clear that many of the discussions concerning devices, circuits, and hybrid systems are general and remain valid for other wave-based computing approaches.
	
\subsection{Spin-wave interconnects}
\label{Sec_interconnect}
	
In the previous section, the basic principles of spin-wave interference have been discussed and it has been shown that they can be used for logic operations. However, in a computing system, data need to be transmitted to the inputs of the logic circuit, exchanged between gates, and finally output data need to be transmitted to \emph{e.g.}~a memory. This is the task of the interconnect, which may also transmit clock signals as well as power. In conventional digital integrated circuits, the logic states $0$ and $1$ are encoded in voltages, which allows for data transmission by metal wires. While interconnect performance is today often limiting the overall performance of integrated circuits, solutions are mature and well understood from the point of view of their capabilities and associated overhead. 

\begin{figure}[t]
\includegraphics[width=8.2cm]{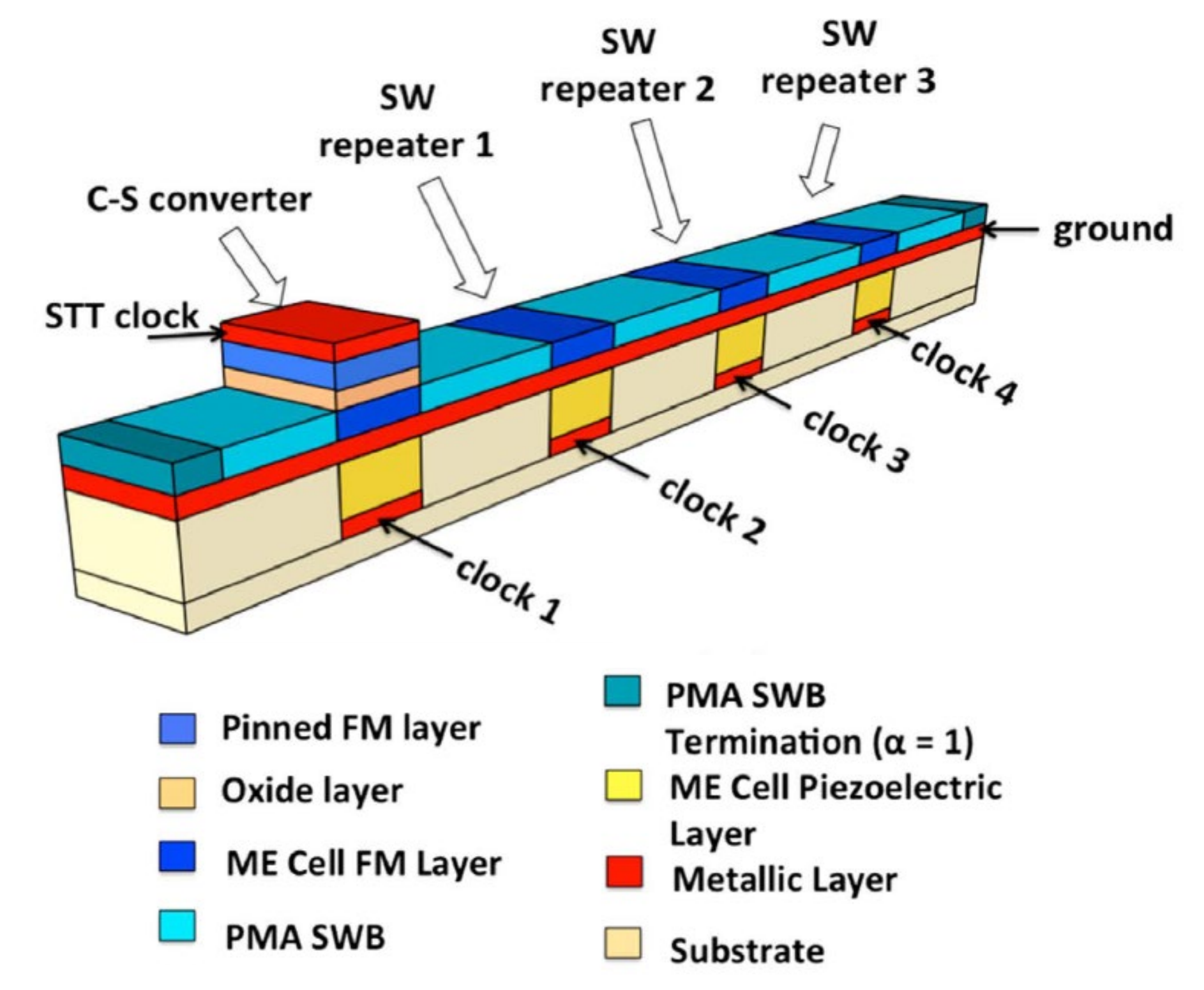}
\caption{Schematic of a clocked spin-wave interconnect. Reproduced with permission from S. Dutta, S.-C. Chang, N. Kani, D. E. Nikonov, S. Manipatruni, I. A. Young, and A. Naeemi, Sci. Rep. 5, 9861 (2015). Copyright 2015 Nature.}
\label{fig:interconnect_naeemi}
\end{figure} 

A natural approach to connect spin-wave logic gates is by means of waveguides, in which spin waves propagate from \emph{e.g.}~a gate output to an input of a subsequent gate. Besides cascading issues for specific implementations discussed in more detail in Sec.~\ref{sec:Requirements for Spin Wave Circuit Design}, the rather slow and lossy spin-wave propagation leads to fundamental limitations for spin-wave interconnects.\cite{logic16,khitun_efficiency_2007,liu_interconnect_2015} Since the spin-wave group velocity is much lower than that of electromagnetic waves in (nonmagnetic) metallic wires, interconnection by spin waves propagating in waveguides adds a considerable delay overhead, which depends on waveguide length and material. Some representative numbers for the spin-wave group velocity are listed in Tab.~\ref{table:1}. Typical delays are about 1 ns/\textmu{}m (\textmu{}s/mm), which means that spin waves propagating in waveguides cannot be efficiently utilized for long-range data transmission. Even for short range data communication, the delay introduced by spin-wave propagation may not be negligible. As an example, for a spin-wave circuit with a waveguide length of a few \textmu{}m, the propagation delay may already exceed the duration of a typical clock cycle of a high performance CMOS logic processor of about 300 ps ($\sim 3$ GHz clock frequency). It is worth noting that the overall delay is determined by the longest propagation path in the circuit. Hence propagation delays may limit the computing throughput of a spin-wave circuit. Additional boundaries for the throughput of spin-wave circuits and systems are discussed in Sec.~\ref{sec:Requirements for Spin Wave Circuit Design}.
	
Moreover, the spin-wave amplitude decays during propagation due to intrinsic magnetic damping. Such propagation losses remain limited when spin-wave circuits are much smaller than the attenuation length, which strongly depends on the waveguide material (see Tab.~\ref{table:1} for indicative numbers). This can impose severe limits on the size (and therefore the complexity) of spin-wave circuits. Losses can in principle also be compensated for by spin-wave amplifiers or repeaters. As an example, a clocked interconnect concept based on spin-wave repeaters has been reported in Ref.~\onlinecite{Interconnect4} (see Fig.~\ref{fig:interconnect_naeemi}). While such approaches can mitigate limitations of signal propagation by spin waves, they add a significant overhead to the circuit and need to be carefully considered when the energy consumption and delay of a spin-wave computing system is assessed. Spin-wave repeaters and amplifiers are discussed in more detail in Sec.~\ref{Sec_Repeaters}.
	
\subsection{Spin-wave memory}
	
To date, rather little work has been devoted to specific spin-wave memory elements that are required for computing systems based on spin waves only. Spin waves are volatile dynamic excitations, which decay at timescales of ns to \textmu{}s (see Tab.~\ref{table:1}). There are two different basic approaches to memories for spin waves. The natural spintronic memory element is a nanomagnet, in which the information is encoded in the direction of its magnetization. In such a memory element, an incoming spin wave deterministically sets (switches) the orientation of the magnetization of the nanomagnet. When phase encoding is used, the interaction between the spin wave and the nanomagnet needs to be phase dependent. The clocked interconnect concept\cite{Interconnect4} depicted in Fig.~\ref{fig:interconnect_naeemi} employs the deterministic phase-sensitive switching of nanomagnets with perpendicular magnetic anisotropy in the repeater stages. It therefore also offers some memory functionality. A 2D-mesh configuration of such structures has also been proposed.\cite{memory2,gertz_magnonic_2015} 

An alternative approach is the use of conventional charge-based memories after signal conversion in the hybrid spin-wave--CMOS systems discussed in the next section. An introduction to charge-based memory devices is beyond the scope of this tutorial and can be found \emph{e.g.}~in Refs.~\onlinecite{device2,sharma_advanced_2002,hong_emerging_2014}.
	
\subsection{Hybrid spin-wave--CMOS computing systems}
	
Above, we have argued that spin-wave propagation in magnetic waveguides may add considerable delay and is therefore not competitive over distances of more than a few 100 nm to 1 \textmu{}m. To address this issue, metallic or optic interconnects can be used for long range data transmission after spin-wave signals have been converted to electric or optical signals. Voltages and light travel very fast through metal wires and optical fibers, respectively, with propagation velocities given by the speed of light in the host materials. Such solutions lead naturally to hybrid system concepts, in which spin-wave circuits coexist with conventional CMOS or mixed-signal integrated circuits, including memory. Such solutions rely on (frequent) forth-and-back conversion between spin-wave and charge domains using transducers, which may themselves add substantial delay and energy consumption overhead. To minimize the overhead, the number of necessary transducers should remain limited. The acceptable conversion granularity depends on the relation between delay and energy consumption of spin-wave circuits, transducers, and CMOS/mixed-signal circuits. In practice, it is of course technology dependent. 
	
Today, design guidelines for such hybrid circuits are only emerging. Their development and the benchmarking of the ensuing hybrid circuits constitute a crucial step towards real-world applications for spin-wave computing. Since hybrid systems require efficient and scalable transducers, the approaches to generate and detect coherent spin waves are discussed in the next section. Such transducers form critical elements of the spin-wave devices and circuits that are reviewed in Sec. \ref{sec:General Spin Wave device structure}. 
	
\section{Spin-wave transducers}
\label{Sec_transducers}
	
As argued above, spin-wave computing systems require transducers to convert spin-wave-encoded signals to/from voltage signals. The scalability and the energy efficiency of the transducers can be expected to be crucial for the overall performance of a hybrid system. This section introduces different concepts of spin-wave transducers. As discussed in Sec.~\ref{sec:Basics and background of SW technology}, spin waves are a response of a magnetic material to oscillatory external (effective) magnetic fields. In the linear regime, \emph{i.e.} for weak excitation, excited spin waves have the same frequency as the applied oscillatory field, a well-defined phase, which depends on the specific interaction, and an amplitude proportional to the magnitude of the excitation. In principle, any oscillatory effective field can launch spin waves in a waveguide. From a practical point of view, the need to generate oscillatory effective magnetic fields at GHz frequencies has led to several preferred approaches. It should be mentioned that the scalability and the energy efficiency of such transducers at the nanoscale has not been definitively assessed and is currently actively researched. As argued in Sec.~\ref{sec:Conclusions}, the demonstration of a nanoscale spin-wave transducer with high energy efficiency is one of the key prerequisites for the ultimate goal of hybrid spin-wave--CMOS computing systems.
	
\subsection{External magnetic fields: inductive antennas}
	
The ``reference'' method to excite spin waves is by means of external magnetic fields generated by an AC current in a microwave antenna. The AC current generates an alternating Oersted field via Amp\`ere's law, which in turn exerts a torque on the magnetization in an adjacent ferromagnetic medium. At excitation frequencies above the ferromagnetic resonance, the Oersted field can then excite spin waves in the medium, as outlined in Sec.~\ref{sec:Basics and background of SW technology} and described by the LLG equation (\ref{eq:llg}).
	
Different antenna designs have been used in spin-wave experiments, such as \textit{i.e.}~microstrip antennas, coplanar waveguide antennas, or loop antennas. An overview can be found in Ref.~\onlinecite{antenna1}. The specific antenna design has strong repercussions on the spin-wave spectrum that can be excited. It is intuitive that an oscillating magnetic field that is uniform over a distance $L$ along the waveguide cannot efficiently excite spin waves with wavelengths $\lambda \ll L$. More quantitatively, the excitation efficiency $\Gamma_n$ of a spin wave propagating along the $x$-direction with mode number $n$, wavenumber $k$, and angular frequency $\omega$ by a dynamic magnetic field distribution $\mathbf{h}(\bm{x})e^{-i\omega_h t}$ is proportional to the overlap integral over the magnetic volume $V$,
\begin{equation}
\label{eq:antenna_efficiency}
\Gamma_n \propto \left| \iiint_V  \mathbf{h}\left(\bm{x}\right) \cdot \mathbf{m}\left(\bm{x}\right) d\bm{x} \right|\times \delta\left( \omega - \omega_h\right) \,,
\end{equation}
\noindent with $\mathbf{m}\left(\bm{x}\right)e^{-i\omega t}$ the distribution of the dynamic magnetization. Note that finite spin-wave lifetimes due to magnetic damping broaden the $\delta$-function. For inductive antennas transversal to the direction of the waveguide, the magnetic field points essentially in the $x$-direction along the waveguide. For thin films, the magnetization is uniform over the film thickness and the dynamic magnetization of a plane wave can be written as $\mathbf{m}\left(\bm{x}\right) = \tilde{\mathbf{m}}(y)e^{ikx}$, with $\tilde{\mathbf{m}}(y)$ describing the transverse mode pattern. Equation~(\ref{eq:antenna_efficiency}) then becomes\cite{Demidov15}
\begin{equation}
\Gamma_n \propto \left|\int h_x(x)e^{ikx} dx\right| \left|\int h_x(y)\tilde{m}_x(y) dy\right|\times \delta\left( \omega - \omega_h\right) \,,
\end{equation}
\noindent The first integral indicates that the the wavelength dependence of the spin-wave excitation efficiency is determined by the Fourier spectrum of the driving Oersted field along the waveguide. The second term leads to a dependence on the symmetry of the spin-wave mode. For symmetric dynamic magnetic field distributions (as in the case of an inductive antenna), only spin-wave modes with symmetric transverse mode patterns can be excited and the excitation efficiency is zero for antisymmetric modes. For an inductive antenna with width $w$, this leads to 
\begin{equation}
\label{eq:sinc_antenna_efficiency}
\Gamma_n \propto 
\begin{cases}
\frac{w}{n} \mathrm{sinc}\left(\frac{kw}{2} \right)\times \delta\left( \omega - \omega_h\right), & \text{for odd}\ n \\
0, & \text{for even}\ n \\
\end{cases}\,.
\end{equation}
\noindent Here, the dependence on the sinc function stems from the Fourier transform of the uniform magnetic field underneath the antenna, whereas the explicit dependence on the mode number $n$ is caused by the transverse integral over the mode pattern. This discussion shows that the shape and the dimensions of inductive antennas have strong impact on the spin-wave excitation bandwidth. Reducing the dimensions (width, gap) of an antenna increases its bandwidth and the peak magnetic field strength underneath. 
	
While inductive antennas can be rather efficient at ``macroscopic'' scales $\gg 10$ \textmu{}m, scaling their dimensions into the \textmu{}m and sub-\textmu{}m range strongly reduces the antenna quality factor, \emph{i.e.}~the ratio between inductance and resistance, $Q = L/R$, and the spin-wave excitation efficiency. In general, since the Oersted field is proportional to the \emph{current} via Amp\`ere's law, antennas do not scale favorably, with strongly increasing \emph{current densities} (and thus degraded reliability) at smaller dimensions. More details on the relation between antenna design, spin-wave excitation efficiency, and bandwidth can be found in Ref. \onlinecite{antenna1}. It has also been shown that a magnetic near field resonator in the vicinity of the antenna can enhance the spin-wave excitation efficiency.\cite{Au_antenna_2012,Au_SWvalve_2012}

\begin{figure}[t]
\includegraphics[width=7.5cm]{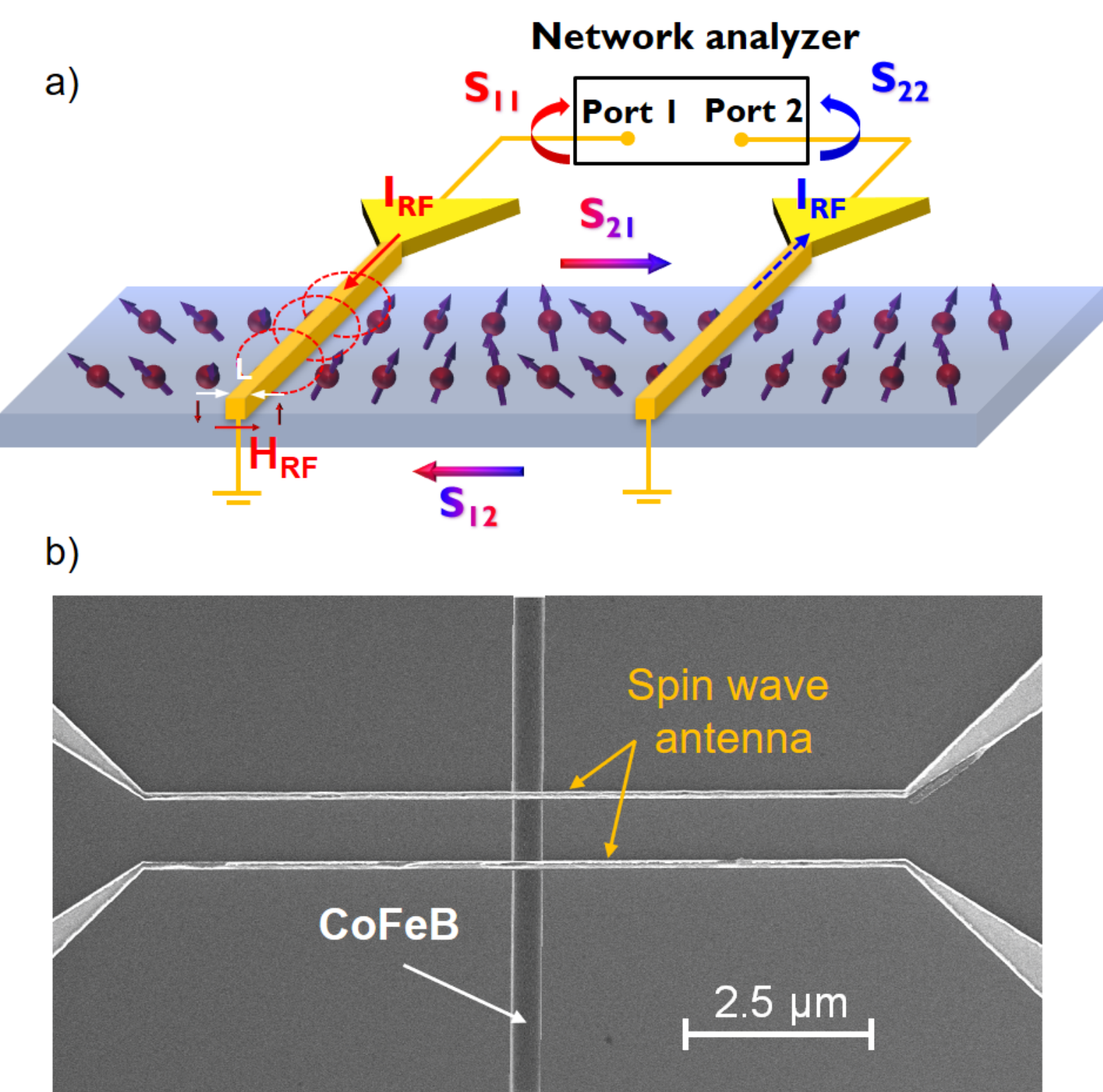}
\caption{\label{Fig_SW_trans}(a) Sketch of a typical experimental setup for spin-wave transmission based on a waveguide and two inductive antennas. Spin waves are excited by the Oersted field created by a microwave current in one of the antennas and detected inductively by the second. The power transmitted by the spin waves is measured using a vector network analyzer and extracted from $S$-parameters. The arrows inside the waveguide symbolize spin precession during propagation. (b) Scanning electron micrograph of a 500 nm wide CoFeB waveguide and two 125 nm wide inductive antennas.}
\label{fig:antenna_device}
\end{figure}

Inductive antennas can also detect spin waves. The dynamic dipolar field generated by the spin waves induces a current in an adjacent antenna via Faraday's law. Thus inductive antennas can be used both as input and output ports in all-electrical spin-wave transmission experiments.\cite{Serga10,bailleul_propagating_2003,vlaminck_spin-wave_2010,hyun_kwon_spin_2011,huber_reciprocal_2013,sato_propagating_2014,ciubotaru_all_2016,bhaskar_backward_2020} A schematic of such an experiment is shown in Fig.~\ref{Fig_SW_trans}. A first inductive antenna launches spin waves in a ferromagnetic waveguide, which are subsequently detected by a second antenna. The microwave power transmitted by the spin waves can be measured with phase sensitivity using a vector network analyzer. Both the fraction of transmitted ($S_{21}$, $S_{12}$) and reflected ($S_{11}$, $S_{22}$) microwave power can be used to analyze the measurements. 

\subsection{Spin--transfer and spin--orbit torques}
\label{Sec_STT_SOT}

\begin{figure*}[t]
\centering
\includegraphics[width=15cm]{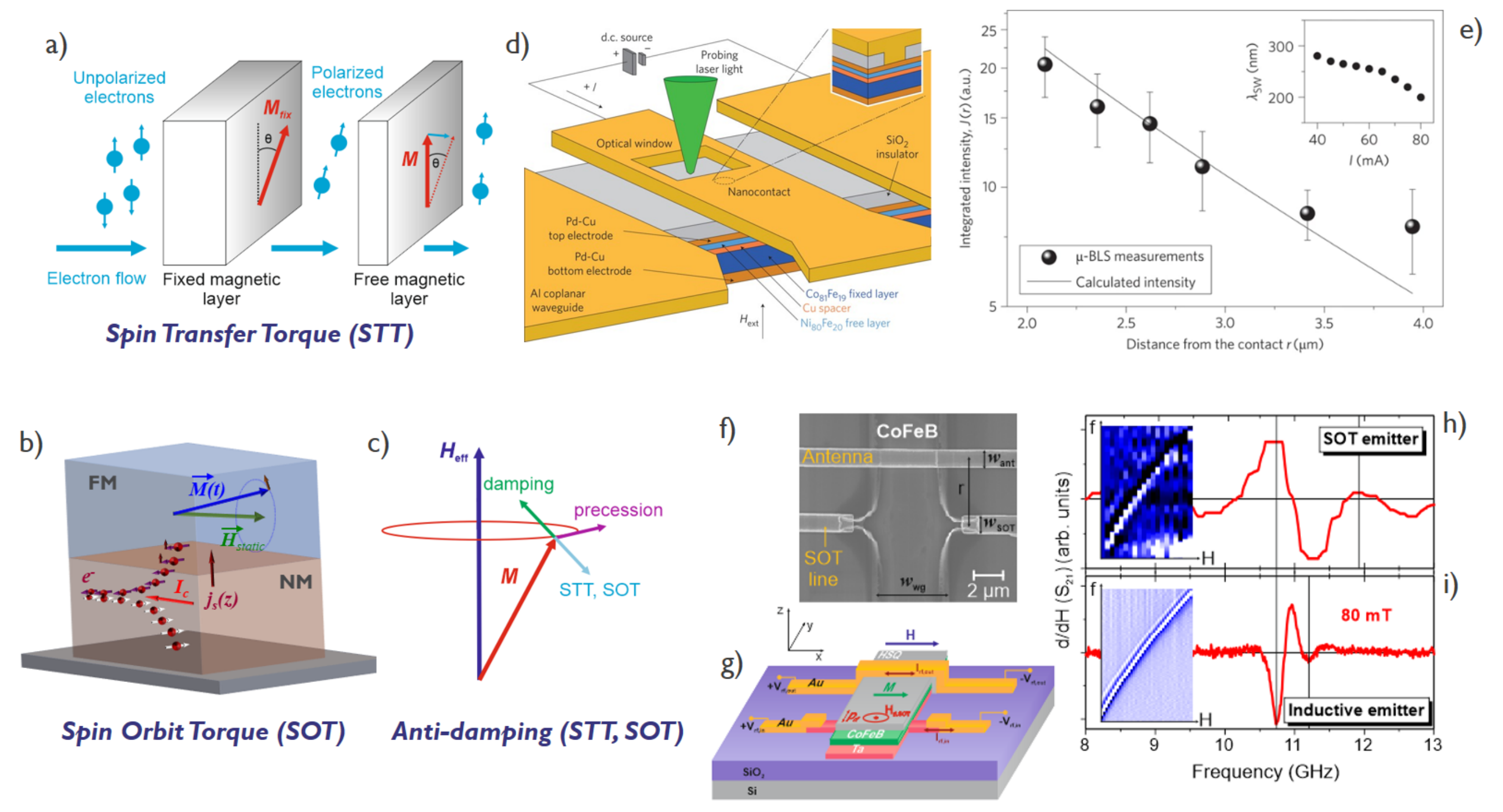}
\caption{ Schematic illustrations of (a) spin--transfer torque (STT) and (b) spin--orbit torque (SOT) processes. (c) Magnetization dynamics in an effective field including precession, damping, as well as both STT and SOT. (d) Device layout used for the excitation of spin waves by STT. Reproduced with permission from M. Madami, S. Bonetti, G. Consolo, S. Tacchi, G. Carlotti, G. Gubbiotti, F. B. Mancoff, M. A. Yar, and J. Åkerman, Nature Nanotech. 6, 635(2011). Copyright 2020 Springer Nature. (e) Attenuation of the excited spin waves during propagation Reproduced with permission from M. Madami, S. Bonetti, G. Consolo, S. Tacchi, G. Carlotti, G. Gubbiotti, F. B. Mancoff, M. A. Yar, and J. Åkerman, Nature Nanotech. 6, 635(2011). Copyright 2020 Springer Nature. (f) Scanning electron micrograph and (g) schematic layout of a device based on an SOT emitter and an inductive antenna detector. (h) and (i) Intensity of spin waves generated by a SOT antenna [(h), magnified $20\times$] and an inductive antenna [(i)] Reproduced with permission from G. Talmelli, F. Ciubotaru, K. Garello, X. Sun, M. Heyns, I. P. Radu, C. Adelmann, and T. Devolder, Phys. Rev. Appl. 10, 044060 (2018). Copyright 2018 Physical Review Applied. The plots show the magnetic-field derivative of the forward-transmission $S$-parameter, $dS_{21}/dH$ (emitter-to-detector distance 4 \textmu{}m, applied magnetic field $\mu_0H = 80$ mT). The insets show field--frequency signal maps corresponding to spin waves emitted by the two types of antennas (magnetic fields $\mu_0H = 52$--145 mT, frequencies 8--15 GHz).}
\label{fig:RF_SOT}
\end{figure*}

In the previous section, it was discussed that spin waves can be excited by the oscillatory Oersted field created by a microwave AC current in an inductive antenna. In addition, DC currents can also generate spin waves or switch nanomagnets as long as they are spin polarized.\cite{slonczewski_current-driven_1996,berger_emission_1996, tsoi_excitation_1998, katine_dc-driven_2000,urazhdin_dc-driven_2003,lee_excitations_2004,houssameddine_spin-torque_2007,ruotolo_phase-locking_2009,demidov_direct_2010,madami_direct_2011,divinskiy_excitation_2018} When an electric current passes through a uniformly magnetized layer, the electron spins align themselves with the magnetization direction, generating a spin-polarized current [Fig. \ref{fig:RF_SOT}(a)].\cite{slonczewski_current-driven_1996,berger_emission_1996} When such a polarized current flows through a second magnetic layer, the spins reorient again if the direction of the magnetization is not aligned with the spin polarization. This leads to the transfer of angular momentum to the magnetization of the second layer, which can change its orientation if the layer is thin enough (a few nm). The transfer mechanism of angular momentum by spin-polarized currents to the magnetization is known as spin--transfer torque (STT). The spin--transfer torque that acts on the magnetization $\mathbf{M}$ of a ``free layer'' due to a spin-polarized current from a ``fixed'' reference layer with magnetization $\mathbf{M}_\mathrm{fix}$ is given by \cite{slonczewski_current-driven_1996,hillebrands_spin_2006}
\begin{equation}
\label{STTorque} \left(\frac {d\mathbf{M}} {dt}\right)_\mathrm{STT} =
\frac{-\lvert g \lvert}{2} \frac{\mu_\mathrm{B}}{M_\mathrm{s}} \frac{1}{d}
\frac{J}{e}\mathcal{P}\left[\mathbf{M}\times\left(\mathbf{M}\times\mathbf{M}_\mathrm{fix}\right)\right]\,,
\end{equation}
\noindent where $J$ is the current density and $d$ the thickness of the free layer. $\mathcal{P}$ represents the current polarization, $\mu_\mathrm{B}$ is the Bohr magneton, $g$ is the Land\'{e} factor, and $e$ denotes the elementary charge.
	
The effect of a spin-polarized current on the magnetization dynamics can be calculated by introducing an STT term in the LLG equation (\ref{eq:llg}). Using the notations
\begin{equation}
\label{NotLLGS1} \boldsymbol{\zeta}=\frac{\mathbf{M}}{M_\mathrm{s}}  \,\text{,} \quad \boldsymbol{\zeta}_\mathrm{fix}=\frac{\mathbf{M}_\mathrm{fix}}{M_\mathrm{s,fix}} \,\text{,}
\quad \boldsymbol{\eta}=\frac{\mathbf{H}}{M_\mathrm{s}} \,\text{,} \quad \tau =
\gamma_{0}M_\mathrm{s}t\,,
\end{equation}
\noindent and
\begin{equation} \label{NotLLGS2} \chi =\frac{\hbar}{2}
\frac{1}{\mu_{0}M^{2}_\mathrm{s}} \frac{1}{d} \frac{J}{e}\mathcal{P}\,,
\end{equation}
\noindent the LLG equation including the STT term can be written in a dimensionless form as\cite{hillebrands_spin_2006}
\begin{equation}
\label{LLGS} \frac{d\boldsymbol{\zeta}} {d\tau} =-
\underbrace{(\boldsymbol{\zeta}\times\boldsymbol{\eta})}_{\text{precession}}-\underbrace{\chi\left[\boldsymbol{\zeta}\times\left(\boldsymbol{\zeta}\times\boldsymbol{\zeta}_\mathrm{fix}\right)
\right]}_{\text{spin transfer torque}}+\underbrace{\alpha\left(\boldsymbol{\zeta}\times\frac{d\boldsymbol{\zeta}}{d\tau}\right)}_{\text{damping}}\,.
\end{equation}
	
Depending on the direction of current flow, the torque exerted by the spin-polarized current can enhance or compensate for the intrinsic damping. When the damping is exactly compensated for, the STT enables a steady precession of the magnetization. Even larger polarized current densities lead to a negative the damping torque and the magnetization precession is strongly amplified. The critical current required to excite the magnetization in the free layer (from an initially parallel orientation of both magnetic layers) is given by \cite{hillebrands_spin_2006,grollier_field_2003}
\begin{equation}
\label{Icr} I_\mathrm{crit}= \frac{2e}{\hbar}\frac{\alpha}{\mathcal{P}}
V\mu_{0}M_\mathrm{s}\left(H+H_{k}+\frac{M_\mathrm{s}}{2}\right)
\end{equation}
\noindent where $V$ represents the volume of the magnetic free layer, and $H$ and $H_{k}$ denote the external and the anisotropy magnetic fields, respectively.

To limit the critical currents necessary for stable magnetization precession, the volume of the magnetic layers $V$ is typically reduced by patterning pillars with sub-\textmu{}m diameters. These devices have been termed spin-torque nano-oscillators (STNOs) and can also emit spin waves if the free layer is coupled to a waveguide. It has been demonstrated that spin waves emitted by STNO can travel for several \textmu{}m and that their propagation direction can be controlled by a magnetic bias field [see Figs.~\ref{fig:RF_SOT}(d) and \ref{fig:RF_SOT}(e)].\cite{demidov_direct_2010, madami_direct_2011}
	
Another mechanism to generate spin currents is based on the spin Hall effect (SHE). This effect originates from the spin-dependent electron scattering in a charge current flowing through a nonmagnetic metal or a semiconductor with (large) spin--orbit interaction. \cite{dyakonov_current-induced_1971,hirsch_spin_1999} The resulting spin current is perpendicular to the charge current and can therefore be transferred to an adjacent ferromagnetic material even if the charge current is only flowing in the nonmagnetic metal. The spin current exerts a torque on the magnetization of the ferromagnet, as illustrated in Fig.~\ref{fig:RF_SOT}(b).  In addition, the spin--orbit interaction of the conduction electrons in a two-dimensional system can also generate an effective magnetic field---the so-called Rashba effect.\cite{bychkov_2007,miron_current-driven_2010} The torque on the magnetization due to spin--orbit effects can be expressed by\cite{kovalev_current-driven_2007,allen_experimental_2015}
\begin{equation}
\label{TSOT} \left(\frac {d\boldsymbol{\zeta}} {dt}\right)_\mathrm{SOT} =
\gamma \beta_{\parallel} \left[ \boldsymbol{\zeta} \times \left( \bm{p}\times \boldsymbol{\zeta}\right) \right] + \gamma \beta_{\perp} \left(\bm{p}\times \boldsymbol{\zeta}\right)\,,
\end{equation}
\noindent with 
\begin{equation}
\label{DampFieldTerm} \beta_{\parallel} = \varepsilon_{\parallel} \frac{\hbar}{2e} \frac{J_\mathrm{s}}{t_\mathrm{FM}}  \,\text{,} \quad \beta_{\perp} = \varepsilon_{\perp} \frac{\hbar}{2e} \frac{J_\mathrm{s}}{M_\mathrm{s}t_\mathrm{FM}}\,.
\end{equation}
\noindent Here, $\beta_{\parallel}$ and $\beta_{\perp}$ are the coefficients for the antidamping (in-plane) and field-like (out-of-plane) components of the spin--orbit torque (SOT), whereas the factors $\varepsilon_{\parallel}$ and $\varepsilon_{\perp}$ account for the efficiency of the spin-transfer process. $\hbar$ is the reduced Planck constant, $\bm{p}$ represents the spin-polarization orientation of the injected spin current, $t_\mathrm{FM}$ is the thickness of the ferromagnetic layer, and $J_\mathrm{s}$ represents the spin current density.
	
The first experimental observation of SOT effects on spin waves was a damping reduction due to a spin current generated via the SHE in permalloy/Pt bilayers.\cite{ando_electric_2008} The excitation of spin waves by SOT has been demonstrated in YIG/Pt heterostructures, \cite{demidov_excitation_2016,evelt_high-efficiency_2016,kajiwara_transmission_2010} whereas device nanopatterning allowed for the demonstration of spin Hall nano-oscillators (SHNOs), \cite{demidov_magnetic_2012,ando_electric_2008,liu_magnetic_2012,durrenfeld_20_2017} their synchronization to external microwave signals, \cite{demidov_synchronization_2014} and the mutual synchronization of SHNOs by pure spin currents.\cite{urazhdin_mutual_2016} Recently, it has also been shown that SOT antennas can excite spin waves when driven by microwave currents. It was estimated that the generated antidamping spin--Hall and Oersted fields contributed approximately equally to the total effective field, providing an improvement over conventional inductive antennas.\cite{talmelli_spin_2018}
	
\subsection{Magnetoelectric transducers}
\label{Sec:ME_transducers}
	
Magnetoelectric transducers are a more recent addition to the approaches to excite and detect spin waves. They are based on magnetoelectric compounds, which consist of piezoelectric and magnetostrictive bi- or multilayers. In such transducers, effective magnetoelastic fields are generated in the magnetostrictive ferromagnetic layer(s) via application of stress/strain due to the inverse magnetostriction (Villari) effect. The stress/strain itself can be generated by an electric field applied across the piezoelectric layer(s). Magnetoelectric transducers thus couple voltages with magnetic fields indirectly via mechanical degrees of freedom. Reviews of the magnetoelectric effect can be found in Refs.~\onlinecite{srinivasan_composite_2015, fiebig_revival_2005, srinivasan_magnetoelectric_2010, ma_recent_2011, martin_multiferroic_2012, fernandes_vaz_artificial_2013, fusil_magnetoelectric_2014, chu_review_2018, spaldin_advances_2019}.
	
In a magnetostrictive material, the application of a strain with tensor $\mathbf{\varepsilon}$ generates an effective magnetoelastic field. The magnetoelastic field is given by
\begin{equation}
\mathbf{H}_\mathrm{mel} = -\frac{1}{\mu_0} \frac{d \mathcal{E_\mathrm{mel}}(\mathbf{M})}{d\mathbf{M}} \,.
\end{equation}
\noindent as outlined above in Eq.~(\ref{eq:H_eff}). In general, the magnetoelastic field depends on the crystal symmetry of the magnetic material. An explicit formula can be derived for cubic crystal symmetry. In this case, the magnetoelastic field is given by\cite{kittel58}
\begin{equation}
\label{eq:mel-field}
\mathbf{H}_\mathrm{mel} = -\frac{2}{\mu_0M_\mathrm{s}}  \begin{pmatrix}
B_1\varepsilon_{xx}\zeta_{x} + B_2\left(\varepsilon_{xy}\zeta_{y}+\varepsilon_{xz}\zeta_{z}\right) \\
B_1\varepsilon_{yy}\zeta_{y} + B_2\left(\varepsilon_{xy}\zeta_{x}+\varepsilon_{yz}\zeta_{z}\right) \\
B_1\varepsilon_{zz}\zeta_{z} + B_2\left(\varepsilon_{xz}\zeta_{x}+\varepsilon_{yz}\zeta_{y}\right) 
\end{pmatrix}.
\end{equation}
\noindent Here, $B_1$ and $B_2$ are the magnetoelastic coupling constants of the waveguide material and $\boldsymbol{\zeta} = \mathbf{M}/M_\mathrm{s}$. Equation~(\ref{eq:mel-field}) also describes the case of an isotropic material. In this case, the magnetoelastic coupling constants are equal, \emph{i.e.}~$B_1=B_2$. This indicates that the magnetoelastic field depends on both the magnetization orientation and the strain tensor geometry. For uniform magnetization, Eq.~(\ref{eq:mel-field}) indicates that normal strain parallel or perpendicular to the magnetization does not exert a torque $\mathbf{T} = \mathbf{H}_\mathrm{mel}\times\mathbf{M}$ on the magnetization since the magnetoelastic field is either parallel to the magnetization or zero. By contrast, torques on the magnetization are exerted by oblique normal strain (with respect to the magnetization) or shear strain.\cite{duflou} 

So far, experimental studies have focused mainly on spin-wave excitation by propagating surface acoustic waves. \cite{Zhou14,Cherepov14,Foerster17,Weiler11,Dreher12,Gowtham15,Labanowski16,Thevenard14,Gowtham16,Bhuktare17,Li17,Verba18,Verba19} However, the interdigitated transducers used to excite surface acoustic waves are difficult to scale to small dimensions and resonance frequencies are typically well below ferromagnetic resonance even in low-$M_\mathrm{s}$ ferrites. In all cases, the excitation of spin waves requires strain fields oscillating at GHz frequencies. The, the strain tensor is not static but determined by the dynamic oscillating strain field generated by the transducer, which is typically characterized by a series of electromechanical resonances (standing waves) in the transducer itself and propagating elastic (acoustic) waves in the magnetic waveguide.\cite{logic1,Balinskiy18,bichurin_present_2010} Whereas the magnetoelastic coupling at low-frequency electromechanical resonances, below ferromagnetic resonance, is well understood,\cite{bichurin_theory_2005,laletin_frequency_2005,filippov_resonance_2007} few studies have addressed the coupling to acoustic waves at GHz frequencies (hypersound). When the transducer launches propagating acoustic waves, spin-wave excitation is generally nonlocal and occurs in the waveguide after acoustic wave propagation also.\cite{barra_voltage_2017} For mechanical resonators with high quality factors, the emission of elastic waves is however weak and thus spin waves are generated locally at the transducer. As for antennas, the spin-wave excitation efficiency is proportional to the overlap integral of the spatial distribution of the dynamic excitation field due to the standing waves in the transducer and the dynamic magnetization of the spin-wave mode, as described by Eq.~(\ref{eq:antenna_efficiency}). For $\mathbf{H}_\mathrm{mel} = \mathbf{h}_\mathrm{mel}\left({x,y}\right)e^{i\omega_\mathrm{mel}t}$, the excitation efficiency of a spin-wave mode with dynamic magnetization $\tilde{\mathbf{m}}({x,y})e^{i\omega_\mathrm{sw} t}$ in a thin waveguide can be written as
\begin{equation}
\label{eq:dirac_mel_efficiency}
\Gamma \propto \left| \iint {\mathbf{h}}({x,y})\cdot\tilde{\mathbf{m}}({x,y})\,  dx\,dy\right| \times \delta\left( \omega_\mathrm{mel} - \omega_\mathrm{sw}\right)\,.
\end{equation}
\noindent Here, the integral is carried out over the waveguide volume that is mechanically excited by the transducer. In contrast to the Oersted field generated by an inductive antenna, the magnetoelastic field is not necessarily uniform along the transverse $y$ direction, so modes with both odds and even mode numbers can in principle be excited. For small spin-wave amplitudes, the magnetization $\bm{\zeta}$ in Eq.~(\ref{eq:mel-field}) is equal to the static magnetization and does not change with time. By contrast, large spin-wave amplitudes can lead to considerable nonlinearities when $\bm{\zeta}$ precesses in time. For linear elastic systems, the integral in Eq.~(\ref{eq:dirac_mel_efficiency}) can be evaluated for each strain tensor component individually. As an example, the excitation efficiency of spin wave propagating along the $x$-direction with mode number $n$, wavenumber $k$, and angular frequency $\omega_\mathrm{sw}$ by an oscillating shear strain $\varepsilon_{xy}(x,y)e^{i\omega_\mathrm{mel}t}$ in a waveguide uniformly magnetized along the transverse $y$-direction is 
\begin{equation}
\Gamma_{\varepsilon_{xy}} \propto \left| B_2 \int \varepsilon_{xy}(x) e^{ikx}\,  dx\right| \times \delta\left( \omega_\mathrm{mel} - \omega_\mathrm{sw}\right)\,.
\end{equation}
\noindent Thus, the excitation efficiency is in this case given by the Fourier transform of the mechanical (strain) mode of the transducer in waveguide direction and can thus feature resonances that are linked to the mechanical response of the transducers. However, the mechanical behavior of realistic devices is expected to be rather complex and the understanding is currently only emerging.\cite{duflou,barra_voltage_2017,tierno_strain_2018,vanderveken_magnetoelectric_2019}

\begin{figure}
\includegraphics[width=7.5cm]{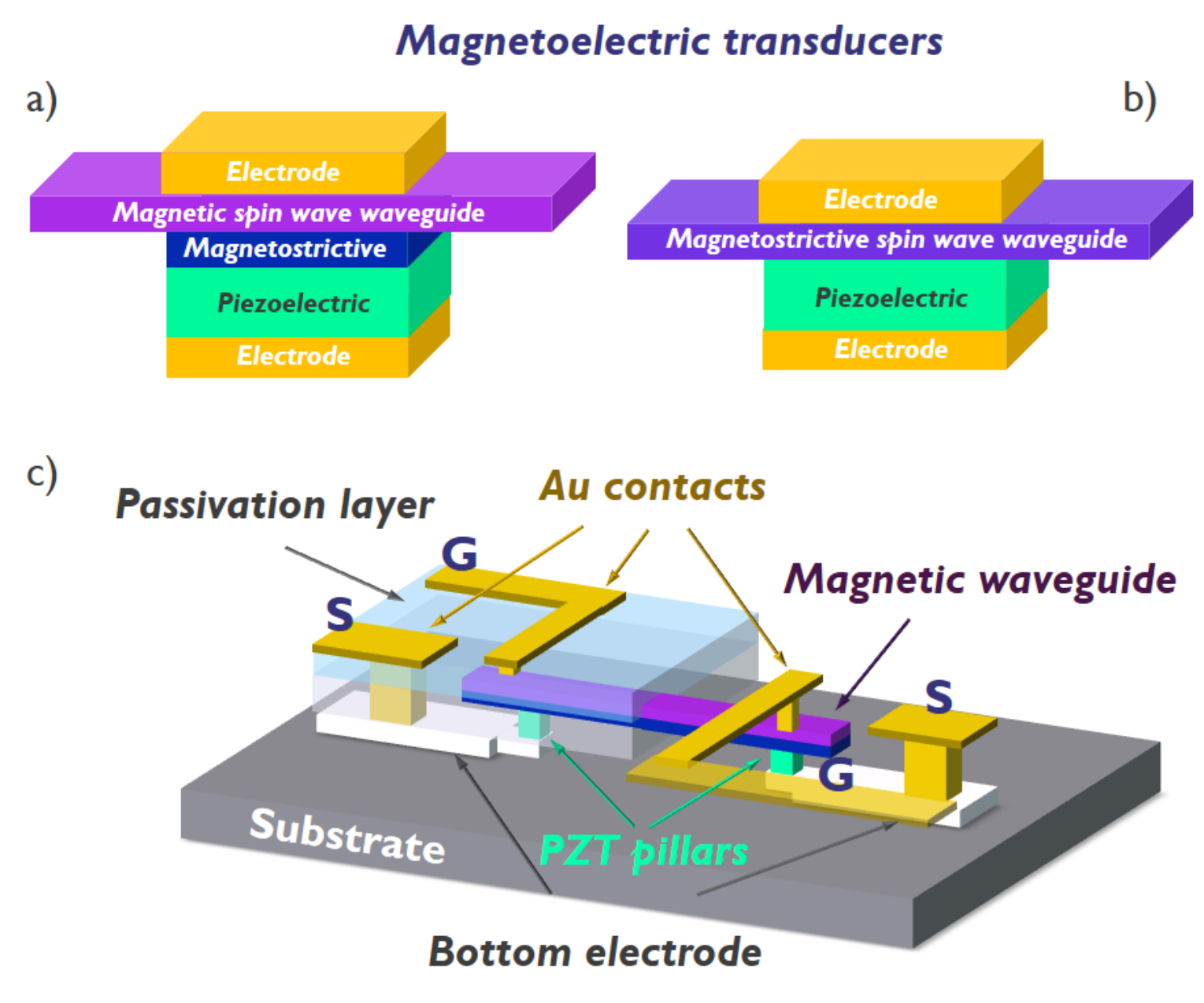}
\caption{Schematics of magnetoelectric transducers consisting of a piezoelectric element and a magnetic spin-wave waveguide formed by a) a ferromagnetic and magnetostrictive bilayer system, and b) by a simultaneously ferromagnetic and magnetostrictive single layer. c) Schematic of a spin-wave transmission experiment including on a magnetic waveguide for spin-wave propagation and two magnetoelectric transducers. Similar to the case of two antennas in Fig.~\ref{Fig_SW_trans}, the power transmitted by spin waves can be measured by a vector network analyzer connected to the ground (G) and signal (S) microwave electrodes of the devices.}
\label{fig:MEcell}
\end{figure}

In magnetoelectric compounds based on linear piezoelectric materials, the strain is proportional to an applied voltage---and therefore also the magnetoelastic field. Schematics of magnetoelectric transducers are depicted in Fig.~\ref{fig:MEcell}. Because typical charging energies of scaled magnetoelectric capacitors can be orders of magnitude lower than Ohmic losses in inductive antennas or STT devices, magnetoelectric transducers are potential candidates to enable low-power and high-efficiency transduction. Moreover, since the mechanism depends on electric fields, it shows favorable scaling properties with larger magnetoelectric voltage coupling for thinner piezoelectric films. 

Beyond the generation of spin waves by the magnetoelectric effect, also an inverse magnetoelectric effect exists, which can be used to detect spin waves. A spin wave in a magnetostrictive material creates a dynamic displacement field and thus an elastic wave.  This inverse effect therefore acts as an energy conversion mechanism from the magnetic to the elastic domain. The effect can cause additional losses of propagating spin waves by emission of elastic waves. These magnetoelastic losses can be limited by reducing the ``inverse'' overlap integral between the dynamic magnetization of the spin wave and the displacement field of the elastic wave as well as the overlap with elastic resonances. However, this inverse coupling can also be applied to design spin wave detectors. When the displacement field of the elastic wave induces strain in an adjacent piezoelectric capacitor, it creates an oscillatory charge separation and an oscillating electric polarization in the piezoelectric material. The polarization can then be read out as a microwave voltage. 

The mutual interactions between spin waves and elastic waves (action and back action) in magnetostrictive media can lead to the formation of strongly coupled magnetoelastic waves when the respective dispersion relations cross. The physics of magnetoacoustic waves is well understood in bulk materials,\cite{Akhiezer59,Fedders74,Tucker72,kittel58} although their behavior in thin films and waveguides has only recently been studied.\cite{vanderveken_magnetoelastic_2020} When magnetoelectric transducers are employed, the excitation of magnetoacoustic waves may allow for the maximization of the transduction efficiency although concrete device proposals based on magnetoacoustic waves are still lacking.

\subsection{Voltage control of magnetic anisotropy (VCMA)}
\label{Sec:VCMA}

A different type of magnetoelectric effects relies on the voltage control of magnetic anisotropy (VCMA).\cite{weisheit_electric_2007,duan_surface_2008,maruyama_large_2009,khalili_amiri_electric-field-controlled_2015} VCMA describes the modulation of the perpendicular magnetic anisotropy (PMA) of ultrathin magnetic films in a magnetic tunnel junction by an electric field. In many cases, PMA can be induced in ultrathin films and multilayers of $3d$ ferromagnets (\emph{e.g.}~Fe, Co, Ni, or their alloys) by forming interfaces with nonmagnetic metals (\emph{e.g.}~Pt, Pd, W, Au)\cite{pal_tunable_2011} or metal oxides (\emph{e.g.}~Al$_2$O$_3$, MgO, Ta$_2$O$_5$, HfO$_2$).\cite{pal_tunable_2011,Miyazaki1995,moodera_large_1995,yuasa_characterization_2005,niranjan_electric_2010,yang_first-principles_2011,vermeulen_ferroelectric_2019} As an example, the interfacial PMA in CoFeB/MgO heterostructures originates from the strong bonding of the $3d$ orbitals of Fe with the $2p$ orbitals of O. The electric field induced by applying a voltage across the interface between the MgO and CoFeB layers changes the electron density in the $3d$ orbitals of Fe, and implicitly their coupling strength with the $2p$ orbitals, impacting thus the interfacial PMA.\cite{kawabe_electric_2017}

\begin{figure*}[t]
\includegraphics[width=15cm]{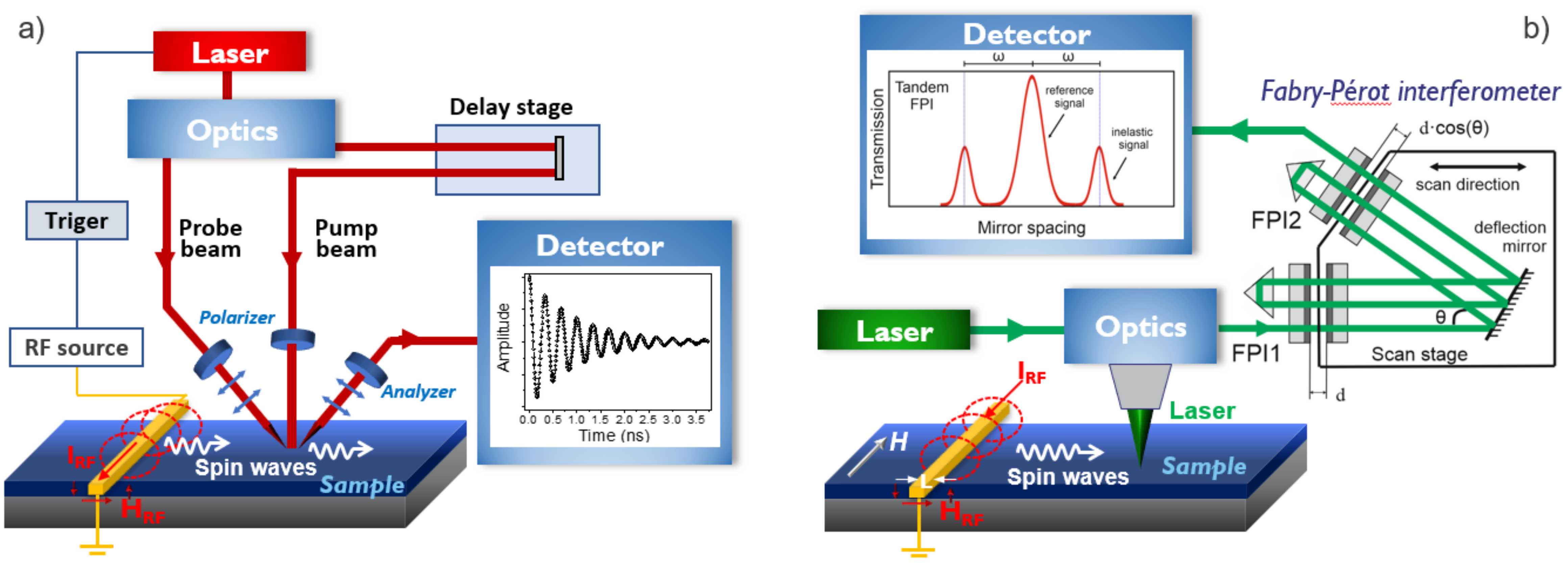}
\caption{Simplified scheme of a time-resolved magneto-optic Kerr effect setup (a), and of a Brillouin light scattering setup (b), respectively. }
\label{fig:BLS_Setup}
\end{figure*}

Recent studies have demonstrated that the dynamic VCMA effect by microwave (GHz) electric fields can excite ferromagnetic resonance (FMR) in \textmu{}m-scale\cite{nozaki_electric_2012} to nm-scale\cite{zhu_voltage_2012} magnets with a power consumption of at least two orders of magnitude less than the direct current induced STT excitation.\cite{nozaki_electric_2012} Furthermore, it was demonstrated that VCMA-based transducers can emit propagating spin waves.\cite{verba_excitation_2016,Rana_VCMA_2017,rana_towards_2019} A disadvantage of VCMA-based transducers is however that no spin waves or FMR-like magnetic excitations can be excited in magnets that are uniformly magnetized either in-plane ($\theta=\frac{\pi}{2}$) or out-of-plane ($\theta=0$).\cite{verba_parametric_2014} However, spin waves can still be generated in these configurations by means of nonlinear parallel parametric pumping, in which the VCMA transducer is driven at twice the frequency of the excited spin-wave modes (\emph{cf}.~Sec.~\ref{Sec_NL_SW_phys}).\cite{verba_parametric_2014,chen_parametric_2017}

While VCMA-based transducers are established for the generation and amplification of spin waves with promise for scalability and low power consumption, the detection of spin waves by VCMA-like effects is an emerging topic.\cite{Shukla_VCMA_2020} In addition, homodyne detection schemes may be used, in which the microwave signal from a spin wave is rectified and generates a DC voltage.\cite{nozaki_electric_2012,zhu_voltage_2012} A drawback for such detection schemes is however the low output voltage, typically a few \textmu{}V, which needs to be amplified to be read by conventional CMOS circuits. Furthermore, the phase information is lost since the output is converted to a DC signal.

\subsection{Optical excitation and detection of spin waves}
	
While the integration of optical transducers into hybrid spin-wave systems is not practical, optical spin-wave excitation and measurement schemes are widely used in many magnonic experiments. Moreover, optical methods are capable of accessing the magnetization dynamics at ultrashort timescales of ps down to fs, which are difficult or impossible to assess by microwave electronics. Therefore, in this section, different optical methods to excite and detect spin waves are briefly reviewed. 

It is well known that ultrashort optical pulses with durations of ps down to fs can generate spin waves in magnetic samples by different mechanisms. For example, the inverse Faraday effect can be exploited to generate an effective magnetic field in a transparent ferromagnet, generated from a circularly polarized light pulse. The effective magnetic field is parallel to the direction of the laser beam and can exert a torque on the magnetization. Hence, it can cause the emission of spin waves.\cite{satoh_directional_2012,savochkin_generation_2017} In addition, laser-induced thermal effects can either decrease the magnetic anisotropy\cite{ogawa_photodrive_2015, khokhlov_optical_2019} or lead to an ultrafast demagnetization process with the generation of spin waves.\cite{van_kampen_all-optical_2002,au_direct_2013,kirilyuk_laser-induced_2013,iihama_quantification_2016} Furthermore, it was demonstrated that the properties of the emitted spin waves, such as their wavelength and energy flow direction, can be steered by shaping the laser spot or tuning the sequence of the excitation pulses.\cite{au_direct_2013,jackl_magnon_2017,savochkin_generation_2017} 
	
The study of the magnetization dynamics induced by (sub)-ps laser pulses relies typically on pump-and-probe techniques. The first (pump) pulse triggers the magnetization oscillation whereas the probe pulse interacts with the sample after a delay (Fig.~\ref{fig:BLS_Setup}(a)). The magnetization orientation can be measured by the change in the polarization of a reflected probe pulse due to the magneto-optic Kerr effect (MOKE). Alternatively, the Faraday effect can be used in a transmission geometry.\cite{kirilyuk_laser-induced_2013,hiebert_direct_1997,freeman_advances_2001,bauer_time_2001} The time resolution of the measurement is provided by the delay between pump and probe pulses and can easily reach ps time scales. High spatial resolution can be obtained by focusing the pulses on the sample. The resolution is limited by diffraction effects and the numerical aperture of the used microscope. The time-resolved MOKE can also be used to detect spin waves emitted by electric transducers. In this case only the probe beam is in operation.

In addition, Brillouin light scattering (BLS) spectroscopy (Fig.~\ref{fig:BLS_Setup}) is a powerful technique to investigate magnetization dynamics because of its very high sensitivity to small spin-wave amplitudes (including \emph{e.g.}~thermal spin waves),\cite{sandercock_light_1978,hillebrands_situ_1987} and high versatility.\cite{demidov_generation_2011}  BLS allows to study magnetization dynamics with spatial, \cite{demidov_radiation_2004,banholzer_visualization_2011} temporal, \cite{buttner_linear_2000} and phase resolution,\cite{serga_phase-sensitive_2006} as well as with wavevector selectivity.\cite{sandweg_wide-range_2010}

The physical mechanism of BLS is based on the interaction of monochromatic light with a material whose optical density varies with time and changes the light energy (frequency) and path. The optical density may vary due to the presence of acoustic excitations (phonons), magnetic excitations (spin waves), or thermal gradients in the medium. The presence of spin waves in the material creates a phase grating in the dielectric permittivity, which propagates with the spin-wave phase velocity. The incident light is Bragg reflected by the phase grating and its frequency undergoes a Doppler shift corresponding to the spin-wave frequency. The change in the direction of the scattered light is related to the periodicity of the phase grating. Thus, Brillouin scattered light contains information about magnetization dynamics in solids and can be used to probe the characteristics of magnetic excitations. The frequency analysis of the scattered light can be realized by a tandem Fabry--P\'erot interferometer [Fig.~\ref{fig:BLS_Setup}(b)].\cite{mock_construction_1987,hillebrands_progress_1999} The frequency range of the interferometer is typically several hundred GHz, whereas the frequency resolution depends on the frequency range and can reach a few 10 MHz at frequencies of a few GHz. The minimum detectable spin-wave wavelength is given by half the wavelength of the used laser light (\emph{e.g.}~$\lambda_\mathrm{SW,min}= 266$ nm for a green laser with $\lambda = 532$ nm). BLS microscopy integrates a microscope objective with a high numerical aperture to focus the light onto the sample. Scanning the focus position can then be used to image the spin-wave intensity with a spatial resolution of about 250 nm.\cite{Sebastian15}

\section{Spin-Wave Devices}
\label{sec:General Spin Wave device structure}

After introducing basic concepts of spin-wave computing and the transducers at the input and output ports of spin-wave devices, we now discuss practical implementations of logic elements and gates that can be used to design spin-wave logic circuits. While nonlinear devices such as spin-wave transistors and directional couplers are also reviewed, the section focuses on passive linear logic gates based on spin-wave interference. Linear passive gates take the most advantage of the wave computing paradigm and bear the highest promise for ultralow-power electronics. The repercussions of such approaches for circuit design are then discussed in Sec.~\ref{sec:Requirements for Spin Wave Circuit Design}.

\subsection{Spin-wave conduits}

\begin{figure}
	\includegraphics[width=7.5cm]{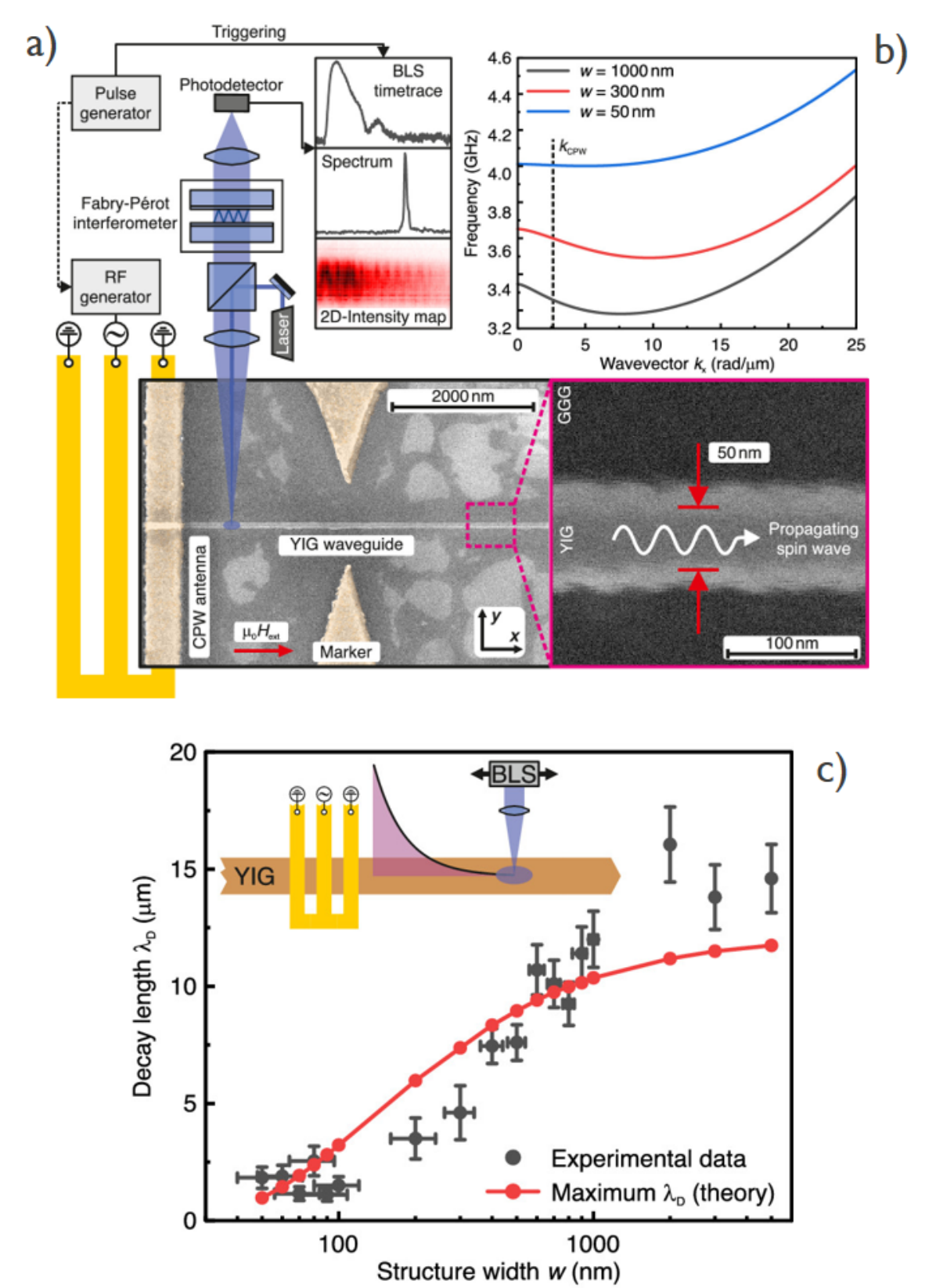}
	\caption{a) Schematic of the BLS experimental configuration and scanning electron micrographs of a 50~nm wide YIG conduit. b) Spin-wave dispersion relations for YIG waveguides with different widths ($w = 1000$, 300, 50~nm) in the backward volume geometry. c) Experimental spin-wave attenuation (decay) length \emph{vs.} structure width. Reproduced with permission from Heinz, T. Brächer, M. Schneider, Q. Wang, B. Lägel, A. M. Friedel, D. Breitbach, S. Steinert, T. Meyer, M. Kewenig, C. Dubs, P. Pirro, and A. V. Chumak, Nano Lett. 20, 4220 (2020). Copyright 2020 Nano Letter.} 
	\label{fig:Scaled_YIG}
\end{figure}

The most fundamental element for information processing and transfer by spin waves is a waveguide: the spin-wave conduit. In the conduit, information encoded in the spin-wave amplitude or phase propagates at the spin-wave group velocity, which depends on material, frequency, and the effective static magnetic bias field in the waveguide. When the spin wave wavelength is comparable to the conduit length, the phase of the spin wave oscillates along the conduit. An ideal conduit material combines low Gilbert damping and high Curie temperature. Large saturation magnetization $M_\mathrm{s}$ maximizes the spin wave power transmission and increases the output signal by inductive antennas but also reduces the magnetoelastic coupling [\emph{cf.} Eq.~(\ref{eq:mel-field})]. Typical materials include YIG with very low Gilbert damping in single-crystal form or more CMOS-compatible polycrystalline or amorphous metallic ferromagnets such as CoFeB or permalloy (Ni$_{80}$Fe$_{20}$), with Heusler alloys such as Co$_2$(Mn$_\mathrm{x}$Fe$_{1-\mathrm{x}}$)Si emerging.\cite{Sebastian12,felser_heusler_2015, palmstrom_heusler_2016,wollmann_heusler_2017} Basic magnetic properties of these materials are listed in Tab.~\ref{table:1}. Spin-wave conduits show excellent scalability at the nanoscale and propagation of backward volume spin waves in YIG waveguides as narrow as 50 nm has been demonstrated (Fig.~\ref{fig:Scaled_YIG}), albeit with reduced attenuation length.\cite{heinz_propagation_2020}

The routing of spin waves in conduits is however complicated by the anisotropic dispersion relation in the dipolar regime (see Sec.~\ref{Magnetization dynamics and spin waves}). For example, for a given frequency and an in-plane magnetization direction, the wavelength and group velocity of spin waves in orthogonal planar waveguides are generally different. The anisotropy also affects spin-wave propagation around corners and in curved waveguides, in addition to the effects of inhomogeneous magnetization and demagnetizing field in such structures. Although spin waves can be guided along curved waveguides, this typically results in additional losses.\cite{vogt_spin_2012, tkachenko_propagation_2012, xing_how_2013, sadovnikov_spin_2017} Although special waveguide designs alleviate the issue to some extent,\cite{davies_towards_2015,haldar_reconfigurable_2016,vogel_control_2018,albisetti_nanoscale_2018} the routing capabilities of spin waves at the nanoscale are limited, with repercussions on the spin-wave devices layout and scalability. 

In planar conduits, these issues can be avoided when the magnetization is perpendicular to the plane since the in-plane spin-wave properties are in this case isotropic.\cite{haldar_isotropic_2017} While the use of forward volume spin waves in such a configuration is clearly advantageous with more flexible device design options,\cite{logic14} the implementation is hampered by the lack of magnetic materials with simultaneous strong perpendicular anisotropy and low damping. In thin waveguides, the demagnetization field (see Sec.~\ref{Magnetization dynamics and spin waves}) leads to a strong magnetic anisotropy with an in-plane easy axis. To rotate the easy axis out of plane, the in-plane shape anisotropy must be overcompensated by a perpendicular anisotropy. While this can be achieved using \emph{e.g.}~magnetocrystalline \cite{ohandley_magnetic_1999,soumah_ultra-low_2018,bauer_dysprosium_2020} or interfacial anisotropies,\cite{pechan_interfacial_1987,johnson_magnetic_1996,dieny_perpendicular_2017} the integration of such materials with low damping in real devices is still challenging. 

Beyond patterned waveguides, spin waves can also be routed in ferromagnetic domain walls.\cite{garcia-sanchez_narrow_2015,wagner_magnetic_2016} While this may allow in principle for high-density conduits structures, the fabrication of stable domain-wall networks connecting logic gates is challenging. Concepts for routing spin-wave information in three-dimensional networks including multiple layers connected by vias are emerging only very recently.\cite{beginin_spin_2018} Multilevel spin-wave interconnects allow for more flexible routing and potentially smaller spin-wave devices and circuits, although this is not a \emph{sine qua non} requirement for spin-wave circuit design. Such approaches are however again strongly affected by the anisotropic spin-wave dispersion relation. 

Similar to the noisy voltage signal propagation in metallic wires,\cite{johnson_thermal_1928} the spin-wave propagation in ferromagnetic waveguides is affected by thermal noise.\cite{thermal} At nonzero temperature, spin waves are thermally excited according to the Bose-Einstein distribution since the quanta of spin waves, \emph{i.e.}~magnons, are bosons.\cite{van_kranendonk_spin_1958} Thermally-excited spin waves are incoherent and produce a background superimposed to coherent spin-wave signals used for computation. Moreover, adjacent waveguides may also suffer from crosstalk. The dipolar magnetic fields generated by propagating spin waves extend beyond the waveguide and can excite spin waves in adjacent waveguides. This leads to signal crosstalk between waveguides as well as to additional propagation losses. Ultimately, this effect may limit the density of spin-wave conduits and devices in a circuit. More details on noise, crosstalk, and mitigation techniques can be found in Refs.~\onlinecite{crosstalk} and \onlinecite{thermal}.

\subsection{Magnonic crystals}

\begin{figure}
\centering
  \includegraphics[width=6.5cm]{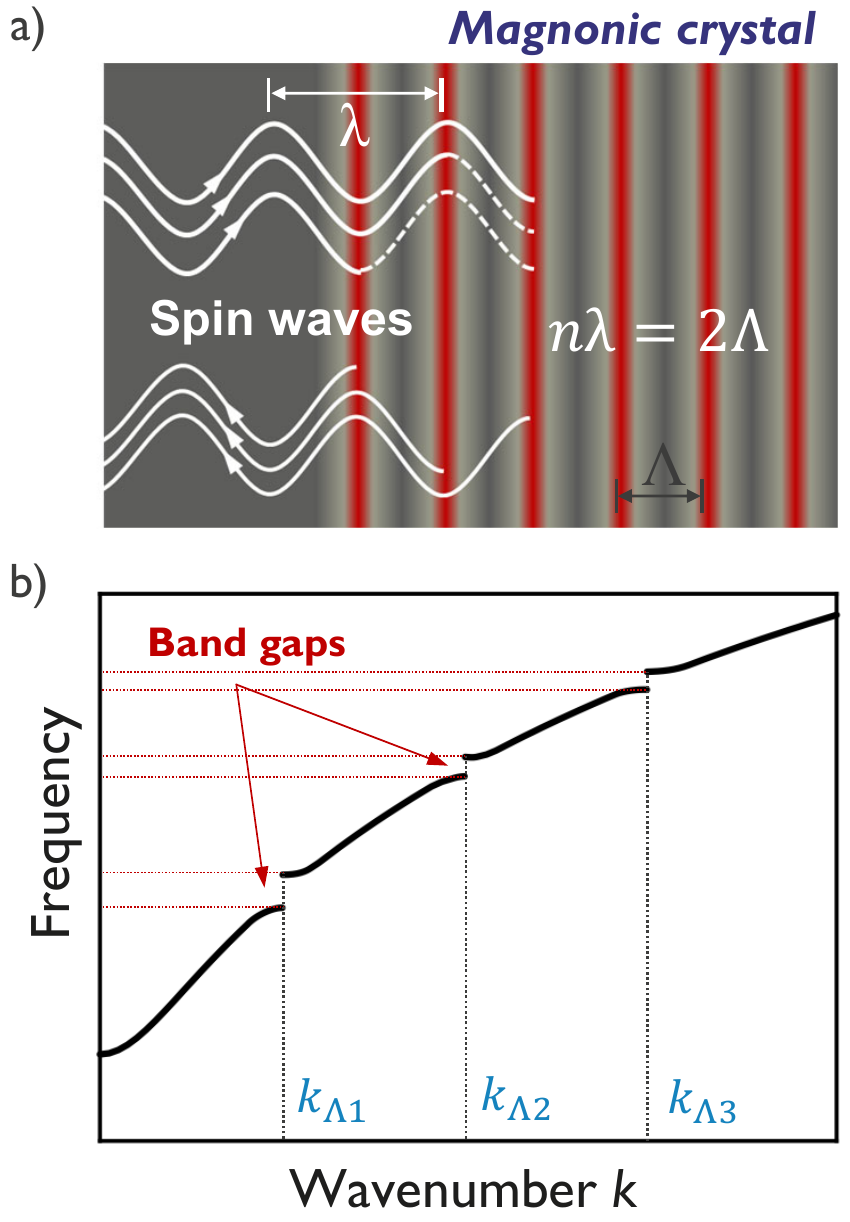}
  \caption{(a) Schematic representation of Bragg reflection of spin waves (wavelength $\lambda$) from a one-dimensional magnonic crystal with periodicity $\Lambda$. (b) Dispersion relation of a SSW mode in a magnonic crystal. The frequency band gaps corresponding to wavenumbers $k_{\Lambda n} = n\pi/\Lambda$ with $n=1,2,3$ are indicated.}
  \label{fig:SW_crystal}
\end{figure}

The spin-wave propagation can be further manipulated by engineering locally the magnetic properties or the shape of the waveguide. Periodic manipulations lead to magnonic crystals. Magnonic cystals are magnetic media whose magnetic properties change periodically in one, \cite{deng_magnon_2002,kostylev_collective_2004} two, \cite{gubbiotti_collective_2012,kumar_magnonic_2014} or three dimensions.\cite{krawczyk_plane-wave_2008,romero_vivas_investigation_2012,Gubiotti_3D_MC} They can be considered as the magnonic equivalents to optical Bragg mirrors. The transmission spin-wave spectra through a magnonic crystal show rejection band gaps, \emph{i.e.} frequency intervals, in which spin waves are forbidden to propagate.\cite{Gulyaev_Bragg,vysotskii_magnetostatic_2005} 

The formation of such band gaps can be attributed to Bragg reflections of the spin waves by the artificial spatial grating created in the magnetic properties of the structure [Fig.~\ref{fig:SW_crystal} (a)]. Thus, the spectral positions of the band gaps are determined by the spatial modulation periodicity of the crystal [see Fig.~\ref{fig:SW_crystal}(b)]. The Bragg condition for the forbidden spin-wave modes can be written as
\begin{equation}
2\Lambda  = n \lambda \quad n \in \mathbb{N}\, ,
\end{equation} 
where $\Lambda$ is the periodicity of the magnonic crystal modulation and $\lambda$ is the wavelength of the spin wave. The depth and the width of the band gaps are controlled mainly by the amplitude modulation of the magnetic or geometric parameters. The spin-wave transmission and the spectral position of the band gaps have been investigated for various types of magnonic crystals based on numerous magnetic materials, \cite{deng_magnon_2002,Kruglyak_magnonics_2006} different shapes of the waveguide,\cite{chumak_scattering_2008,chumak_spin-wave_2009} local modulation of the saturation magnetization,\cite{obry_micro-structured_2013,ciubotaru_magnonic_2013} or local variations of the bias field.\cite{chumak_current-controlled_2009} 

Magnonic crystals can be potentially used in a number of applications, such as spectral filters, delay lines, or phase shifters (inverters, see below). They also form a central part of some spin-wave transistor approaches, as discussed in the next sections. More details on magnonic crystals can be found \emph{e.g.} in Refs.~\onlinecite{krawczyk_review_2014, Magnon_transistor}.

\subsection{Spin-wave transistors}
\label{Sec:SW_transistors}

The basic building block of CMOS circuits is a transistor. Given success of CMOS, one may find it thus natural to mimic the transistor functionality using spin waves. A conventional transistor can act both as a switch as well as an amplifier and shows nonlinear characteristics. Spin-wave  transistors thus typically employ nonlinear effects (see Sec.~\ref{Sec_NL_SW_phys}) beyond the linear small-signal approximation in Sec.~\ref{Magnetization dynamics and spin waves}.\cite{mag1,cottam_linear_1994,Bauer15,Kalinikos13}

\begin{figure}
\centering
  \includegraphics[width=8cm]{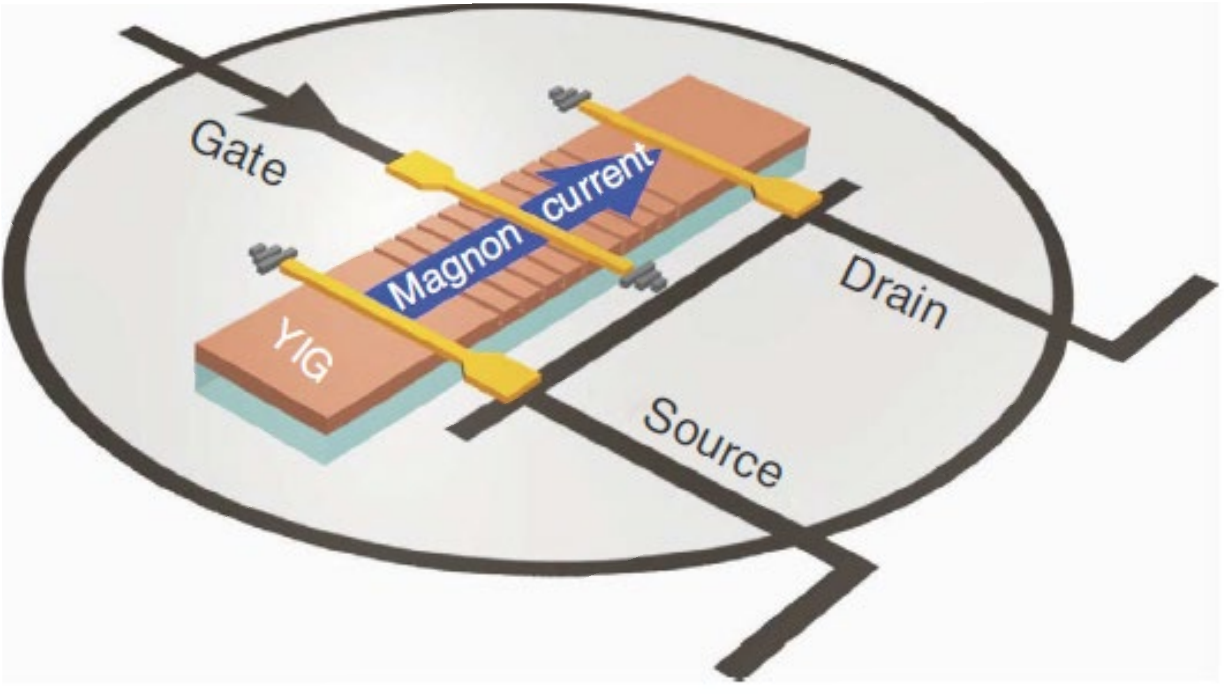}
  \caption{Schematic of a spin-wave transistor. Reproduced with permission from V. Chumak, A. A. Serga, and B. Hillebrands, Nature Communication 5, 4700 (2014). Copyright 2014 Nature Communication.}
  \label{fig:SW_transistor}
\end{figure}

A proposal of a nonlinear spin-wave transistors has been published in Refs.~\onlinecite{chumak_magnon_2014,Magnon_transistor}. They are based on nonlinear interactions of spin waves propagating in a waveguide from ``source'' to ``drain'' with spin waves that are injected in a ``gate'' section of the waveguide (see Fig.~\ref{fig:SW_transistor}). The presence of spin waves in the gate modulates the spin-wave transmission along the ``channel'' via four-magnon scattering. To optimize the modulation and to confine the spin waves in the gate, the central section of the transistor consists of a magnonic crystal, as discussed in the previous section. 

Recently, a ``linear'' transistor that does not require nonlinear interactions between spin waves has been demonstrated.\cite{Magnon_transistor1} In this device, spin waves propagate in a waveguide from source to drain and interfere constructively or destructively with spin waves with variable phase from the gate. In this way, the spin-wave flow from source to drain can be modulated by the gate spin waves.

The modulation of spin-wave transmission between source and drain by spin-wave injection into the gate allows for the operation of such a device as a switch. By contrast, the proposed spin-wave transistors show no (or at best weak) gain and thus cannot be operated as amplifiers, which complicates their usage in spin-wave circuits (\emph{cf.}~Sec.~\ref{sec:Requirements for Spin Wave Circuit Design}). Together with the rather weak modulation of the spin current (well below the typical on--off current ratios of $10^6$ in CMOS transistors), this entails that spin-wave transistors are no direct alternative to CMOS transistors. Nevertheless, the spin-wave transistor prototype \cite{chumak_magnon_2014} opened a new research avenue for all-magnon data processing. In this concept the spin-wave nonlinearity is used to process as much information as possible in the magnetic system instead of conversion of spin-wave energy in electric signals after each gate. This approach was used for the realization of a directional coupler based on spin waves,\cite{cross} and a first integrated magnonic circuit in a form of a half-adder.\cite{wang_realization_2019} These concepts will be discussed in one of the following sections. 

\subsection{Spin-wave logic gates}
\label{Sec_SW_Logic_gates}

Conventional logic CMOS circuits are not designed directly on a transistor level but rather constructed based on a set certain universal building blocks (standard cells), such as \emph{e.g.} NAND or NOR logic gates or SRAM cells. Therefore, it is interesting to develop an equivalent set of spin-wave-based logic gates. As argued above, constructing logic gates from spin-wave transistors does currently not appear promising. A better approach is the design of logic gates using the interference-based paradigm discussed in Sec.~\ref{Computing_schemes_interference}. Different concepts for the implementation of spin-wave logic gates have been proposed, using the different encoding schemes introduced in Sec.~\ref{Information_encoding}. A main advantage is that these gates are linear passive devices and do not require any energy beyond the energy in the spin waves themselves, which renders such approaches promising for ultralow-power computing applications assuming that the involved spin waves can be efficiently excited. 

\subsubsection{Inverters and phase shifters}

\begin{figure}[b]
\centering
  \includegraphics[width=7cm]{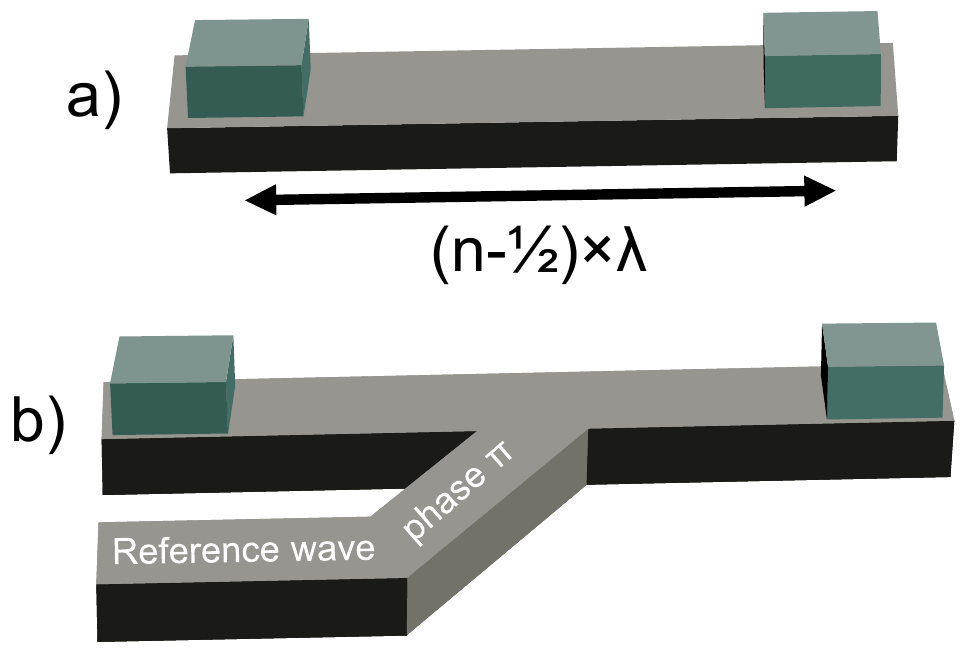}
  \caption{Implementation of spin-wave inverters. (a) Phase encoding: inversion occurs by propagation along a ``delay line'' with a length of $\left( n - \frac{1}{2}\right)\times\lambda$ with $\lambda$ the spin-wave wavelength and $n = 1,2,3,\ldots$ an integer. (b) Amplitude encoding: inversion occurs by interference with a reference wave with phase $\pi$.}
  \label{fig:Inverter}
\end{figure}

Before discussing more complicated logic gates, it is instructive to review inverter concepts for different encoding schemes. The simplest inverter is obtained by using phase encoding since in this case, logic inversion corresponds simply to a phase shift of $\pi$. Such a phase shift can be achieved by propagation in a waveguide with a length of $L = \left( n - \frac{1}{2}\right)\times\lambda$ with $\lambda$ the spin-wave wavelength and $n = 1,2,3,\ldots$ an integer. The advantage of such inverters is that they are passive and do not require additional external power. A schematic of such an inverter is graphically depicted in Fig.~\ref{fig:Inverter}(a).

\begin{figure*}[t]
\centering
  \includegraphics[width=16.4cm]{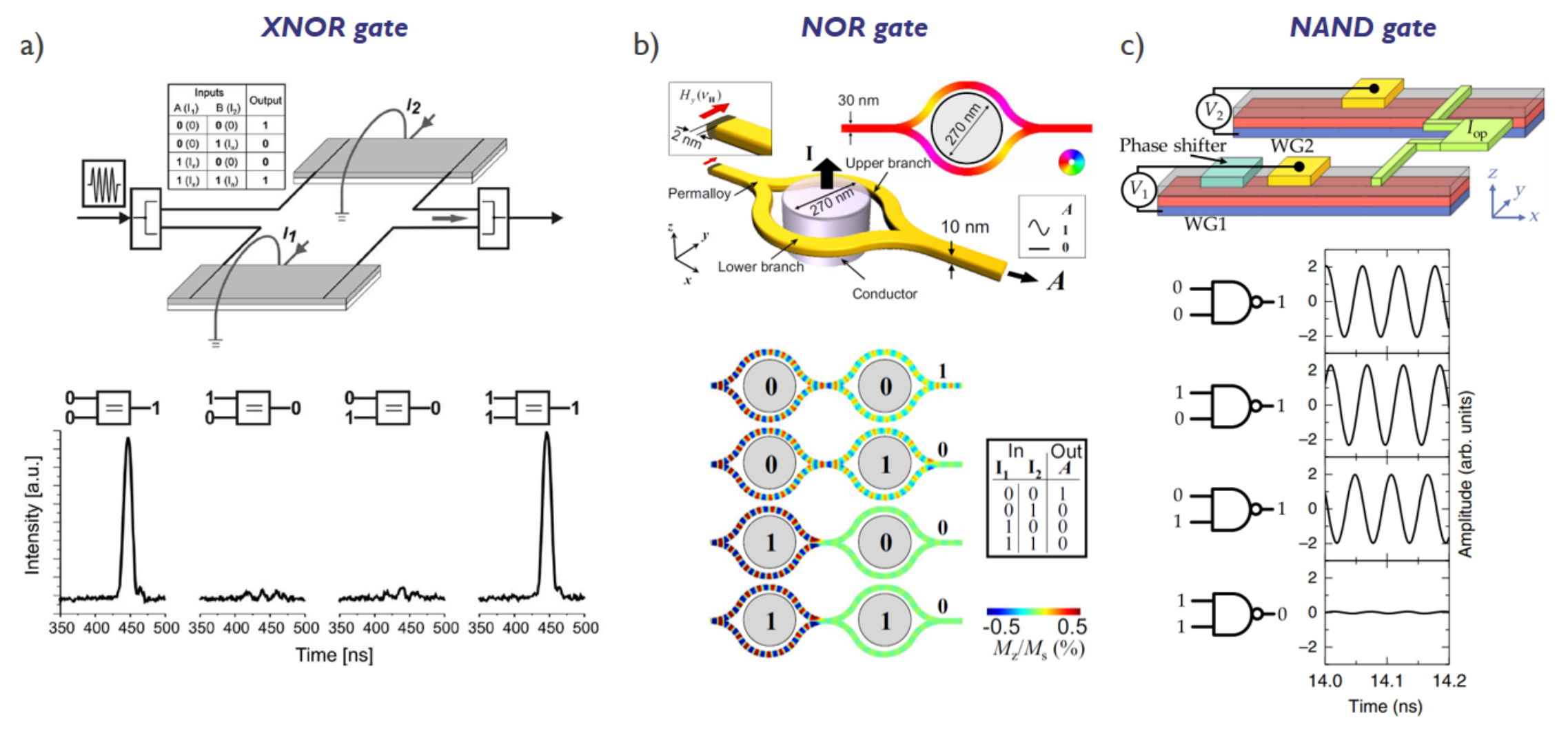}
  \caption{Implementation of spin-wave logic gates based on Mach--Zehnder interferometers. (a) XNOR gate consisting of two yttrium iron garnet (YIG) waveguides. The currents $I_{1}$ and $I_{2}$ represent the logical inputs whereas the logical output is given by the spin-wave interference signal. Reproduced from Schneider, A. A. Serga, B. Leven, B. Hillebrands, R. L. Stamps, and M. P. Kostylev, Appl. Phys. Lett. 92, 022505 (2008), with the permission of AIP Publishing. (b) NOR gate consisting of the two Mach--Zehnder interferometers in a serial configuration. Reproduced from S. Lee and S.-K. Kim, J. Appl. Phys. 104, 053909 (2008), with the permission of AIP Publishing. (c) Voltage-controlled universal NAND gate consists of two parallel waveguides. Reproduced with permission from B. Rana and Y. Otani, Phys. Rev. Appl. 9, 014033 (2018). Copyright 2018 Physical Review Applied. }
  \label{fig:MachZehnder}
\end{figure*}

In addition, phase shifting concepts can be based on the local modification of the spin-wave dispersion relation. Such inverters can potentially be even more compact than delay lines.\cite{vasiliev_spin_2007,demidov_control_2009,liu_electric_2011} Local changes in saturation magnetization or waveguide width can lead to a local change in wavelength, leading to an additional phase shift with respect to an unperturbed waveguide. Alternatively, external magnetic bias  fields can also be used, including effective fields generated by magnetoelectric effects (\emph{cf.}~Sec.~\ref{Sec:ME_transducers}) or VCMA (\emph{cf.}~Sec.~\ref{Sec:VCMA}), which promise to be more energy efficient than Oersted fields generated by a current. An advantage of such concepts is that they can be reconfigurable, \emph{e.g.}~when a VCMA capacitor is used to generate the effective magnetic field. Magnonic crystals can also be used to generate phase shifts and invert a phase-coded signal. A disadvantage is the more complex device structure as well as potentially the required additional power, \emph{e.g.} when an electromagnet is used. A highly beneficial property of such inverters is that they do not need to be separate logic gates but can be integrated in the design of \emph{e.g.} the spin-wave majority gates discussed below. Extending the length of an input or output waveguide by $\frac{\lambda}{2}$ renders the input or output inverting. In general, this can be expected to reduce the size of spin-wave circuits considerably. 

In case of amplitude level encoding, inverters can be obtained by interference with a reference wave of phase $\pi$. For a suitably chosen geometry [Fig.~\ref{fig:Inverter}(b)], the reference wave interferes destructively with a potential signal wave. If a wave is present, its amplitude is reduced to zero, \emph{i.e.}~an output of 0 is obtained for an input of 1. For an input of 0, the reference wave reaches the output, leading to a logic 1. Such inverters are not passive, unlike the above delay lines, and therefore require additional power to generate the reference wave.

\subsubsection{Amplitude level encoding: logic gates based on interferometers}
\label{Sec_ALEnc_gates}

Initial work on spin-wave logic gates has mainly focused on amplitude level coding in combination with a device design based on an analog of a Mach–-Zehnder interferometer.\cite{logic11,logic12,logic21,logic24,rana_towards_2019} In such a spin-wave interferometer, an incoming spin wave is split into two waves in the interferometer arms. A current flowing through a wire perpendicular to the plane of the interferometer generates an Oersted field, which leads to a relative phase shift of the spin waves in the two interferometer arms. Subsequently, the waves are recombined and interfere. The relative phase shift, and therefore the amplitude of the output wave, depends thus in an oscillatory way on the current in the wire.

This approach can be used to design different logic gates, such as XNOR, NOR, or NAND. Basic gate structures and their operation principles are depicted in Fig.~\ref{fig:MachZehnder}. It should be mentioned that such logic gates are inherently hybrid devices since input signals are encoded in currents whereas output signals employ spin waves for information encoding. For logic gate operation, the parameters are chosen so that an input current leads to destructive spin-wave interference in the interferometer (logic 0), whereas no current leads to constructive interference (logic 1). Additional interference between spin waves emanating from different interferometers can in principle be used for more complex logic gates or circuits. Alternative proposals use voltages rather than currents, \emph{e.g.}~via VCMA or magnetoelectric effects, to modulate the spin-wave phase during propagation.\cite{logic24,rana_towards_2019}

Several logic gates---\emph{e.g.} NOT, NAND, or XNOR---have been demonstrated experimentally, as illustrated in Fig.~\ref{fig:MachZehnder}(a) for XNOR.\cite{logic12,logic21} Device sizes were a few mm. Since the device operation is based on Oersted fields generated by \emph{currents}, scaling the devices leads to a strongly increasing current densities in the wires and to reliability (\emph{e.g.}~electromigration) issues. If the distance between the wire and the waveguide is also scaled, a part of the increase in current density can be avoided. Nonetheless, such current-based devices scale significantly worse than devices operating with voltages or current densities. In addition, the hybrid character of the logic gates leads to cascading issues since the output of a logic gate (spin-wave amplitude/intensity) cannot be used as an input for a subsequent gate, which requires encoding in a current. Therefore, practical spin-wave circuits entail additional electric circuits for signal conversion. Such issues are discussed in more detail in Sec.~\ref{sec:Requirements for Spin Wave Circuit Design}. 

\begin{figure*}
\centering
  \includegraphics[width=16cm]{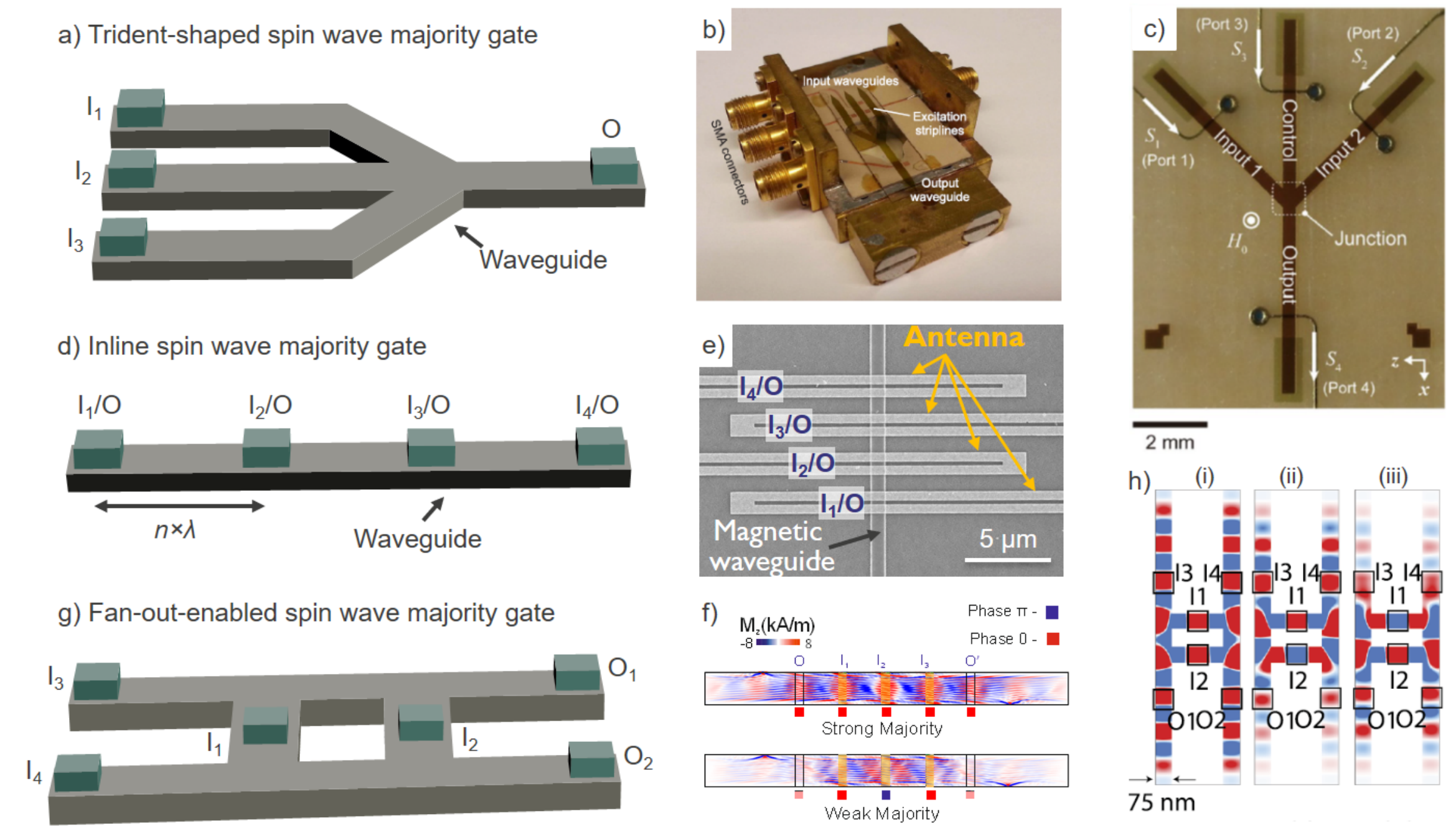}
  \caption{Overview over interference-based spin-wave majority gates. (a) Schematic of a trident-shaped spin-wave majority gate. $I_1$ to $I_3$ indicate the three input ports, whereas $O$ indicates the output port. (b) and (c) Photographs of experimental trident-shaped spin-wave majority gates using YIG. Reproduced from T. Fischer, M. Kewenig, D. A. Bozhko, A. A. Serga, I. I. Syvorotka, F. Ciubotaru, C. Adelmann, B. Hillebrands, and A. V. Chumak, Appl. Phys. Lett. 110, 152401 (2017), with the permission of AIP Publishing, and Kanazawa, T. Goto, K. Sekiguchi, A. B. Granovsky, C. A. Ross, H. Takagi, Y. Nakamura, H. Uchida, and M. Inoue, Sci. Rep. 7, 7898 (2017). Copyright 2017 Nature, respectively. (d) Schematic of an inline spin-wave majority gate. Since the gate is reconfigurable, every port can serve as input ($I_n$) or output ($O$). (e) Scanning electron micrograph of an 850 nm wide inline spin-wave majority gate (by courtesy of G. Talmelli) (f) Micromagnetic simulations of the operation of an 850 nm wide spin-wave majority gate. (g) Schematic of a fan-out-enabled spin-wave majority gate and (h) demonstration of the  majority functions by micromagnetic simulations: (i) input (0,0,0); (ii) input  (0,0,$\pi$); and (iii) (0,$\pi$,0) on ports ($I_3$/$I_4$, $I_2$, $I_1$). Reproduced from A. Mahmoud, F. Vanderveken, C. Adelmann, F. Ciubotaru, S. Hamdioui, and S. Cotofana, AIP Adv. 10, 035119 (2020), with the permission of AIP Publishing.}
  \label{fig:SWMG_overview}
\end{figure*}

\subsubsection{Phase encoding: spin-wave majority gates}

Beyond the initial hybrid devices, recent work has focused on spin-wave logic gates that encode both input and output signals in spin waves. Conventional AND and OR logic gates have been demonstrated using colinear\cite{logic25,logic16} or cross junction\cite{logic19} geometries. Multivalued logic gates have also been proposed by combining phase and amplitude coding.\cite{SWmodulator2,nonboolean1} The most studied device is however the spin-wave majority gate, originally proposed by Khitun and Wang.\cite{logic1} Majority gates have recently elicited much interest due to potential reductions of circuit complexity with respect to conventional Boolean-based circuit design. It is rather natural to employ phase encoding for spin-wave majority gates since the interference of three (or any other larger odd number) input waves with phases 0 or $\pi$ generates an output wave with the phase that corresponds to the majority of the input waves. 

Spin-wave majority gates consist in general of transducers and input waveguides that provide input spin waves to the logic gate, a region where the spin waves can interfere, and an output port where the phase of the output wave is detected or transferred to an input waveguide of a subsequent gate. The input spin waves must have the same wavelength $\lambda$ and amplitude in the interference region. When the amplitudes of the three spin waves decay differently during propagation, it may be necessary to compensate for the unequal decay at the input level. For correct operation, the spin waves representing the same logic level need to be in phase at the output. This is best realized in logic gates, in which the path lengths of the three spin waves between their respective inputs and the output, $D_i$ ($i = 1,2,3$), differ only by integer multiples of $\lambda$, \emph{i.e.}~$D_i - D_j = n\times \lambda$ with $n = 0,1,2,\ldots$. Such ``resonant'' conditions are preferred since they allow for the utilization of the same input phases for all three waves. When such conditions are not met, the spin waves accumulate different phases during propagation to the output port, which need to be compensated for at the transducer or external signal level. 

Alternatively, an inverting input $I_i$ can be obtained when the path length of the corresponding spin wave, $D_i$, is extended or shortened so that the spin wave accumulates an additional phase of $\pi$ with respect to the others, \emph{i.e.}~$D_i - D_j = \left(n - \frac{1}{2}\right)\times \lambda$ with $n = 1,2,3,\ldots$. Moreover, shifting the output port by the same distance leads to an inverted output signal $\overline{\mathrm{MAJ}}$, \emph{i.e.}~to an inverted logic majority (or ``minority'') function. This indicates that inverters do not have to be distinct logic gates as in the case of CMOS but can be integrated into the majority gate design in a straightforward way.

The initial proposals of spin-wave majority gates were based on a trident-shaped (also referred to as $\Psi$-shaped) device layout [Fig.~\ref{fig:SWMG_overview}(a)].\cite{logic1,logic13,logic14,logic10} In this layout, three parallel input waveguides are combined into a single output waveguides in a region where the spin waves interfere. It should be kept in mind that the three waveguides are generally not equivalent and thus the lengths of the trident prongs must be adapted to the spin-wave wavelength and the relative phase shifts that are accumulated during propagation.\cite{logic13,logic20,logic14} Reducing the dimensions of such a structure to the nanoscale requires careful design and parameter selection to avoid strong spin-wave attenuation at the bends of the trident.\cite{logic13,logic20} As discussed above, using forward volume spin waves in devices with perpendicular magnetization can alleviate these constraints.\cite{logic20,logic14,mahmoud_fan-out_2020,kanazawa_role_2017}

The operation of a trident-shaped spin-wave majority gate has been demonstrated experimentally at the mm scale using YIG waveguides.\cite{logic2,kanazawa_role_2017} Figures~\ref{fig:SWMG_overview}(b) and \ref{fig:SWMG_overview}(c) show photographs of the devices. The device in Fig.~\ref{fig:SWMG_overview}(b) used in-plane magnetized YIG and backward volume spin waves,\cite{logic2} whereas the device in Fig.~\ref{fig:SWMG_overview}(c) operated with forward volume spin waves in out-of-plane magnetized YIG.\cite{kanazawa_role_2017} The phase of the output wave was extracted from time-domain measurements and used to assemble the full truth table of the majority function. These proof-of-concept demonstrations clearly indicate the feasibility of the approach. However, to become competitive with CMOS, these gates need to be miniaturized to the nanoscale and their throughput needs to be improved, \emph{e.g.}~by selecting different spin-wave configurations with high group velocity.

To tackle the scaling challenge, colinear (inline) designs of majority gates [Fig.~\ref{fig:SWMG_overview}(d)] have been proposed, which are more compact, more scalable, and easier to fabricate than the trident-shaped gates.\cite{logic25,logic16,ganzhorn_magnon-based_2016,zografos_exchange-driven_2017,ciubotaru_first_2018} In inline majority gates, spin-wave transducers are placed along a single straight waveguide.\cite{sato_electrical_2013} When the transducer distance $d_t$ is equal to an integer multiple of the spin-wave wavelength $\lambda$, \emph{i.e.}~$d_t = n\times \lambda$ with $n = 1,2,3,\ldots$, in-phase electrical signals at the transducers generate in-phase spin waves throughout the device, which is ideal for spin-wave interference. Snapshots of micromagnetic simulations of the steady-state magnetization dynamics in an 850 nm wide CoFeB waveguide are depicted in Fig.~\ref{fig:SWMG_overview}(f) and indicate that strong and weak majority can be clearly distinguished (red representing logic $0$, blue representing logic $1$) despite rather complex spin-wave modes and wave patterns. Based on the position of the output port, both a majority gate or, after additional propagation over $\frac{\lambda}{2}$, an inverted majority (minority) gate can be obtained. The output port can also be positioned between the input ports, which renders the design reconfigurable.\cite{ciubotaru_first_2018,talmelli_reconfigurable_2019} The operation of an inline majority gate has been recently demonstrated experimentally using CoFeB as the waveguide material and surface spin waves with high group velocity.\cite{talmelli_reconfigurable_2019,ciubotaru_first_2018} This approach has also allowed for the scaling of the waveguide width down into the sub-\textmu{}m range [see Fig.~\ref{fig:SWMG_overview}(e)].\cite{talmelli_reconfigurable_2019}

An additional advantage of inline spin-wave majority gates is the possibility of a fan-out of 2 since spin waves can travel in both directions in the waveguide.\cite{talmelli_reconfigurable_2019} The importance of fan-out for the realization of spin-wave circuits is discussed in more detail in Sec.~\ref{sec:Requirements for Spin Wave Circuit Design}. To improve the fan-out of the majority gates, a modified design has been recently proposed using forward volume spin waves in perpendicularly magnetized waveguides.\cite{mahmoud_fan-out_2020} A schematic of such a gate is depicted in Fig.~\ref{fig:SWMG_overview}(g). Again, adding distances of $\frac{\lambda}{2}$ can be used for logic inversion in specific sections of the device with the possibility to design \emph{e.g.}~inverting inputs or outputs. Micromagnetic simulations of the operation of such majority gates are shown in Fig.~\ref{fig:SWMG_overview}(h) for an excitation frequency of 9 GHz ($\lambda=\frac{2\pi}{k} = 60$ nm) and CoFeB material parameters (\emph{cf.} Tab.~\ref{table:1}).\cite{mahmoud_fan-out_2020} The snapshots of the resulting magnetization dynamics (blue representing logic $0$, red representing logic $1$) represent different sets of input phases that demonstrate that the entire majority function can be obtained. The snapshots also clearly demonstrate that shifting the output position by $\frac{\lambda}{2}$ leads to the inverted majority (minority) function. The advantage of such a gate is that it has two distinct output ports with equal spin-wave signals. Since forward volume waves can be guided around bends in the waveguide, such a design can be used to generate circuits of connected majority and minority gates. They can thus be used as ``standard cells'' for spin-wave circuits, which is the starting point of Sec.~\ref{sec:Requirements for Spin Wave Circuit Design}. It should also be noted that the device design concept can be extended to different output geometries and a fan-out $>2$.\cite{mahmoud_fan-out_2020}

\subsection{Directional spin-wave couplers}
\label{Sec_Dir_Couplers}

Directional couplers are passive devices commonly used in radio technology or photonics. They couple a defined amount of the electromagnetic power in a transmission line into a port, which allows for the use of the signal in another circuit. In magnonics, the dipolar coupling between two adjacent spin-wave conduits\cite{cross1,sadovnikov_nonlinear_2016} has been used to design and realize directional couplers for spin waves.\cite{cross,wang_realization_2019} For spin-wave computing, directional couplers can provide multiple functionalities. In the linear regime, directional couplers can act as power splitters, frequency dividers, or signal multiplexers. In the nonlinear regime, the coupling depends on the spin-wave amplitude and directional couplers can be used for amplitude normalization\cite{Dir_couplers_renorm} and the realization of logic gates.\cite{wang_realization_2019} 

\begin{figure}
\centering
  \includegraphics[width=7.8cm]{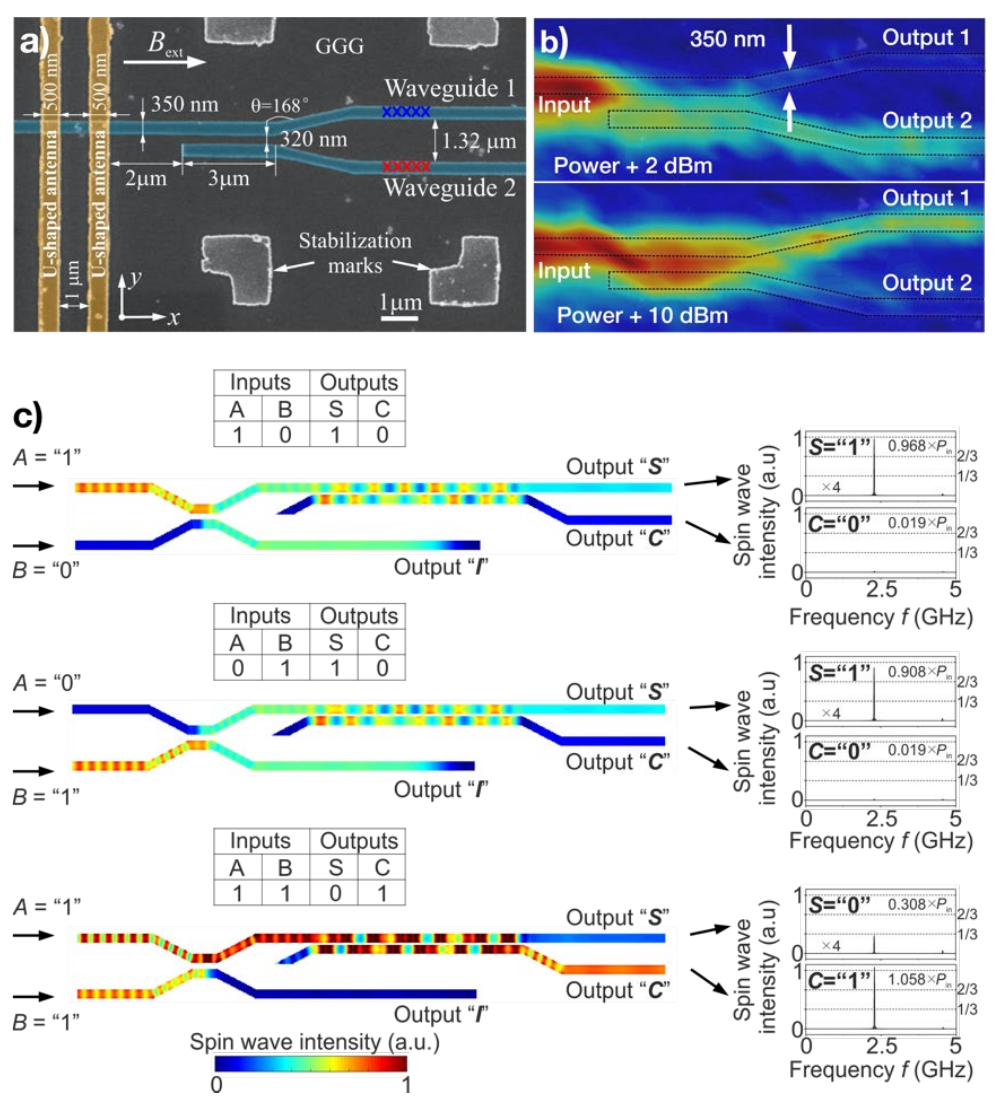}
  \caption{(a) Scanning electron micrograph of a directional coupler (shaded in blue). A small external magnetic field is applied along the YIG waveguide in the $x$-direction to saturate the directional coupler in a backward volume geometry. Reproduced with permission from Wang, M. Kewenig, M. Schneider, R. Verba, B. Heinz, M. Geilen, M. Mohseni, B. Lägel, F. Ciubotaru, C. Adelmann, C. Dubs, S. D. Cotofana,
  T. Brächer, P. Pirro, and A. V. Chumak, Nature Electron. (2020), in print. licensed under a Creative Commons Attribution (CC BY) license. (b) Nonlinear transfer characteristics of a nanoscale directional coupler. Reproduced with permission from Wang, M. Kewenig, M. Schneider, R. Verba, B. Heinz, M. Geilen, M. Mohseni, B. Lägel, F. Ciubotaru, C. Adelmann, C. Dubs, S. D. Cotofana,
  T. Brächer, P. Pirro, and A. V. Chumak, Nature Electron. (2020), in print. licensed under a Creative Commons Attribution (CC BY) license. The color maps represent the two-dimensional spin-wave intensity distributions measured by BLS microscopy for input powers of 2 dBm (top) and 10 dBm (bottom). (c) Operating principle of a magnonic half-adder: two-dimensions spin-wave intensity maps from micromagnetic simulations for different input combinations. Normalized spin-wave spectra at the output ports $S$ and $C$ are shown on the right-hand side.}
  \label{fig:Dir_Coupler}
\end{figure}

Figure~\ref{fig:Dir_Coupler}(a) depicts a scanning electron micrograph of a nanoscale (350 nm wide waveguides, 320 nm wide gap) directional coupler fabricated from an 85 nm thick YIG film. Spin waves were excited by inductive antennas and their intensity distribution in the device was mapped by BLS microscopy [Fig.~\ref{fig:Dir_Coupler}(b)].\cite{wang_realization_2019} Due to the presence of the second waveguide nearby, the spin-wave dispersion in the first waveguide splits into antisymmetric (as) and symmetric (s) modes due to the dipolar interaction between the waveguides. This results in an oscillation of the spin-wave energy between the two coupled waveguides. This means that after the propagation for a ``coupling length'', the energy of spin waves in one waveguide is completely transferred to the adjacent other. The coupling length is defined by the wavenumber of the spin-wave mode and thus strongly depends on the spin-wave dispersion. The ratio of the waveguide and coupling lengths determines the power transmission ratio and decides, into which output waveguide the spin wave is guided.\cite{cross,wang_realization_2019} Controlling the spin-wave dispersion, \emph{e.g.} by an external magnetic bias field, can lead to multifunctionality and reconfigurability of the device. 

The general transfer characteristics of directional couplers are nonlinear and therefore complimentary to the linear logic gates based on interference that were introduced above. As discussed below in Sec.~\ref{sec:Requirements for Spin Wave Circuit Design}, logic circuits require nonlinear elements. In CMOS circuits, the nonlinearity is provided by the current--voltage characteristics of the transistors themselves. Analogously, directional couplers may provide the necessary nonlinearity in spin-wave circuits. In the nonlinear regime, an increase in spin-wave amplitude results in a downward shift of the spin-wave dispersion relation and, consequently, in the change of the coupling length. Figure~\ref{fig:Dir_Coupler}(b) shows that the output spin-wave intensity strongly depends on the input microwave power: at lower excitation power (here 2 dBm), the spin-wave energy is transferred to the second waveguide, whereas a higher excitation power (10 dBm) leads to a transfer of the energy back to the first waveguide.\cite{wang_realization_2019}

The behavior of the directional couplers can be exploited to design a spin-wave half adder as an example of a simple spin-wave logic circuit that consists of two directional couplers, the first working in the linear regime and the second in the nonlinear regime [Fig.~\ref{fig:Dir_Coupler}(c)]. The functionality of the half adder has been verified by micromagnetic simulations\cite{cross} as well as by experiments.\cite{wang_realization_2019} In such devices, the data are encoded in the spin-wave amplitude. The first linear directional coupler is designed so that it divides the incoming spin-wave energy equally into two parts when spin waves are present in only one of the waveguides [top two panels in Fig.~\ref{fig:Dir_Coupler}(c)]. In this case, the second directional coupler remains in the linear regime and transfers the energy to output $S$. However, when spin waves propagate in both input waveguides, constructive interference leads to a $4\times$ stronger spin-wave intensity that is transferred entirely to the upper waveguide. In this case, the second directional coupler enters the nonlinear regime and transfers the energy to output $C$, leading to the full half adder truth table. Further details of the operation mode of directional spin-wave couplers can be found in Refs.~\onlinecite{cross,wang_realization_2019}.

\subsection{Spin-wave amplifiers and repeaters}
\label{Sec_Repeaters}

In addition to logic devices, spin-wave circuits may also require ``auxiliary'' elements, such as repeaters or amplifiers. As discussed above, spin waves have a lifetimes of ns to \textmu{}s and thus lose energy during computation or information transfer. Spin-wave amplifiers are thus crucial to compensate for such losses. Similarly, propagation losses can be compensated for by repeaters, which are devices that receive signals and retransmit them. Amplifiers and active repeaters can also provide gain in otherwise passive linear interference-based logic circuits. 

The amplification of spin-wave signals can be realized by different mechanisms. In principle, the transducer concepts discussed in Sec.~\ref{Sec_transducers} can also be used for amplification. The spin-wave signal can be enhanced by decreasing the magnetic damping in a waveguide using STT or SOT\cite{divinskiy_excitation_2018} generated by a DC current (see Sec.~\ref{Sec_STT_SOT}). Alternatively, spin waves can be amplified parametrically though a temporally periodic variation of a system parameter. For spin waves, two cases of parametric amplification can be distinguished: (i) parallel and (ii) perpendicular pumping. Perpendicular parametric pumping is often described in terms of multi-magnon (three- or four-magnon) scattering processes that are discussed in Sec.~\ref{Sec_NL_SW_phys}. This process requires the generation of large-amplitude spin waves to reach the nonlinear regime and is therefore potentially not energetically efficient for logic applications. In the case of parallel pumping, the spin-wave signal can be amplified by generating an alternating magnetic field with twice the spin-wave frequency parallel to the longitudinal component of the magnetization. This can \emph{e.g.} be realized using inductive antennas, \cite{SWA5,SWA6,SWA7,SWA8,SWA10, ciubotaru_mechanisms_2011} but also STT,\cite{SWA9} VCMA,\cite{logic24,rana_towards_2019} or magnetoelectric effects,\cite{SWA1,SWA3} which intrinsically support the coupling to the longitudinal component of the magnetization. The similarity between transducers and amplifiers has the advantage that these components do not require very different integration schemes to be embedded in the same circuit and chip.

Spin-wave repeaters are an alternative to amplifiers and can provide additional memory or clocking functionality. A schematic of a proposed repeater based on magnetoelectric transducers in combination with out-of-plane nanomagnets is depicted in Fig.~\ref{fig:interconnect_naeemi}.\cite{Interconnect4} As an alternative, the use of nanomagnets with canted magnetic anisotropy has been proposed.\cite{logic1,zografos_exchange-driven_2017} For suitably designed devices, spin waves propagating in a waveguide can switch a nanomagnet in a magnetoelectric element when synchronized electric signals are applied to the latter. Based on the orientation of the magnetization of the nanomagnet, spin waves can then be re-emitted into the waveguide by a second clock cycle. In this way, a spin-wave signal can be transferred from one stage to the next within a clock cycle. Micromagnetic simulations have indicated that the relative phase of the incoming and outgoing spin wave can be controlled. Such repeaters can compensate for losses or even provide gain, as well as regenerate and normalize spin-wave signals. This functionality is discussed in Sec.~\ref{sec:Requirements for Spin Wave Circuit Design}.

\subsection{Spin-wave multiplexers}

\begin{figure}[b]
\centering
  \includegraphics[width=8.5cm]{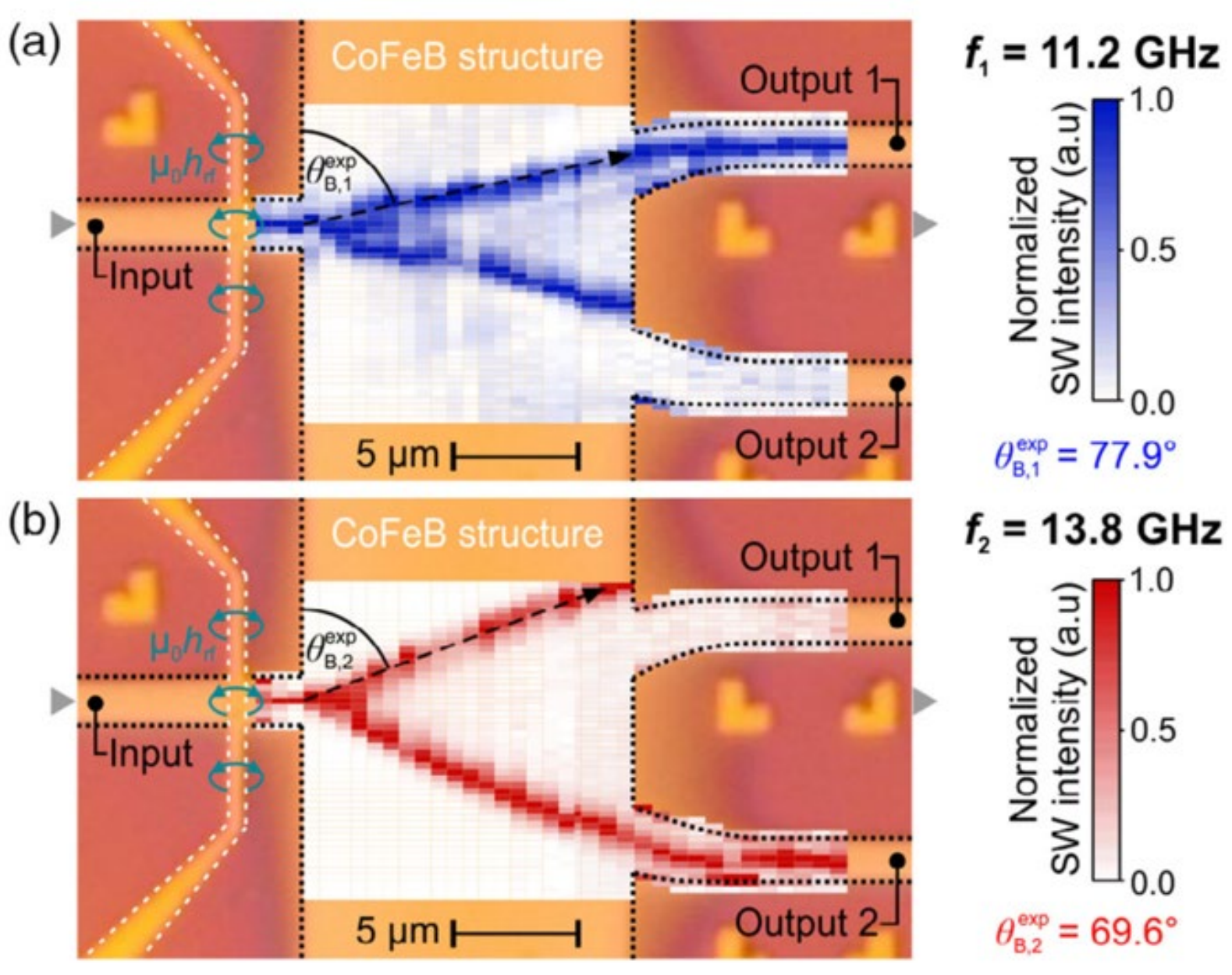}
  \caption{Device structure and experimental demonstration of spin-wave demultiplexing using caustic beams. Reproduced with permission from Heussner, G. Talmelli, M. Geilen, B. Heinz, T. Brächer, T. Meyer, F. Ciubotaru, C. Adelmann, K. Yamamoto, A. A. Serga, B. Hillebrands, and P. Pirro, Phys. Status Solidi RRL 14, 1900695 (2020). Copyright 2020 physica status solidi. The images show the distribution of the spin-wave intensity mapped by BLS microscopy for two different frequencies. a) The spin-wave intensity is guided into output 1 at 11.2 GHz and b) into output 2 at 13.8 GHz.}
  \label{fig:Multiplex}
\end{figure}

A multiplexer is a device that selects from several analog or digital input signals and forwards the chosen one to a single output line. Multiplexers are mainly used to increase the amount of data that can be sent over a network with a fixed bandwidth. Conversely, a demultiplexer is a device that disentangles a single input signal into several output signals. Parallel data transmission can \emph{e.g.} be enabled using different (spin-wave) frequencies in frequency-division multiplexing. Several approaches have been reported for the realization of a spin-wave (de-)multiplexer. A number operates by guiding spin waves into one arm of Y- or T-shaped structures by controlling the magnetization using magnetic fields,\cite{Sadovnikov_splitter_2015,Davies_mux_2015} including current-induced local magnetic field control.\cite{MUX} A drawback of these approaches is that they increase the power consumption.

By contrast, passive devices, which do not require electric currents, may offer much lower energy consumption. Two proposals for such passive (de-)multiplexers have been published to date. The first one is based on the directional spin-wave couplers\cite{cross,wang_realization_2019} discussed in the Sec.~\ref{Sec_Dir_Couplers}. The second one is based on the utilization of caustic spin-wave beams.\cite{heussner_frequency-division_2018,heussner_experimental_2020} Such caustic beams are nondiffractive spin-wave beams with stable subwavelength transverse aperture\cite{schneider_nondiffractive_2010} and are a consequence of the strong anisotropy of the spin-wave dispersion relation in in-plane magnetized films (\emph{cf.}~Sec.~\ref{Magnetization dynamics and spin waves}). In an anisotropic medium, the direction of the group velocity does not generally coincide with the direction of the phase velocity and the wavevector. For sufficiently strong anisotropy, the direction of the group velocity can become independent of the wavevector in a certain part of the spectrum. In such a case, wave packets excited with a broad (angular) spectrum of wavevectors in the specific part of the dispersion relation are channeled along the direction of the group velocity.\cite{schneider_nondiffractive_2010,heussner_frequency-division_2018,heussner_experimental_2020} These caustic beams are linear and do not interact with each other, allowing in principle for the realization of complex two-dimensional spin-wave networks in unpatterned magnetic films.

These effects have been used to route spin waves in unpatterned thin magnetic films. The direction of such beams depends on the spin-wave frequency and can be controlled by an external magnetic field. Thus, caustics can selectively transfer information encoded in spin waves. The frequency dependence of the phenomenon was successfully used to realize multiplexer and demultiplexer functionalities first by micromagnetic simulations\cite{heussner_frequency-division_2018} and recently experimentally.\cite{heussner_experimental_2020} The device concept and the operating principle are illustrated in Fig.~\ref{fig:Multiplex}. The device consists of a 30 nm thick narrow CoFeB waveguide as input and two output waveguides. In the unpatterned central part of the device, caustic beams are propagating under different angles for different spin-wave frequencies. As a result, the spin-wave intensity is transferred to different output waveguides, depending on the frequency. This behavior can be used to separate information encoded in spin waves at different frequencies in frequency-division multiplexing schemes to enhance the computational throughput. In provides an ``all-magnonic'' alternative to demultiplexing in the electric domain after detection of the complex multifrequency signal by the output transducer, leading to reduced bandwidth requirements at individual output ports. 

\section{The road from logic gates to spin-wave circuits}
\label{sec:Requirements for Spin Wave Circuit Design}

In the previous section, numerous spin-wave devices have been introduced that can be used as building blocks for spin-wave circuits. In spin-wave circuits, spin-wave logic gates are combined to calculate more complex logic functions. An example of such a more complex circuit is an arithmetic logic unit that can perform different operations on binary integer numbers, such as addition, subtraction, multiplication, or bit shift operations. 

For CMOS, the circuit design methodology has been developed for decades and highly sophisticated design and routing software tools (electronic device automation, EDA) are available to enable the very large-scale integration (VLSI, also ultra-large-scale integration, ULSI) of billions of transistors on a chip.\cite{weste_cmos_2011,lavagno_electronic_2016} Such EDA tools typically use standard cells to design (and layout) specific circuits based on their logic representations. Standard cells can provide logic (\emph{e.g.}~NAND, NOR) or memory functions (\emph{e.g.}~a flip-flop). This hierarchical design approach has been developed in the late 1970s by Mead and Conway\cite{mead_introduction_1979} and has allowed to separate technology and system development.

By contrast, few attempts to design spin-wave circuits have been made,\cite{logic1,logic8,logic16,logic25,logic10,Excitation_table_ref16,hybrid_spin_CMOS} and a methodology for spin-wave circuit design has not yet been established. While circuit design based on MAJ and INV is well understood\cite{MAJ1,MAJ2} and can be automated, the implementation of complex circuits by spin-wave logic gates and interconnects is still challenging and has not yet been demonstrated. In this section, we discuss the current understanding as well as the main hurdles on the road to spin-wave circuits, with a focus on gate interconnection, fan-out achievement, and input--output consistency. The goal of the section is to provide insight in the requirements for spin-wave devices from the viewpoint of circuit design.

Fundamental devices, such as transistors or logic gates, have to fulfill several criteria so that they can be used to design logic circuits:\cite{keyes_what_1985,miller_are_2010} 

\begin{itemize}
\item Cascadability, \emph{i.e.}~the possibility to use the output signal of a logic gate as input signal for a subsequent gate. 

\item Fan-out, \emph{i.e.}~the capability to drive several gates with an output signal of a single gate. 

\item Logic-level restoration and robust logic levels, \emph{i.e.}~the logic signals should not degrade during data transfer between individual cascaded stages in the circuit; in particular, the separation between 0 and 1 logic levels should remain large. 

\item Input/output isolation, \emph{i.e.}~the input logic signals should only physically affect the output logic signal but not \emph{vice versa}.

\end{itemize}

The combination of the above criteria are currently still a major challenge for the practical realization of spin-wave circuits. The output of a spin-wave logic gate must be capable to drive several inputs of subsequent logic gates in the circuit. In CMOS, this is achieved by representing logic values of $0$ and $1$ by voltages of $0$ and $V_{DD}$, respectively, at both the logic gate inputs and outputs. Thus, an output signal can directly drive the input of a cascaded logic gate. Since transistors provide gain, a single transistor (or logic gate) output can drive several other inputs of transistors or logic gates, providing fan-out. Moreover, in digital integrated CMOS circuits, solutions exist for communication and data exchange between gates, for power distribution, and for local and/or global synchronization via a clock signal. These functions are currently provided by the interconnect system using metal wires, with optical/photonic or plasmonic interconnects being actively researched. These interconnection as well as power and clock distribution solutions are mature and well understood from the point of view of their capabilities and the associated overhead. 

\begin{figure*}[t]
\centering
  \includegraphics[width=14cm]{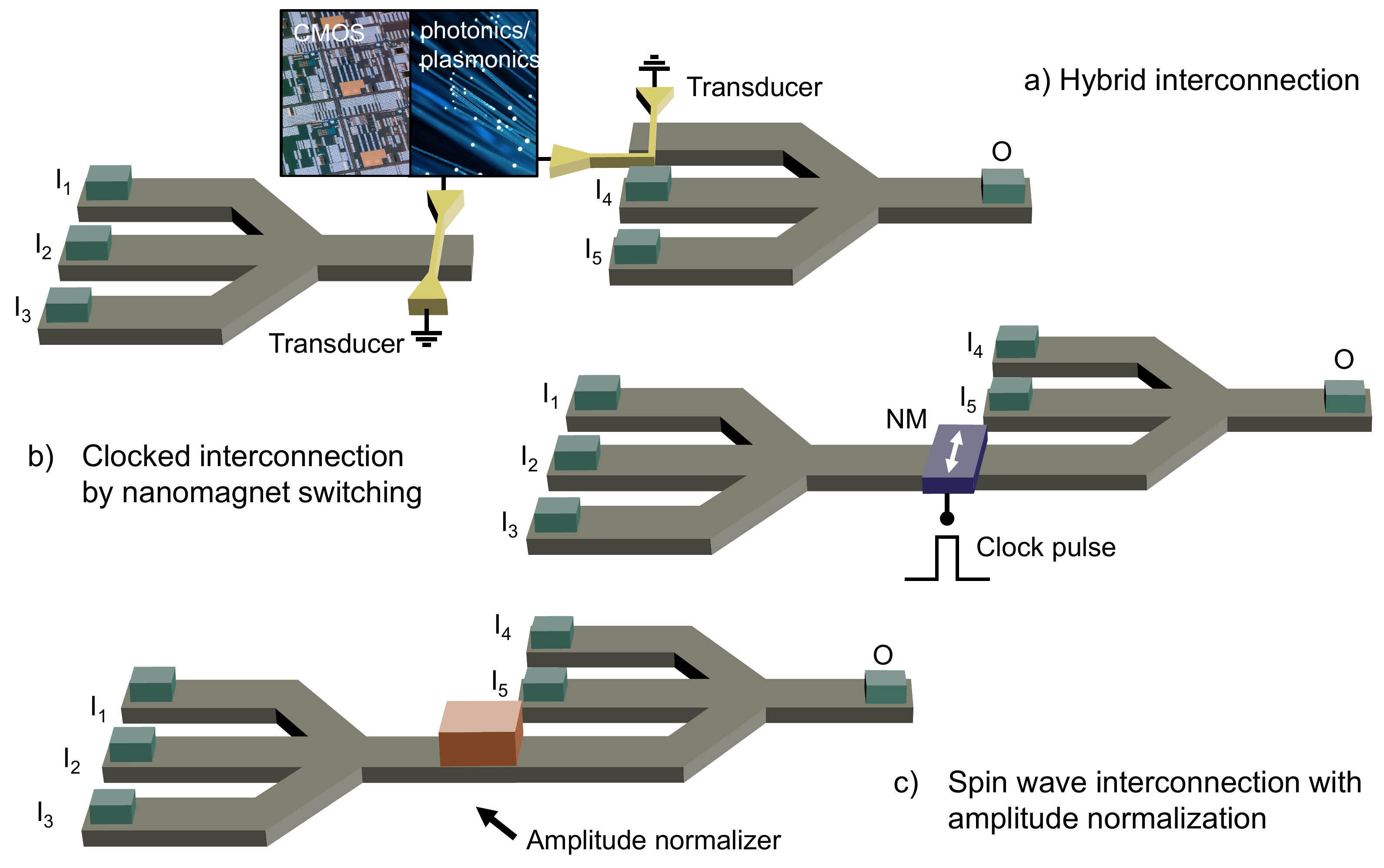}
  \caption{Spin-wave gate interconnection schemes. (a) Hybrid interconnection with signal conversion from the spin-wave to the electronic or photonic/plasmonic domain. The signal is then regenerated, transmitted to the next gate inputs by electronic or photonic/plasmonic interconnects and converted again to the spin-wave domain. (b) Clocked interconnection is possible by phase-sensitive switching of a nanomagnet (NM) by a spin wave. In a next clock cycle, a secondary spin wave is launched again from the nanomagnet with a defined relative phase. (c) All-spin-wave interconnections require a nonlinear device that normalizes the amplitude of the output spin wave. Directional spin-wave couplers can provide such functionality.}
  \label{fig:Interconnect}
\end{figure*}

Unfortunately, this is not the case for spin-wave logic gates. Straightforward cascading can be based on signal conversion between spin-wave and electronic domains at the gate level [Fig.~\ref{fig:Interconnect}(a)]. This means that the spin-wave signal at the output of a logic gate is read out by a transducer (see Sec.~\ref{Sec_transducers}), treated if needed, and converted to a spin-wave input signal of a subsequent logic gate by a second transducer. Such an approach appears mandatory for mixed-signal devices, specifically for the amplitude-level encoded gates discussed in Sec.~\ref{Sec_ALEnc_gates}. The advantage of this approach is that it fulfills all criteria. Gain can be provided after transduction in the CMOS domain, so this scheme also allows for fan-out.

In such an approach, the overhead due to signal conversion and CMOS data treatment needs to be considered carefully. A key parameter that determines the overhead is the signal level generated by the transducer. On one hand, the signal level determines the complexity of the CMOS circuit required to detect it. Signal levels of a few 100 mV may be large enough to directly drive a transistor for amplification. Lower voltages require the usage of \emph{e.g.}~sense amplifiers. Phase-sensitive detection entails even more complex circuits.\cite{egel_design_2017} These CMOS circuits consume power and occupy area and therefore contribute significantly to the overall circuit performance. While a complete benchmark of hybrid interconnection schemes has not been carried out yet, it is questionable whether such an approache can operate at sufficiently low energy to outperform the direct implementation of the desired circuit in (low-power) CMOS.

Moreover, the signal level may limit the conversion throughput. As an example, the Johnson-Nyquist voltage noise in the resistive component $R$ of a transducer (\emph{e.g.}~in an inductive antenna) is given by\cite{nyquist_thermal_1928,johnson_thermal_1928}
\begin{equation}
v_\textrm{rms} =  {\sqrt {4k_{\textrm{B}}TR\Delta f}}\, ,
\end{equation}
\noindent with $v_\textrm{rms}$ the root-mean-square noise of the voltage, $k_\textrm{B}$ the Boltzmann constant, $T$ the temperature, and $\Delta f$ the bandwidth of the measurement. For resistances $R$ of a few k$\Omega$ and a readout bandwidth of 10 GHz, the noise is about 1 mV. The signal thus should be at least (several) 10 mV to enable fast readout even with sensitive circuits. Similar arguments apply for capacitive (\emph{e.g.}~magnetoelectric) transducers. Hence, hybrid interconnection schemes may add also significant delay to the circuit.

It is therefore strongly preferred to cascade and interconnect logic gates in the spin-wave domain without conversion to electronic signals. However, additional issues arise for spin-wave logic gates using phase-encoded information. While the interference of spin waves with phases 0 or $\pi$ and amplitude $m_\mathrm{in}$ in a majority gate generates the correct output phase, the amplitude of the resulting spin wave $m_\mathrm{out}$ is different in cases of strong (fully constructive interference) and weak (partially destructive interference) majority. Concretely, if two input phases of a spin-wave majority gate are identical and the third is different (weak majority), the amplitude of the generated spin wave is $m_\mathrm{out} = m_\mathrm{in}$, whereas it is $m_\mathrm{out} = 3\times m_\mathrm{in}$ in case of strong majority, \emph{i.e.}~when all three input phases are identical. Consequently, if two majority gates are directly cascaded, amplitude differences at the output of the driving gate can lead to wrong results at the driven gate, which has been designed to operate with equal spin-wave input amplitudes $m_\mathrm{in}$. For example, if a driving gate produces a strong $0$ output, whereas the other two input signals of the driven gate are weak $1$ signal, the output of the driven gate is $0$ and not $1$ as expected. Therefore, a certain mechanism to restore or normalize the spin-wave amplitude is required between gates to guarantee proper circuit behavior. Note that since the amplitude normalization is a \emph{nonlinear} operation, it cannot be implemented using linear devices, \emph{e.g.}~based on spin-wave interference.

\begin{figure}[b]
\centering
  \includegraphics[width=8.2cm]{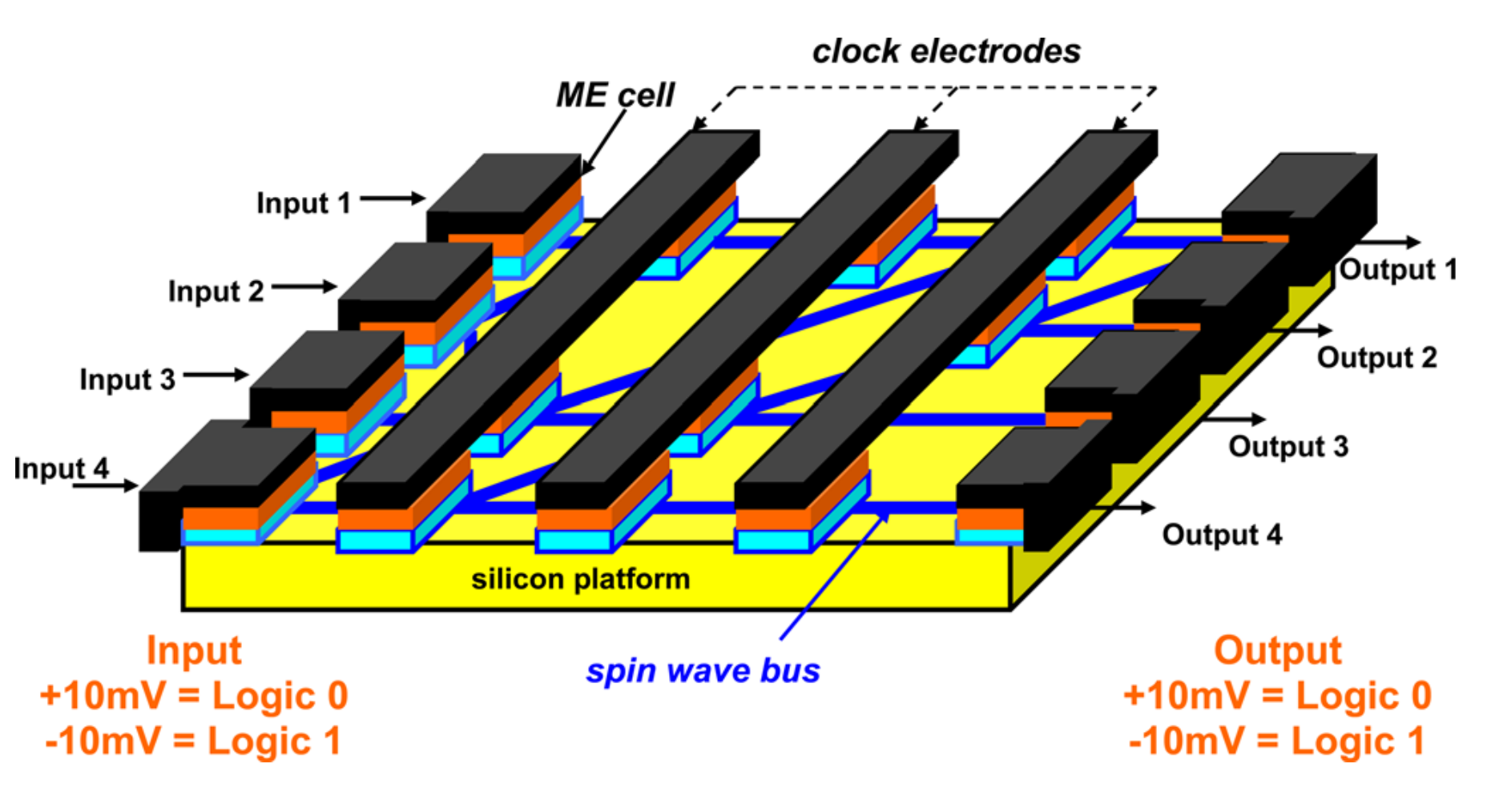}
  \caption{Schematic view of a spin-wave circuit with nanomagnet-based cascading. Spin waves propagating between nodes of the circuit switch the magnetization of bistable nanomagnets. Clock electrodes then provide trigger signals to launch spin waves from one node to the next in the following clock cycle. Reproduced from A. Khitun and K. L. Wang, J. Appl. Phys. 110, 034306 (2011), with the permission of AIP Publishing.}
  \label{fig:SW_circuit_Khitun}
\end{figure}

Two main approaches have been proposed to normalize the amplitude of a spin wave. In spintronics, an obvious nonlinear operation is the switching of a nanomagnet, which provides a threshold function. Moreover, the information storage in nanomagnets is nonvolatile, which provides a route towards nonvolatile logic circuits. This points to the usage of spin-wave repeaters (see Sec.~\ref{Sec_Repeaters}) between logic gates [Figs.~\ref{fig:Interconnect}(b) and \ref{fig:SW_circuit_Khitun}] that can both normalize and restore spin-wave signals. Repeaters can also compensate for propagation losses and provide gain as well as fan-out. Different repeater concepts have been proposed based on canted nanomagnets\cite{logic1,radu_spintronic_2015,zografos_exchange-driven_2017} or magnetoelectric elements with perpendicular anisotropy,\cite{Interconnect4} which both can provide phase-sensitive amplitude normalization and spin-wave signal restoration. In this approach, an incoming spin wave switches the orientation of a nanomagnet depending on its phase, as demonstrated by micromagnetic simulations.\cite{logic1,Interconnect4,zografos_exchange-driven_2017} In the next clock cycle, an electric (pulse) signal relaunches a spin wave from the repeater that is in phase with the initial spin wave. Such schemes require however low gate granularity and complex clocking schemes and the operation of the entire circuit may thus last multiple clock cycles, determined by the longest path in the circuit. This means that during every clock cycle, only one gate gate result can be evaluated, while \emph{e.g.}~current CMOS logic processors employing instruction-level parallelism can execute several full operations per clock cycle. Enhancing the throughput of spin-wave circuits can be achieved by \emph{e.g.} frequency-division multiplexing or pipelining.\cite{logic1,Pipeline} Yet, the energy and delay overhead of such cascading schemes may still be significant. To date, no circuit simulation of such a scheme has been reported and future work is thus required to assess its competitiveness with respect to CMOS. In addition, the switching of a nanomagnet by a spin wave has not been experimentally demonstrated yet, in particular not with phase sensitivity.

Recently, an alternative method of signal normalization has been proposed using directional spin-wave couplers (see Sec.~\ref{Sec_Dir_Couplers}).\cite{Dir_couplers_renorm} Directional couplers operate based on nonlinear spin-wave interactions and can be designed to couple a spin wave with a certain amplitude (\emph{i.e.}~a normalized amplitude $m_\mathrm{in}$) into an adjacent waveguide, independent of the amplitude of the propagating spin wave. As demonstrated by micromagnetic simulations,\cite{Dir_couplers_renorm} this allows for ``passive'' spin-wave amplitude normalization without the need to switch nanomagnets and for clocked signal repetition. 

Yet, approaches to connect spin-wave gates by means of waveguides, repeaters, or directional couplers may still add substantial overhead to the circuit since spin waves propagate rather slowly through waveguides. While the actual gate interconnection delay is circuit dependent, it is in any case much longer than that of metallic or optical interconnects. Indicative numbers for spin-wave group velocities can be found in Tab.~\ref{table:1}. The propagation delay is typically a few 100 ps/\textmu{}m, which can add significant delays for large circuits and impedes the utilization of waveguides and repeaters for long range interconnects. Moreover, when a spin wave propagates along a waveguide, its amplitude is attenuated due to Gilbert damping, which may affect the next logic gate if the amplitude is much lower than the expected value of $m_\mathrm{in}$. This may require the utilization of spin-wave amplifiers or repeaters (Sec.~\ref{Sec_Repeaters}) to compensate for losses with added energy and delay overhead. Hence, all these schemes rely on the availability of a variety of energy-efficient and fast spin-wave devices beyond the logic gates themselves. However, the granularity of the signal conversion, amplification, or repetition is still crucial for the performance of the spin-wave circuit. Cascading in the CMOS domain or by switching nanomagnets entails a granularity at the logic gate level. By contrast, directional couplers may increase the granularity to dimensions comparable to the spin-wave attenuation length with much less associated overhead \emph{e.g.}~from clocking circuits. 

\begin{figure}
\centering
  \includegraphics[width=6.5cm]{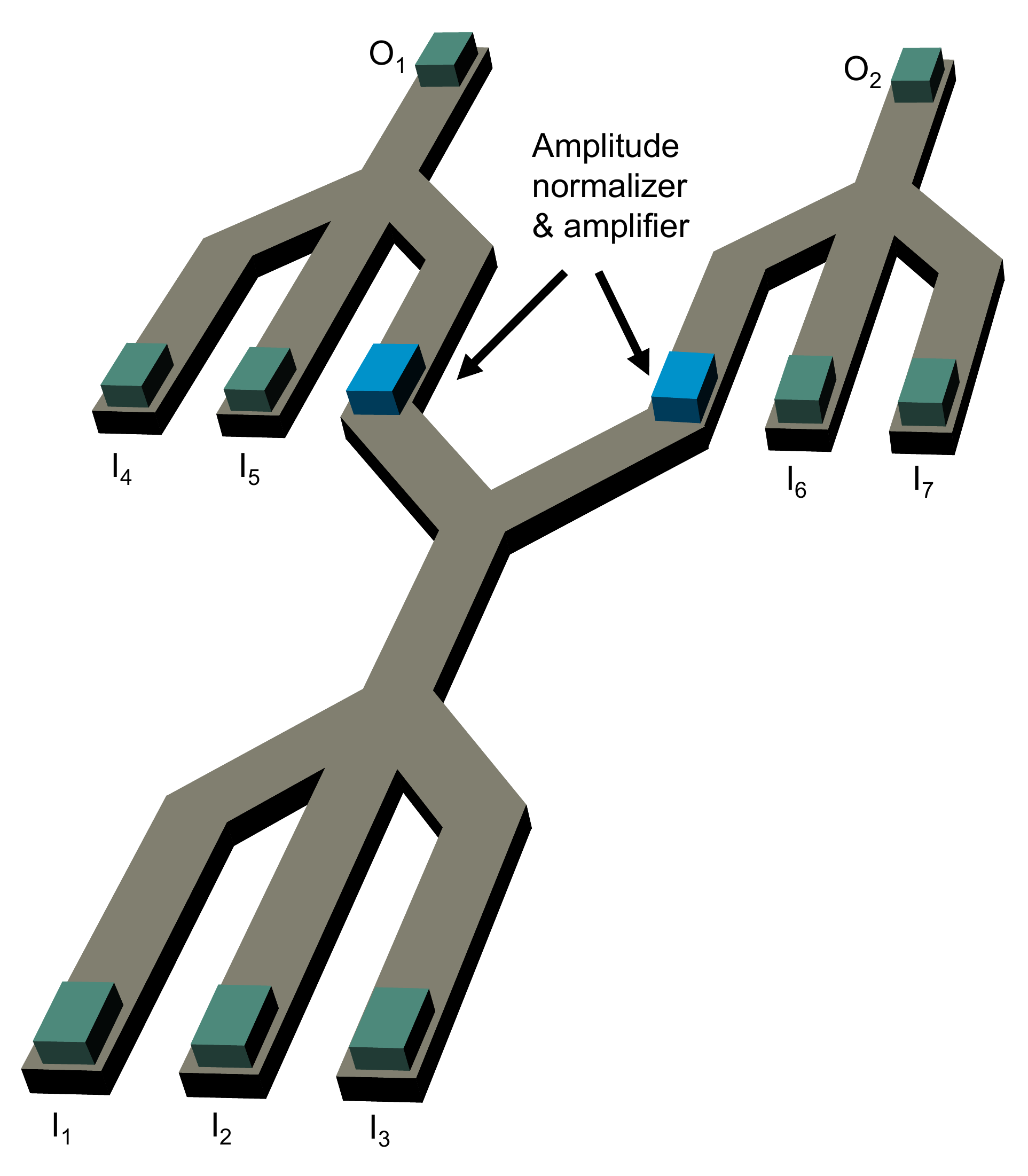}
  \caption{Schematic of cascaded spin-wave majority gates with a fan-out of 2. Amplitude normalizers and amplifiers are required at the inputs of the secondary majority gates.}
  \label{fig:fanout}
\end{figure}

Apart from cascadability, circuits require gate fan-out since one gate output signal is often used as input signal for more than one gate, as illustrated in Fig.~\ref{fig:fanout}. In CMOS, fan-out achievement is straightforward due to the inherent gain of CMOS transistors. Thus, the output voltage of a logic gate can be directly fed into several inputs by metallic wires. By contrast, achieving fan-out in spin-wave circuits is less straightforward as it requires replication of the spin-wave signal. Signal division can be achieved using Y-shaped waveguides (as in Fig.~\ref{fig:fanout}) or directional couplers. However, since the spin-wave energy (intensity) is conserved, splitting a spin wave reduces the amplitude of the two resulting spin waves by $\frac{1}{\sqrt{2}}$ even without additional losses. This needs be compensated for by spin-wave amplifiers (Fig.~\ref{fig:fanout}) with additional energy and possibly delay overhead. In contrast, fan-out enabled majority gates [Fig.~\ref{fig:SWMG_overview}(g)] provide two equivalent outputs without the need to split the spin wave after computation.\cite{mahmoud_fan-out_2020,fanout10} An alternative is the replication of the logic gate or the subcircuit itself to provide two (or more) identical outputs for the realization of fan-out. However, for large circuits, this leads to considerable area and energy overhead. As an example, if the output of a 32-bit adder is required at the input of two or more gates, the entire 32-bit adder needs be replicated twice or more. 

Beyond spin-wave circuits designed by majority-gate and inverter synthesis, the computation with waves opens other possibilities for circuit design, in particular for network-like circuits, such as reconfigurable meshes,\cite{miller_parallel_1993} cellular nonlinear networks,\cite{chua_cellular_1988,chua_cnn_1993} or systolic arrays.\cite{petkov_systolic_1992} These approaches can enable parallel computing using specific algorithms and bridge the gap to neuromorphic computing schemes. Different spin-wave-based implementations have been proposed. \cite{logic1,logic8,logic7,logic10,eshaghian-wilner_emulation_2007, khitun_magnetic_2010} Such circuits can be represented by a set of nodes, \emph{e.g} spin-wave repeaters, connected by a network of waveguides, as represented in Fig.~\ref{fig:SW_circuit_Khitun}. A discussion of such computing architectures is beyond the scope of this tutorial. More details can be found \emph{e.g.}~in Refs.~\onlinecite{sharp_1992,dogaru_2008,bobda_hartenstein_2010}. To date, none of these computing architectures has been experimentally realized. A major obstacle is the rather strong spin-wave attenuation in many magnetic materials that limits the maximum size of such networks, especially since spin waves may have to propagate along complex pathways.

\section{Hybrid spin-wave--CMOS systems}
\label{sec:Hybrid_Systems}

The above section has outlined potential solutions to design spin-wave circuits based on a set of basic devices, namely waveguides, majority gates, inverters, amplitude normalizers, amplifiers, as well as transducers. The extension of such circuits to complete competitive spin-wave-based computing systems is however limited \emph{e.g.}~by the lack of high-performance long-distance interconnection or concepts for spin-wave memory elements. These limitations can be overcome by embedding spin-wave circuits in a CMOS and/or mixed signal environment, resulting in hybrid spin-wave--CMOS systems. The performance of such a system is determined by the individual performances of the spin-wave circuit, the CMOS environment, and last but not least the interdomain transducers.

\begin{figure}
\centering
  \includegraphics[width=8cm]{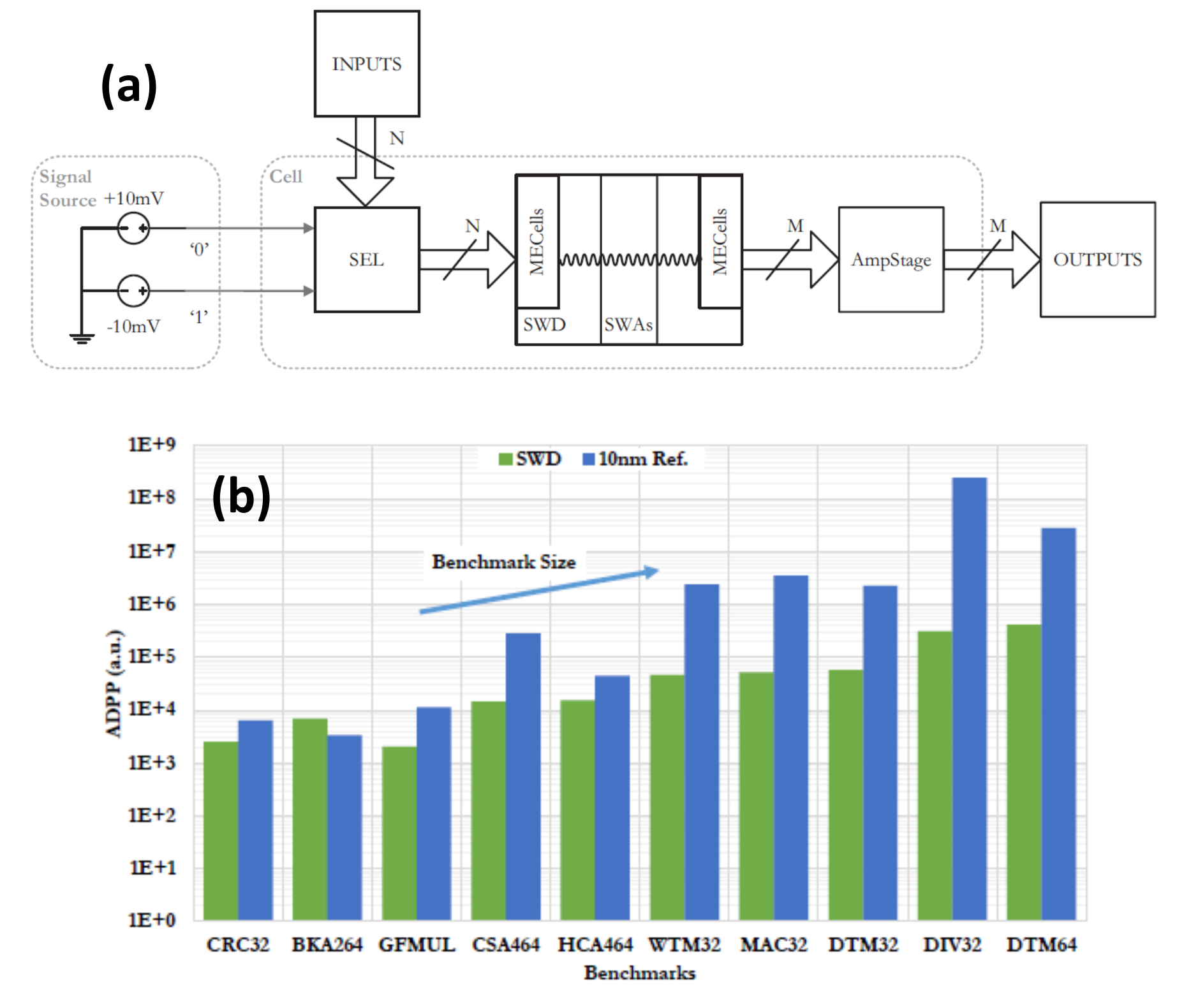}
  \caption{Benchmark of hybrid spin-wave--CMOS systems. (a) Schematic of the hybrid system. Reproduced with permission from Zografos, P. Raghavan, L. Amarù, B. Sorée, R. Lauwereins, I. Radu, D. Verkest, and A. Thean, in 2014 IEEE/ACM International Symposium on Nanoscale Architectures (NANOARCH) (2014) pp. 25–30. Copyright 2014 IEEE. (b) Area-delay-power (ADP) product of several arithmetic circuits (see text) implemented in hybrid spin-wave--CMOS technology as well as 10 nm CMOS as a reference. Reproduced with permission from O. Zografos, B. Sorée, A. Vaysset, S. Cosemans, L. Amarù, P.-E. Gaillardon,G. D. Micheli, R. Lauwereins, S. Sayan, P. Raghavan, I. P. Radu, and A. Thean, in 2015 IEEE 15th International Conference on Nanotechnology (IEEE-NANO) (2015) pp. 686–689. Copyright 2015 IEEE.}
  \label{fig:benchmark1}
\end{figure}

To date, little attention has been devoted to hybrid systems and experiments have been typically carried out using vector network analyzers or optical detection techniques like BLS. Whereas such techniques are useful for fundamental research and proof-of-concept demonstrations, they cannot be employed in real-world applications and need ultimately to be replaced by CMOS-based (mixed-signal) periphery circuits that provide input signals and analyze the output of the spin-wave circuit. It is clear that the benchmarking of spin-wave computing technology must ultimately be accomplished on complete systems including periphery, not only on the spin-wave circuit or at the device level. Although no hybrid circuit has been realized experimentally to date, a benchmark of hybrid spin-wave--CMOS arithmetic circuits has been recently performed and reported, based on the design and simulation of specific logic circuits.\cite{hybrid_spin_CMOS,Excitation_table_ref16,zografos_spin-based_2019} The benchmark suite included adders (BKA264, HCA464, CSA464), multipliers (DTM32, WTM32, DTM64, GFMUL), a multiply-and-accumulate (MAC) module, a divider (DIV32), and a cyclic-redundancy-check (CRC32) module. These logic circuits have been implemented using majority-based design approaches and layouted using majority-gate and inverter primitives (see Refs.~\onlinecite{hybrid_spin_CMOS,Excitation_table_ref16} for specific designs and device footprints). The input signals of the spin-wave circuits were synthesized using a 10-nm-CMOS-based circuit. The output signals were detected using sense amplifiers, also implemented in 10 nm CMOS. The signal conversion at the boundaries between spin-wave and CMOS domains was realized by magnetoelectric transducers. A schematic of such a system is represented in Fig.~\ref{fig:benchmark1}(a). Due to the CMOS-based periphery at the inputs and outputs of the spin-wave circuit, the system can operate with logic levels encoded in voltages and thus interact with conventional charge-based electronic systems, including memory.

For comparison, the same circuits were implemented in 10 nm CMOS using conventional EDA software. The performance of both types of circuits was then simulated by commercial software tools, using spin-wave gate delays and energies obtained from micromagnetic simulations. The performances have then be compared in terms of power, area, and delay. The results in terms of the area-delay-power product (ADPP) are depicted in Fig.~\ref{fig:benchmark1}(b).\cite{Excitation_table_ref16} Currently, no complete methodology to assess the properties and performance of spin-wave circuits and transducers is available, and thus several assumptions were made:\cite{Excitation_table_ref16} (i) the critical dimension (including the spin-wave wavelength) is 48 nm; (ii) the spin-wave excitation and detection is performed by means of magnetoelectric transducers (delay 0.42 ns, energy consumption 14.4 aJ); (iii) the delay and energy loss due to spin-wave propagation within the waveguide are negligible with respect to the overhead due to spin-wave generation and detection; and (iv) the signals provided by the magnetoelectric transducers at the output ports of the spin-wave circuit ($\sim 100$ mV) are read out using a CMOS sense amplifier (delay of 0.03 ns, energy consumption 2.7 fJ). Under such assumptions, the results in Fig.~\ref{fig:benchmark1}(b) indicate that the ADPP of hybrid spin-wave--CMOS arithmetic circuits can be significantly lower than that of their 10 nm CMOS counterparts.

Although several assumptions in the benchmark are certainly not fully realistic and actual area-delay-power product of hybrid spin-wave--CMOS can be expected to be larger, several interesting conclusions can be drawn from this exercise. As an example, the area-delay-power product for a very complex circuit such as DIV32 implemented in hybrid spin-wave--CMOS is roughly about $800\times$ lower than its CMOS implementation; individually, the power consumption is about $1800\times$ lower, the area is about $3.5\times$ smaller, whereas the delay is about $8\times$ longer. The results indicate that (i) hybrid spin-wave--CMOS circuits are promising as ultralow power circuits although at the expense of latency (delay, throughput). Nonetheless, under the above assumptions, the power-delay product may still be lower than that of 10 nm CMOS. (ii) The power dissipation in the CMOS periphery is typically much larger than in the spin-wave circuit itself. This means that the performance advantage of hybrid spin-wave--CMOS circuits typically increases with their size since the CMOS periphery overhead becomes relatively smaller. As an example, Fig.~\ref{fig:benchmark1}(b) indicates that hybrid spin-wave--CMOS implementations of large multipliers (DTM64) or dividers (DIV32) outperform CMOS in this benchmark, whereas smaller adders (BKA264, HCA464) show little to no advantage. (iii) The area of hybrid spin-wave--CMOS circuits can be competitive with CMOS circuits despite the comparatively large critical dimension of 48 nm, which is within the limits of single-exposure immersion lithography. This is due to the efficiency of the majority gate design. Again, advantages increase with the size of the spin-wave circuit since the area overhead occupied by the CMOS periphery becomes relatively smaller. 

While this first benchmark clearly indicates promise of hybrid spin-wave--CMOS circuits, the assumptions appear not yet fully realistic. Future improved benchmarking studies should include \emph{e.g.}~the material-dependent propagation delay of spin waves, as well as the overheads due to gate cascading, signal renormalization, and fan-out achievement. The computing throughput can be enhanced in principle by frequency-division multiplexing, although this also increases the overhead due to the associated multifrequency CMOS periphery and the system-level advantages are not yet clear. The availability of compact models for spin-wave devices and for transducers is essential for the accurate behavior and performance evaluation of hybrid spin-wave--CMOS circuits with a SPICE-based simulation framework.\cite{model,crosstalk} However, despite its limitations, the benchmark clearly indicates that hybrid spin-wave--CMOS systems bear promise for ultralow-power applications. Moreover, it demonstrates that future spin-wave-based technologies need to be assessed at the systems level---and not on the device or (small) circuit level. 

An open question relates to the spin-wave processing island granularity, \emph{i.e.}~the maximum complexity of a practical spin-wave circuit that can be implemented without requiring forth and back conversion to the charge (CMOS) domain. To minimize energy consumption of the full system, signal conversion between spin-wave and charge domains must be sufficiently coarse-grained (well beyond the individual gate level) and the number of transducers and sense amplifiers should be minimal. On the other hand, large spin-wave circuits require frequent signal amplification and restoration to compensate for losses due to magnetic damping and possibly dephasing. Moreover, the layout of large-scale spin-wave circuits is complicated by losses due to bent waveguides as well as the current lack of multilevel interconnects and spin-wave vias. Large circuits may thus have to be partitioned into spin-wave islands embedded in a CMOS periphery. Inside these islands, data processing is performed by cascaded spin-wave gates, whereas the islands themselves are interconnected using electric (charge, voltage) signals after conversion by (magnetoelectric) transducers. These conversion blocks can also restore the signal, reducing the need for signal restoration and amplification in the spin-wave domain. A possible length scale for the spin-wave islands could be the spin-wave attenuation length, which suggests the usage of low-damping magnetic materials. Nonetheless, the conversion blocks contribute to the overall circuit delay and the overall energy consumption,\cite{Excitation_table_ref16} and therefore the optimum spin-wave island granularity depends on the properties of both the spin-wave system as well as the conversion block, consisting of transducers and CMOS periphery. 

Finally, practical circuits require clocking schemes---a necessary evil that most computation platforms cannot properly function without. Clocking spin-wave circuits and systems can also be an important contributor to the circuit complexity and performance. For example, if the information is converted from spin wave to charge and back at the individual logic gate level, a complex clocking circuit is required to control the gate-output sampling process. A similarly complex clocking system is required for nanomagnet-based spin-wave repeaters, which require clock control of each nanomagnet node, potentially with large overheads. By contrast, if cascading can be achieved by ``passive'' spin-wave amplitude normalizers, signals need to be converted only at the island outputs, in the same way as pipeline stage outputs are sampled in a  pipelined processor.\cite{Pipeline} This substantially diminishes the clock distribution network complexity and allows for lower clock frequency and significantly reduced energy consumption. 

Another essential aspect for the energy consumption of spin-wave circuits is the operation mode. When spin waves are excited by continuous-wave microwave signals at the inputs transducers, the overall energy consumption is determined by the input power and the delay by the critical path in the island (\emph{i.e.}~the longest spin-wave propagation distance) and/or the bandwidth of the transducers as well as the read-out circuitry. Therefore, materials with fast spin-wave propagation velocity are clearly favored for waveguides and logic gates. Alternatively, spin waves can be excited by microwave pulses to reduce the energy consumption per operation. This may also allow for pipelined computation schemes using a spin-wave pulse train propagating in the circuit. However, the excitation of propagating stable wave packets by microwave pulses is not trivial due to the nonlinear spin-wave dispersion relation that can lead to pulse distortion. In addition, long spin-wave lifetimes are detrimental to the formation of short wave packets since the magnetization dynamics at the excitation site decay only slowly. Spin-wave solitons\cite{mag1,kalinikos_observation_1983,serga_self-generation_2004} due to nonlinear ``bunching'' or pulse compression effects may offer a potential solution but require specific conditions as well as high excitation power. In such cases, when short wave packets or solitons are used, the control of the propagation and the interference in the structure is challenging and requires complex clocking schemes. Although frequent spin-wave repeaters may alleviate the issue (while adding considerable overhead), the feasibility of hybrid systems built on throughput-optimized spin-wave islands realized with waveguide interconnection without repeaters is still and open issue.

\section{Spin-wave applications beyond logic gates}
\label{Sec_beyond}

\subsection{Unconventional and analog computing approaches}
\label{Sec_analog_comp}

Beyond digital spin-logic circuits and wave computing systems, spin-wave-based ``unconventional'' and analog circuits have also been proposed. A brief discussion has already been presented at the end of Sec.~\ref{sec:Requirements for Spin Wave Circuit Design}. While less universal than digital systems, these concepts take particular advantage of the wave nature of spin waves and can be very efficient for specific tasks such as signal and data processing, \cite{parallel_data_processing4, memory4, parallel_data_processing3} prime factorization,\cite{parallel_data_processing2, prime} or Fourier transforms.\cite{nonboolean2}

Pioneering work on wave-based computing in the 1970s and 1980s has used photons to develop optical computers.\cite{feitelson_optical_1988,streibl_digital_1989,miller_are_2010} While optical data communication is today ubiquitous, optical computing has not become competitive with CMOS. The challenges of optical computing overlap with those of spin-wave computing and the realization of competitive optical computers has been hindered by difficulties to confine photons at ultrasmall length scales and the power efficiency at the transducer level.\cite{miller_are_2010,miller_device_2009} Nonetheless, both digital and analog computing concepts have been developed and the work on optical computing has inspired spin-wave computing.\cite{nonboolean2}

An example for a nonbinary computing architecture is the magnonic holographic memory. It consists of a two dimensional network of crossing waveguides with transducers for spin-wave excitation and detection at the edges.\cite{memory2,memory3,memory4,gertz_magnonic_2015} After spin waves have been excited, they propagate through the structure, interfere with each other, and generate an interference pattern in the network. In such a structure, all inputs directly affect all outputs, which can be used for parallel data processing.\cite{khitun_magnetic_2010,parallel_data_processing4,nonboolean2, parallel_data_processing3,memory4} Cellular nonlinear networks are structurally similar to magnonic holographic memories and consists also of an array of magnetic waveguides.\cite{khitun_magnetic_2010} By contrast, active transducers at every waveguide crosspoint can be used to locally manipulate the magnetization. Wave superposition and interference can again be used for parallel data or image processing.\cite{image_processing3,parallel_data_processing2,parallel_data_processing4}

Spin waves can also be employed for the design of reversible logic gates \cite{rev3}. Here, both reversibility of the logic operation as well as of the physical processes are used to perform ultralow energy operations. Moreover, several spin-wave-based concepts for neuromorphic computing have been proposed.\cite{ neuromorphic_computing, survey3,  ANN,khitun_magnetic_2010,macia_spin-wave_2011,memory4,arai_neural-network_2018} Finally, the asymmetric propagation and nonlinear behavior of spin waves renders them promising candidates for reservoir computing.\cite{reservoir_computing,tanaka_recent_2019,watt_reservoir_2020}

\subsection{Three-dimensional magnonics}

The spin-wave devices described in this tutorial are based on films and multilayers that are prepared by thin film deposition techniques and lithographically patterned into the desired structures. Hence, the resulting structures are all planar and two-dimensional. Recently, research to extended the planar structures into the third dimension has intensified,\cite{gubbiotti_three-dimensional_2019} and several proof-of-concept experiments have been demonstrated.\cite{sangiao_magnetic_2017,dobrovolskiy_spin-wave_2019} The fabrication of such three-dimensional structures was enabled by the recent advances in focused electron beam induced deposition (FEBID).\cite{huth_focused_2018} FEBID is a promising three-dimensional direct-write nanofabrication technique,\cite{huth_focused_2018,fischer_launching_2020} which opens prospects to building magnonic three-dimensional nanoarchitectures with complex interconnectivity and the development of novel types of human brain-inspired neuromorphic networks using spin waves. In addition, the ease of area-selective tuning of the magnetization in spin-wave conduits via their postgrowth irradiation with ions\cite{flajsman_zero-field_2020} or electrons,\cite{dobrovolskiy_tunable_2015} or the proximity to superconductors\cite{kompaniiets_long-range_2014} opens pathway to the fabrication of spin-wave circuits with graded refractive index for the steering of spin waves in curved waveguides or into the third dimension.

\subsection{Towards quantum magnonics}

One of the prominent advantages of magnonics is the possibility to exploit complex data processing concepts at room temperature. Nevertheless, in recent years, increasing attention has been devoted to the behavior of spin waves at cryogenic temperatures for two reasons. First, the physics of hybrid superconductor-ferromagnet structures provides access to fascinating new physics that may potentially be exploited for data processing or quantum computing. Second, decreasing the temperature below 100 mK leads to the freeze-out of thermal magnons, which enables experiments with single magnons. Thus, such conditions give access to quantum magnonics.

The combination of ferromagnetism and superconductivity in hybrid ferromagnet/superconductor (F/S) systems leads to emerging physical phenomena. For instance, in proximity-coupled S/F/S three-layers, a substantial reduction of the ferromagnetic resonance field is attributed to the generation of unconventional spin-triplet superconductivity.\cite{jeon_effect_2019} It has been demonstrated that coupling of spin waves in F with S results in an enhanced phase velocity of the spin waves due to the Meissner screening of AC magnetostatic stray fields by S.\cite{golovchanskiy_ferromagnetsuperconductor_2018} Several novel effects emerge for proximity-decoupled S/F hybrids in out-of-plane magnetic fields.\cite{dobrovolskiy_magnonfluxon_2019} When the S layer is in the mixed state, an external magnetic field can penetrates in the form of a lattice of Abrikosov vortices (fluxons). The stray fields emanating from the vortex cores produce a periodic modulation of the magnetic order in F, such that the S/F bilayer can be viewed as a fluxon-induced magnonic crystal. It has been shown that the Bragg scattering of spin waves on a flux lattice moving under the action of a transport current in the S layer is accompanied by Doppler shifts.\cite{dobrovolskiy_magnonfluxon_2019} An additional promising research direction is related to the experimental examination of a Cherenkov-like radiation of spin waves by fast-moving fluxons when the vortex velocity exceeds a threshold value.\cite{shekhter_vortex_2011} To prevent instability and the collapse of vortices at the velocity of required 5--15 km/s, one can use, \emph{e.g.}, superconductors with fast relaxation of disequilibrium.\cite{dobrovolskiy_ultra-fast_2020}

Hybrid systems based on superconducting circuits allow also for the engineering of quantum sensors that exploit different degrees of freedom. Quantum magnonics,\cite{lachance-quirion_entanglement-based_2020,huebl_high_2013,tabuchi_hybridizing_2014,zhang_strongly_2014,morris_strong_2017,pfirrmann_magnons_2019} which aims to control and read out single magnons, provides opportunities for advances in both the study of spin-wave physics and the development of quantum technologies. The detection of a single magnon in a millimeter-sized YIG crystal with a quantum efficiency of up to 0.71 was reported recently.\cite{lachance-quirion_entanglement-based_2020} The detection was based on the entanglement between a magnetostatic mode and the qubit, followed by a single-shot measurement of the qubit state. The strong coupling of magnons and cavity microwave photons is one of the routes towards quantum magnonics, which is intensively explored nowadays.\cite{huebl_high_2013,tabuchi_hybridizing_2014,zhang_strongly_2014,morris_strong_2017,pfirrmann_magnons_2019,kosen_microwave_2019,li_hybrid_2020}

In addition to single-magnon operations expected to be realized at mK temperatures, macroscopic quantum states such as magnon Bose-Einstein Condensates (BECs) at room temperature have also been considered as potential data carriers. The fundamental phenomenon of Bose-Einstein condensation has been observed in different systems of both real particles and quasiparticles. The condensation of real particles is achieved through temperature reduction while for quasiparticles like magnons, a mechanism of external boson injection by irradiation is required,\cite{Bozhko_BEC_2016,Serga_BEC_2014} or, as demonstrated recently, a rapid-cooling mechanism can be exploited.\cite{schneider_boseeinstein_2020} Moreover, a supercurrent in a room-temperature Bose-Einstein magnon condensate was demonstrated experimentally.\cite{bozhko_supercurrent_2016} The observation of a supercurrent confirms the phase coherence of the observed magnon condensate and may be potentially used in future magnonic devices for low-loss information transfer and processing.

\subsection{Spin-wave sensors}

The on-chip integrability and miniaturization of spin-wave devices can be also be employed for magnetic field sensing applications. CMOS compatible magnetic sensors play a crucial role in a variety of industries, including the automotive industry, biomedical applications, navigation, robotics, \emph{etc}. Especially magnetoresistive sensors, \cite{dieny_opportunities_2019,suess_topologically_2018,zheng_magnetoresistive_2019} based on anisotropic magnetoresistance, giant magnetoresistance, or tunnel magnetoresistance, have found widespread commercial application due to their high sensitivity as well as low noise and power consumption.\cite{dieny_opportunities_2019,suess_topologically_2018,zheng_magnetoresistive_2019} Recently, several pioneer investigations have been performed to explore the possibility to use spin waves for magnetic sensors.\cite{inoue_investigating_2011,talbot_electromagnetic_2015,metaxas_sensing_2015,cao_exceptional_2019,sensor3,sensor2,sensor4} In particular, magnonic crystals, periodic magnetic structures, have been proposed as sensors with very high sensitivity. \cite{inoue_investigating_2011,talbot_electromagnetic_2015,sensor3,sensor2} Magnonic crystals have also been used for the sensing of magnetic nanoparticles.\cite{metaxas_sensing_2015} Finally, magnon polaritons in PT-symmetric cavities have been proposed for sensors with very high sensitivity.\cite{cao_exceptional_2019} Such miniature sensor applications share many properties of the logic circuits discussed in this tutorial and may also strongly benefit from optimized spin-wave transducers and read-out circuitry.

\subsection{Microwave signal processing}

To date, commercial applications of ferromagnetic resonance and spin waves mainly include macroscopic tunable microwave filters, power limiters, circulators, or gyrators based on ferrite materials, especially low-damping YIG.\cite{pozar_microwave_2012,harris_modern_2012} Much research has been devoted to such devices between the 1960s and 1980s.\cite{sig1,helszajn_y_1985,ishak_magnetostatic_1988,sig2,tanbakuchi_magnetically_1989} Several devices are today commercially available, although typically for niche applications. These devices employ typically magnetic elements in the mm size range. For such large quantities of magnetic material, the microwave absorption by ferromagnetic resonance or spin waves is large, leading to efficient power conversion between electric (microwave) and magnetic domains. Reducing the amount of magnetic material in scaled devices degrades the power conversion efficiency and lead to similar issues that need to be overcome for nanoscale logic circuits. Therefore, advances in spin-wave transducer technology may additionally enable nanoscale analog microwave applications with interesting prospects for telecommunication. 

More recently, increasing interest has been devoted to magnetoelectric antennas. Conventional dipolar antennas are difficult to scale due to the large wavelength of electromagnetic waves in air\cite{McLean96,Wheeler47} and often suffer from losses due to near-field interactions with the environment.\cite{Sten01,Pozar92} Lately, an alternative antenna type based on magnetoelectric composites has been proposed,\cite{Yao15,Domann17} which consists of a piezoelectric--magnetostrictive bilayer. Applying a microwave signal to such an antenna produces an oscillating magnetic dipolar field, which acts as a source of electromagnetic radiation.\cite{Xu19,Kubena19,Nan17} The response can be enhanced by acoustic and magnetic resonances. Due to the much shorter wavelengths of acoustic and magnetic waves at microwave frequencies, magnetoelectric antennas can be more compact that conventional dipolar antennas and may require less power.\cite{Yao15,Petrov08,Manteghi14} 

\subsection{Antiferromagnetic magnonics and terahertz applications}

In recent years, antiferromagnetic spintronics have received increasing attention as an extension of established spintronic approaches based on ferromagnets or ferrimagnets.\cite{THz,jungwirth_antiferromagnetic_2016,baltz_antiferromagnetic_2018} The spin-wave frequencies in antiferromagets are in the THz range\cite{kittel_theory_1951,kampfrath_coherent_2011,bossini_macrospin_2016,grishunin_terahertz_2018} and therefore antiferromagnetic magnonics are of interest for THz applications.\cite{dhillon_2017_2017,zakeri_terahertz_2018} In principle, antiferromagnetic media may conceptually enable spin-wave logic at THz frequencies with prospects of better scalability and higher operating speed.\cite{Magnon_spintronics} However, methods of controlling and detecting magnetic excitations in antiferromagnets are only emerging.\cite{kimel_laser-induced_2004,park_spin-valve-like_2011,wadley_electrical_2016,bodnar_writing_2018} To date, logic devices utilizing antiferromagnetic spin waves have not been demonstrated yet. In particular the controlled excitation and the detection of phase-coherent THz spin waves in antiferromagnetic waveguides is still lacking, as are concepts to efficiently generate THz logic signals by CMOS circuits. Yet, if fundamental research on antiferromagnetic spintronics continues at a fast pace, spin-wave logic at THz frequency may become an interesting alternative to the GHz approaches based on ferromagnetic media.

\section{Conclusions, state of the art, and future challenges}
\label{sec:Conclusions}

The state-of-the-art of spin-wave computing has experienced tremendous advances in the last decade with several proof-of-concept realizations of key devices, such as the spin-wave majority gate.\cite{logic2,kanazawa_role_2017,talmelli_reconfigurable_2019} Much progress has been made in particular in the understanding of the properties of spin waves in nanostructures. The overview of spin-wave devices in the Secs.~\ref{Sec_transducers} and \ref{sec:General Spin Wave device structure} clearly indicates that methods to manipulate spin waves at the nanoscale are ever improving. Scaled individual spin-wave logic gates and many features of wave-based computing have been demonstrated.\cite{talmelli_reconfigurable_2019} Hence, the field of magnonics is rapidly evolving. Moreover, benchmarking studies have suggested that hybrid spin-wave--CMOS computing systems can potentially operate at much lower power than conventional CMOS circuits.

Yet, several obstacles still exist on the road towards the realization of competitive hybrid spin-wave--CMOS computing systems. In the following, we present our view on the most critical hurdles. For a number of these obstacles, potential solutions have been proposed but need to be demonstrated and properly assessed in terms of energy and delay overhead, while others have been less addressed in the research literature so far. 

\paragraph{Cascading, fan-out, and signal restoration in spin-wave circuits.} As discussed in Sec.~\ref{sec:Requirements for Spin Wave Circuit Design}, the step from individual spin-wave devices to circuits requires the possibility to cascade logic gates. Signal normalization, restoration, and fan-out achievement are critical requirements that need to be fulfilled for a practical circuit. Cascading using conventional charge-based interconnects is a possibility but the frequent transduction between spin-wave and charge domains almost certainly consumes much energy, which may and render such approaches uncompetitive. Phase sensitive switching of nanomagnets by spin waves remains to be demonstrated experimentally and the energy efficiency of real devices needs to be established. The development of compact models for spin-wave repeaters and clocked interconnects that are calibrated to experimental devices can then quantify energy and delay overheads. Similar arguments apply to cascading approaches in the spin-wave domain using directional couplers. Experimental demonstrations together with calibrated models can allow the assessment of the energy efficiency and throughput of spin-wave circuits. A first breakthrough would be the experimental demonstration of an operational spin-wave circuit based on a flexible scheme for circuit design.

\paragraph{Transducer efficiency.} A major limitation for all applications of spin waves at the nanoscale is the energy efficiency of spin-wave generation and detection. While large mm-scale antennas and magnetic waveguides can be efficient to transfer electrical energy into ferromagnetic resonance and the spin-wave system, the radiated power and the efficiency decreases with the magnetic excitation volume. Hence, energy-efficient nanoscale spin-wave transducers are still lacking. From a systems point of view, the relevant energy is the external electric energy needed to excite spin waves and not the energy of the spin waves themselves. Hence, the transducer efficiency is a key property for ultralow-power applications of spin-wave computing systems. Magnetoelectric transducers currently appear to be most promising. However, energy-efficient spin-wave excitation by magnetoelectric transducers has not been demonstrated experimentally. Moreover, research of magnetoelectric devices at the nanoscale and at GHz frequencies is only starting. The physics of the magnetoelectric coupling in nanoscale spin-wave transducers is not yet fully established and is expected to be complicated by the complex acoustic response of ``real'' nonideal devices.\cite{tierno_strain_2018} Here, a major breakthrough would be the demonstration of a scaled (or scalable) efficient spin-wave transducer based on a magnetoelectric compound material.

Efficient spin-wave detection is also still challenging. As for generation, the microwave power induced in an antenna decreases with the magnetic volume underneath. To efficiently convert the result of a spin-wave computation to a CMOS-compatible signal, the transducer should ideally generate output signals of about 100 mV. Such large signals have been typically an issue for many spintronic logic technologies. Magnetoelectric transducers may provide a potential solution but the detailed coupling of spin waves to strain and acoustic oscillations in realistic devices has not yet been studied in detail. The demonstration of $\gg 1$ mV output signals in magnetoelectric transducers would certainly be a major breakthrough for spin-wave-based computing as well as for other potential applications. 

\paragraph{Device scaling.} As mentioned above, the scaling of the magnetic volume in a spin-wave device reduces the efficiency of transducers, both for generation as well as detection. Scaling device dimensions also has repercussions on the properties of the spin waves themselves. Narrow waveguides exhibit strong internal dipolar magnetic fields due to shape anisotropy. The magnetization is thus preferentially aligned along the waveguide, which means that scaled devices typically operate with backward-volume spin waves. A distinct advantage of this geometry can be the ``self-biasing'' due to the strong anisotropy field, which does not require external magnetic bias fields. By contrast, the excitation of surface waves requires large external fields to rotate the magnetization transverse to the waveguide, which may not be practical.

Device scaling also has strong repercussions on the spin-wave group velocity. Reducing the waveguide thickness diminishes the group velocity. Smaller devices also require the utilization of backward volume spin waves with shorter wavelengths, with complex effects on the group velocity. Reaching the exchange regime can be advantageous since it reduces the anisotropy of the spin-wave dispersion and increases the group velocity. However, the high frequencies of exchange spin waves in large-$M_\mathrm{s}$ ferromagnetic materials may impose severe conditions on mixed-signal periphery circuits.

The benchmarking of hybrid spin-wave--CMOS systems has indicated that the possibility to design compact majority gates can lead to significant area gains with respect to CMOS circuits. In practice, the benchmark suggests that competitive areas can already be achieved for characteristic dimensions (\emph{i.e.}~waveguide width) of the spin-wave circuit of about 50 nm. Such dimensions have been reached experimentally recently.\cite{heinz_propagation_2020} This indicates that scaling the spin-wave wavelength and the device dimensions should not be a major roadblock. However, the scalability of spin-wave devices may be ultimately limited by other effects, such as the dipolar crosstalk or transducer efficiency.\cite{crosstalk}

\paragraph{High-throughput computation.} To date, experimental spin-wave logic gates have been operated in the frequency domain using vector network analyzers. In real applications however, the devices have to be operated in the time domain. For cascading by nanomagnets, clocking schemes enable time-domain operation but still remain to be developed and benchmarked. Moreover, input-output isolation may be a challenge for such schemes. All-spin-wave cascading schemes may require the use of spin-wave wave packets or solitons. While the time-domain response of spin-wave transmission can be studied via the Fourier transform of the spectral response, excitation, interference, dephasing, and detection of wave packets are not fully understood and remain to be studied experimentally. Electric crosstalk between transducers is a major issue for nanoscale spin-wave devices due to the low efficiency of spin-wave generation and detection. More efficient transducers may facilitate such experiments. A major breakthrough would be a time-resolved spin-wave transmission experiment with phase sensitivity. Note that high-throughput applications require single pulse operation.

\paragraph{CMOS periphery circuits.} In hybrid spin-wave--CMOS systems, spin-wave circuits are embedded in mixed-signal CMOS-based periphery circuits that provide a link with cache/memory and input/output devices. However, only very few studies have been reported on concrete periphery circuits.\cite{egel_design_2017,hybrid_spin_CMOS,Excitation_table_ref16,zografos_spin-based_2019} The design of periphery circuits is currently hindered by the lack of equivalent circuit models for spin-wave devices and transducers. The development of calibrated compact models\cite{model} for a complete set of spin-wave devices and transducers is thus a key first step towards the development of low-power periphery circuits and complete hybrid systems. This is an important \emph{conditio sine qua non} for an accurate benchmark of the performance of hybrid spin-wave--CMOS systems and ultimately for a final assessment of their potential in commercial applications.

\paragraph{New materials.} Spin-wave computing is also an interesting field for material scientists. Many spin-wave experiments have been performed using single-crystal YIG. Epitaxy of high quality YIG on Si (100) has not been demonstrated and thus YIG is incompatible with integration alongside CMOS. Ferromagnetic metals, such as CoFeB or permalloy, are routinely integrated in MRAM memory cells and are compatible with Si technology. Nonetheless, insulating ferrites remain an interesting alternative since they typically show lower losses at microwave frequencies. However, thin ferrite films with low damping that can be cointegrated with Si-based CMOS still have to be demonstrated. 

Magnetoelectric compound materials are also a fascinating research field in material science. Challenges include the combination of Pb-free high-performance piezoelectrics and ferromagnets with large magnetostriction coefficients and low damping. In particular the piezoelectric response at GHz frequencies is often limited due to dielectric and ferroelectric relaxation, although some progress has recently been reported.\cite{tierno_microwave_2018} 

The above discussion indicates that many obstacles still exist before spin-wave technology can lead to competitive computing systems. Initial benchmarking has however clearly established the promise of such a technology for ultralow-power electronics. The large-scale effort in magnonic research will certainly advance the state of the art further in the near future. Hence, one can anticipate that spin-wave circuits will become a reality in the next decade. The remaining obstacles relate to their embedding into the CMOS periphery, including transduction. This field requires close collaboration between researchers in spin-wave physics as well as device and circuit design. Physics-based compact models of spin-waves devices and transducers\cite{model} may enable circuit simulation, periphery design, and ultimately the refinement of the benchmarking procedure to embolden the promises of spin-wave technology. We hope that the present tutorial can be a keystone in establishing this collaboration and contribute to the realization of the exciting prospect of a competitive hybrid spin-wave--CMOS computing technology.

\begin{acknowledgments}

This work has been funded by the European Union’s Horizon 2020 research and innovation program within the FET-OPEN project CHIRON under grant agreement No.~801055. It has also been partially supported by imec’s industrial affiliate program on beyond-CMOS logic. F.V. acknowledges financial support from the Research Foundation –- Flanders (FWO) through grant No.~1S05719N. A.V.C. acknowledges financial support from the European Research Council within the Starting Grant No.~678309, MagnonCircuits. The authors would like to thank Odysseas Zografos (imec), Davide Tierno (imec), Giacomo Talmelli (KU Leuven, imec), Kevin Garello (imec), Hasnain Ahmad (imec), Diogo Costa (imec), Xiao Sun (imec), Iuliana Radu (imec), Inge Asselberghs (imec), Bart Sor\'ee (imec), Ian Young (Intel), Dmitri Nikonov (Intel), Philipp Pirro (TU Kaiserslautern), Burkard Hillebrands (TU Kaiserslautern), Thibaut Devolder (U Paris Sud), Silvia Matzen (U Paris Sud), Umesh Bhaskar (U Paris Sud), Madjid Anane (CNRS), Paolo Bortolotti (Thales), Matthijn Dekkers (Solmates), Thomas Aukes (Solmates), George Konstantinidis (FORTH Heraklion), Alexandru Muller (IMT Bucharest), Oleksandr Dobrovolskiy (U Vienna), and Azad Naeemi (Georgia Tech) for many valuable discussions. 

\end{acknowledgments}

\section*{Data Availability Statement}
All data that support the findings of this study are available within the article.

\bibliography{references_v3}

\end{document}